\documentclass[12pt]{article}

\usepackage{amsmath,amssymb,amsfonts,amsthm} 
\usepackage{times}
\usepackage{graphics} 
\usepackage{graphicx}            
\usepackage{color}     
\usepackage{caption}
\usepackage{enumitem} % Used to reduce itemize/enumerate spacing
\usepackage{bm}
\usepackage{mathrsfs}
\usepackage{color}
\usepackage[utf8]{inputenc}
\usepackage[T1]{fontenc}
\usepackage{sidecap}
\usepackage{float}
\usepackage{etoolbox}
\usepackage[justification=raggedright, format=plain]{caption}
\usepackage{subcaption}
\usepackage{mathtools}
\usepackage[numbers]{natbib}
\usepackage{geometry}
\usepackage{multicol}
\usepackage{booktabs}
\usepackage{siunitx}
\usepackage{tabularx}
\usepackage{wrapfig}
\usepackage{epstopdf}

\newcolumntype{C}{>{\centering\arraybackslash}X}
\newcolumntype{R}{>{\raggedleft\arraybackslash}X}
\newcolumntype{L}{>{\raggedright\arraybackslash}X}

\graphicspath{{./Pictures/}} % Specifies the directory where pictures are stored

\let\OLDthebibliography\thebibliography
\renewcommand\thebibliography[1]{
  \OLDthebibliography{#1}
  \setlength{\parskip}{0pt}
  \setlength{\itemsep}{0pt plus 0.3ex}
}

\begin{document}

% Abstract
%\begin{frontmatter}

\begin{center} 
%~\vspace{0.5in}
~\vspace{24pt}

% Keep title as \Large and authors as \large

\textbf{\Large{Motion of a finite composite cylindrical annulus comprised of nonlinear elastic solids subject to periodic shear}} \\
\vspace{48pt}

% Authors
\large{C.C.~Benjamin, M.~Myneni, A.~Muliana, K.R.~Rajagopal}

Department of Mechanical Engineering\\Texas A\&M University
\medskip

Texas A\&M University, College Station, TX, 77843, USA
\\$^*$Corresponding Author: ccbenjamin@tamu.edu, 979-862-4201
\end{center}
\vspace{36pt}
% Input the abstract
\begin{abstract}

In this paper we study the motion of a finite composite cylindrical annulus made of generalized neo-Hookean solids that is subject to periodic shear loading on the inner boundary. Such a problem has relevance to several problems of technological significance, for example blood vessels can be idealized as finite anisotropic composite cylinders.  Here, we consider the annulus to be comprised of an isotropic material, namely a generalized neo-Hookean solid and study the effects of annular thickness on the stress distribution within the annulus.  We solve the governing partial differential equations and examine the stress response for the strain hardening and strain softening cases of the generalized neo-Hookean model.  We also solve the problem of the annular region being infinite in length, that reduces the problem to a partial differential equation in only time and one spatial dimension.  When the thickness of the annulus is sufficiently large, the solutions to the problems exhibit very interesting boundary layer structure in that the norm of the strain has a large gradient in a narrow region adjacent to one of the boundaries, with the strain being relatively uniform outside the narrow region.  We also find that beyond a distance of two times the annular wall thickness from the ends of the cylinder the solutions for the infinite length cylinder match solutions for the finite length cylinder implying that end effects are not felt in most of the length of a sufficiently long annulus. 

\end{abstract}

Keywords: Generalized Neo-Hookean, Boundary Layers in Elasticity, Unsteady Motion, Axial Shearing

%\end{frontmatter}

% \linenumbers

\section{Introduction}

%Start here
\textcolor{black}{Until recently, by an elastic body one meant either a Green elastic material\cite{green1848laws}, a material that possesses a stored energy function from which the stress can be derived, or the more general Cauchy elastic material\cite{cauchy1823recherches,cauchy1828exercises} wherein the Cauchy stress is assumed to depend on the deformation gradient (Green elastic bodies are a sub-class of Cauchy elastic bodies).  Recently, based on the fundamental notion that an elastic body is incapable of dissipation, Rajagopal generalized the class of elastic bodies to include implicit relations between the stress and the deformation gradient (see Rajagopal\cite{rajagopal2003implicit,rajagopal2007elasticity,rajagopal2011conspectus} and Rajagopal and Srinivasa\cite{rajagopal2006response,rajagopal2008class}) and there has been interest in the response of such bodies as they allow one to explain phenomena which were not possible within a Cauchy elastic body (see Rajagopal\cite{rajagopal2018note} for a discussion of the same).}  
% Stop here
Early models to describe the response of rubber-like materials were the neo-Hookean (see Rivlin\cite{rivlin1948large}) and the Mooney-Rivlin models (see Mooney \cite{mooney1940theory}, Rivlin[\cite{rivlin1948large}).  
%Start Here
\textcolor{black}{Gent\cite{gent1996new} proposed a simple two constant model, wherein both the constants have clear physical underpinnings, that is capable of corroborating experimental data as well or better than most of the empirical models that were available that also invariably required not only more fitting constants but also suffered from a lack of physical interpretation (Ogden\cite{ogden1986recent}, Morgan and Pan\cite{morman1988application}, Yeoh\cite{yeoh1990characterization}). Later Pucci and Saccomandi\cite{pucci2002note} recognized that the model due to Gent was unable to fit the data generated by Treloar\cite{treloar1975physics} in the small and medium range of deformations and suggested a modification to the Gent’s model that would fit well for a more comprehensive range of deformations. After discussing some of the limitations of Gent’s model, Puglisi and Saccomandi\cite{puglisi2015gent} consider a generalization that can be used to study damage of rubber-like materials.}  
%Stop here
In both the neo-Hookean and the Mooney-Rivlin model, the material moduli are constants, and such models are incapable of describing shear softening and shear hardening. A non-linear model that can describe shear softening and shear hardening\footnote{\textcolor{black}{A paper that is relevant to the study carried out herein, from two perspectives, namely a different class of models that exhibit shear hardening and the consideration of static axial shearing is that by Horgan, Saccomandi and Sgura\cite{horgan2002two} where the problem is studied carefully from a numerical standpoint.  Also, the paper by Muliana et al.\cite{muliana2018determining} provides a constitutive relation wherein a nonlinear expression is provided for the Cauchy-Green tensor in terms of the stress which correlates very well with experiments with fewer material constants than those used in popular Cauchy elastic models.}} is the power-law model introduced by Knowles\cite{knowles1977finite}. Knowles, however, introduced the model with a different purpose in mind, namely to see if the model would lead to a prediction of non-singular strains at crack tips. Knowles found that for certain values of the power-law exponent, the equations of plane strain lose ellipticity. There have been many studies concerning the response of power-law neo-Hookean solids (see Hang and Abeyaratne\cite{hang1992cavitation}, Horgan and Saccomandi\cite{horgan2003helical}, Hou and Zhang\cite{hou1990effect}, McLeod and Rajagopal\cite{mcleod1999inhomogeneous}, Wineman\cite{wineman2005some}). 

One interesting aspect of the response of power-law neo-Hookean solids is the development of sharp boundary layers for certain values of the power-law exponent.  Rajagopal and Tao\cite{rajagopal1992inhomogeneous}, Tao, et al.\cite{tao1992circular} showed that for special values of the power-law exponent, pronounced boundary layers, namely regions of concentration of the strain, develop in the deforming body under consideration. Zhang and Rajagopal\cite{zhang1992some} studied the deformation of slabs and annular cylinders comprised of a power-law neo-Hookean solid. They studied the annular region subject to the Lagrange multiplier that enforces the constraint of incompressibility to be time varying\footnote{The Lagrange multiplier that enforces the constraint is often referred to as the “pressure”, but the Lagrange multiplier is not the mean value of the stress, which is the “mechanical pressure” (see Rajagopal \cite{rajagopal2015remarks} for a discussion of the use and misuse of the term “pressure”)}. They also found the development of boundary layers for the strain for certain values of the power-law exponent.

The problem that we study here is different in two fundamental ways, first the external traction that is applied in our study is different from that carried out by Zhang and Rajagopal\cite{zhang1992some}, we apply a longitudinal time-dependent shear, second we consider an infinite or a finite  annular cylinder being comprised of a single material or a composite.  Our studies have relevance to the motion undergone by blood vessels as well as many other technological applications. Real materials like blood vessels are inhomogeneous, and anisotropic and their inhomogeneity is by no means capable of being idealized as a cylindrically layered annular body.  In our study, we consider the idealized problem of a composite cylinder comprised of annular sub-domains that are isotropic (see Fig \ref{NH:Total}). We also assume that the body under consideration is incompressible. To start with we consider the idealized problem of the unsteady motion of an idealized infinite annulus of a generalized neo-Hookean material, comprised of a single material, which is followed by a numerical study of the unsteady deformations of a finite annular cylinder. Then, we study the unsteady longitudinal shearing of both an infinite as well as finite composite annular solid. We find several interesting solutions to the problems that we study, one being the development of boundary layers.

Boundary layers develop in Navier-Stokes fluids in external flows at high Reynolds numbers (Schlichting\cite{schlichting1968boundary}), a thin layer adjacent to a solid boundary wherein the vorticity is concentrated. In the case of flows of non-Newtonian fluids, boundary layers can develop even in internal flows and even when flows take place at very low Reynolds number, and a variety of boundary layers are possible (see Rajagopal\cite{rajagopal1995boundary}). In elasticity, boundary layers once again manifest themselves in unbounded domains (Rajagopal\cite{rajagopal1995boundary}). They are also possible if the elastic body can sufficiently shear soften or shear harden, in domains, as in the case of our current study, when the thickness of the annular region is sufficiently large.  In our case, we find for certain values of the power-law exponent, the presence of strain boundary layers, narrow regions where the norm of the strain has a large gradient, the strain being nearly constant outside of this narrow boundary layer. We also find stress boundary layers, namely narrow regions where norm of the stress has a large gradient but is nearly constant outside the boundary layer region.

Zhang and Rajagopal\cite{zhang1992some} studied the deformation of an annular cylinder to a pulsating pressure (Lagrange multiplier) and found that boundary layers develop with regard to the strain for certain values of the power-law exponent. Our study is in keeping with this study in that we also find that boundary layers develop for certain values of the power-law exponent. 

The organization of the paper is as follows. In the next section, we introduce the constitutive relation that we shall use and document the basic kinematics. In section 3, we introduce the specific deformation for the static longitudinal shear of an annular cylinder, and this is followed by a setting up the problem of the unsteady longitudinal shearing of the annular region. Section 5 is devoted to a discussion of the numerical scheme that is used to solve the problem and the solution to the problem in the case of an infinite annular cylinder. In section 6, we consider the problem of the unsteady longitudinal shearing of a finite cylinder. We notice that if the finite cylinder is sufficiently long, then the results away from the two edges of the annular region agree well with the results obtained in the case of the infinite annular cylinder. In sections 7 and 8, we study the problem of a finite composite annular region subject to unsteady longitudinal shear. A general discussion of the study is provided in the final section.

% Preliminary setup
% ==========================================
% Basic Definitions
% ==========================================

% =============================================================
% Kinematics
% =============================================================

\section{Kinematics}

Let $\kappa_R (\mathscr{B})$ be defined as the reference configuration of a body $\mathscr{B}$, and let the current configuration at time t be denoted by $\kappa_t(\mathscr{B})$.  Let $\mathbf{X} \in \kappa_R(\mathscr{B})$ denote the position of a particle $\mathscr{P} \in \mathscr{B}$ in the reference configuration and let  $\mathbf{x}\in\kappa_t\mathscr{(B)}$ denote the position of the same particle in the current configuration.  The motion of the body is given by,

\begin{equation}\label{Motion}
\mathbf{x} = \chi_R(\mathbf{X},t),
\end{equation}

\noindent
where the subscript $R$ represents a mapping from the reference configuration.  The deformation gradient is defined by,

\begin{equation}
\mathbf{F_{\kappa_R}}\coloneqq \frac{\partial\chi_R(\mathbf{X},t)}{\partial\mathbf{X}}.
\end{equation}

\noindent
We assume the motion $\chi_R$ is invertible and we drop the subscript $\kappa_R$ for notational convenience.  The right and left Cauchy-Green tensors are defined as,

\begin{gather}
 \mathbf{B} \coloneqq \mathbf{FF^T} \textrm{, \hspace{0.5cm}} \mathbf{C} \coloneqq \mathbf{F^TF} 
\end{gather}

\noindent
The Frobenius norm of the strain tensor is defined as, 

\begin{equation}
\vert\vert\mathbf{E}\vert\vert \coloneqq \sqrt{tr(\mathbf{E^TE})},
\end{equation}

\noindent
where, 

\begin{equation}
\mathbf{E} \coloneqq \frac{1}{2}(\mathbf{C-I}).
\end{equation}

% =============================================================
% Governing Equations
% =============================================================

\section{Governing equations for an infinite annular cylinder subject to longitudinal shear}

Before we study the problem of a finite composite cylindrical annulus comprised of an  elastic body subject to time dependent longitudinal shear, we shall consider a much simpler problem, that of an infinite cylinder of a single elastic body subject to time dependent shear.  The purpose of such a study is two-fold.  The governing equations in the case of an infinite cylinder reduces to partial differential equations, albeit nonlinear, that can be solved easily as a problem in one spatial dimension and time.  The solution is of interest in itself, but more importantly it can be used to test the numerical solution to the partial differential equations that govern the finite length annulus problem that will be solved computationally using the finite element method.  In the case of an infinitely long annular cylinder, it would be appropriate to consider a deformation such that

\begin{gather}\label{Motion_Zero}
r=R\textrm{, \hspace{0.5cm}}\theta = \Theta\textrm{, \hspace{0.5cm}} z = Z+f(R,t).
\end{gather}

All the kinematic quantities listed in the previous section can be calculated easily. In particular, the norm of the strain tensor which we use to represent the intensity of strain \cite{rajagopal1994boundary,rajagopal1996deformations,chen2001boundary} is given by

\begin{equation}
    \vert\vert\mathbf{E}\vert\vert = \frac{1}{2}\sqrt{(f')^4+2(f')^2}
\end{equation}
\noindent where $f'$ represents the patial derivative of $f$ with respect to R.

% =============================================================
% Constitutive Relations
% =============================================================

\subsection{Constitutive Relations}

The Cauchy stress in a generalized incompressible neo-Hookean material is given by,

\begin{equation}\label{Const:Law:A}
\mathbf{T} = -p\mathbf{I}+\mathbf{S},
\end{equation}

\noindent
where $p$ is the indeterminate part of the stress due to the constraint of incompressibility, $\mathbf{S}$ is the constitutively prescribed stress given by,

\begin{equation}\label{Const:Law}
\mathbf{S} = \Phi\mathbf{B},
\end{equation}

\noindent
where, 

\begin{equation}\label{Phi_Motion}
\Phi = \mu\left[1+\frac{b}{n}(I_1-3)\right]^{n-1},
\end{equation}

\noindent
$I_1$ is the first invariant of $\mathbf{B}$ and is given as $I_1 = tr\mathbf{B}=(f')^2+3$, and given this relationship, equation \eqref{Phi_Motion} reduces to,

\begin{equation}\label{Phi_II}
\Phi = \mu\left[1+\frac{b}{n}(f')^2\right]^{n-1},
\end{equation}

\noindent
and in equation \eqref{Phi_II} $\mu$, $b$ and $n$ are positive constants with $\mu$ representing the  shear modulus of the linearized model\footnote{Even though $\Phi$ is the shear modulus of the material, we use the word "shear modulus" for $\mu$}.  We shall consider the annular cylinder to be comprised of an incompressible generalized neo-Hookean solid.  From \eqref{Const:Law:A}, components of the Cauchy stress can be calculated as,

\begin{subequations}
\begin{align}
T_{rr} = T_{\theta\theta} & = -p+\Phi, \\
T_{zz} & = -p+(1+(f')^2)\Phi, \\ \label{Shear:Stress}
T_{rz} & = \Phi f'.
\end{align}
\end{subequations}

% =============================================================
% Momentum Balance
% =============================================================

\subsection{Balance of Linear Momentum}
From equation \eqref{Motion_Zero} we obtain the following relationships,

\begin{gather}
\frac{\partial}{\partial r} = \frac{\partial}{\partial R}-\frac{\partial f}{\partial R}\frac{\partial}{\partial Z}\textrm{, \hspace{0.75cm}} \frac{\partial}{\partial\theta}  = \frac{\partial}{\partial\Theta}\textrm{, \hspace{0.75cm}}
\frac{\partial}{\partial z} = \frac{\partial}{\partial Z}.
\end{gather}

The balance of linear momentum, in the absence of body forces, for the special form of the motion assumed, reduces to,

\begin{subequations}
\begin{align}\label{Pre_R}
0 = & \frac{\partial T_{rr}}{\partial r} = \left(\frac{\partial}{\partial R}-\frac{\partial f}{\partial R}\frac{\partial}{\partial Z}\right)(\Phi - p), \\ \label{Pre_Theta}
0 = & \frac{1}{r}\frac{\partial T_{\theta\theta}}{\partial\theta} = \frac{1}{R}\frac{\partial}{\partial\Theta}(\Phi - p), \\ \label{Pre_Z}
\rho\frac{\partial^2f}{\partial t^2} = & \frac{\partial T_{rz}}{\partial r}+\frac{T_{rz}}{r} + \frac{\partial T_{zz}}{\partial z} = \frac{1}{R}\frac{\partial}{\partial R}(R\Phi{f'})+\frac{\partial p}{\partial Z}.
\end{align}
\end{subequations}

From equation \eqref{Pre_Theta} we note that, because of axial symmetry, $p$ can only depend on $R$,$Z$ and time. Thus the $p$ assumes the form,

\begin{equation}
p = p(R,Z,t).
\end{equation}

Additionally from equation \eqref{Pre_R} we obtain the following result,

\begin{equation}\label{Pressure_One}
    \frac{\partial\Phi}{\partial R} = \frac{\partial p(R,Z,t)}{\partial R}-\frac{\partial f}{\partial R}\frac{\partial p(R,Z,t)}{\partial Z}.
\end{equation}

\noindent
The form of $\Phi$ on the left hand side of \eqref{Pressure_One} is completely known and does not depend on $Z$.  It follows that the p can be a function of $R,t$ and can at most be a linear function $Z$. In order to impose the boundedness on stress, we assume that $p$ does not depend on $Z$. We obtain the following two equations,
\begin{subequations}
\begin{align}\label{Radial}
\frac{\partial\Phi}{\partial R} & = \frac{\partial p}{\partial R}, \\ \label{Axial}
\rho\frac{\partial^2f}{\partial t^2}  & = \frac{1}{R}\frac{\partial}{\partial R}(R\Phi{f'}).
\end{align}
\end{subequations}

From equation \eqref{Radial} it can easily be shown that $p = \Phi - C_o$ thus the stresses $T_{rr}$ and $T_{\theta\theta}$ are constant.  In this paper, we consider the case when $C_0$ is zero or a constant . Under these assumptions, we note that the differential equations containing normal stresses and shear stresses are completely decoupled.
 We define the following non-dimensional quantities.

\begin{gather}
\label{non_dim_variables}
    f^*=\frac{f}{R_o-R_i} \textrm{, \hspace{0.2cm}} R^* = \frac{R}{R_o-R_i} \textrm{, \hspace{0.2cm}} t^{*} = \omega t \textrm{, \hspace{0.2cm}} A^* = \frac{A}{R_o-R_i}
\end{gather}

\noindent
where $R_o$ and $R_i$ are the outer and inner radius of the annulus respectively.  Equation \eqref{Axial} reduces to,

\begin{equation}\label{NonDimen:Motion}
\mathscr{C}^2\frac{\partial^2f^*}{\partial {t^*}^2} = \frac{1}{R^*}\frac{\partial}{\partial R^*}(R^* \Phi^*{(f^*)'}),
\end{equation}

\noindent
where,

\begin{gather}\label{NonDimen:Beta}
\Phi^* = \left[1+\frac{b}{n}((f^*)')^2\right]^{n-1} \textrm{ \hspace{0.5cm}} \text{and} \textrm{ \hspace{0.5cm}} \mathscr{C}^2 = \frac{\rho\omega^2}{\mu}(R_o-R_i)^2.
\end{gather}

The parameter $\mathscr{C}^2$ can be written as $\mathscr{C} = \omega/\omega_L$ where $\omega_L^2 = \mu/(\rho(R_o-R_i)^2)$ and is analogous to the natural frequency.  Using the following parameters, $\mu=50$ kPa, $\rho=1000$ kg/m$^3$ and $R_o=12$ mm we can vary the inner radius $R_i$ and examine the effect on $\omega_L$.

\begin{table}[H]
\centering
\begin{tabularx}{1\textwidth}{C||C}
\toprule
\multicolumn{2}{c}{\textbf{Analysis of $\omega_L$}} \\
\midrule
$R_i$ (mm) &  $\omega_L$ (kHz) \\
\hline
10      & 3.53 \\
9       & 2.37 \\
8       & 1.76 \\
3       & 0.79 \\
1.2     & 0.65 \\
\bottomrule
\end{tabularx}
\caption{Values for $\omega_L$ at a fixed outer radius and varying inner radius. The parameters $\mu=50$ kPa and $\rho=1000$ Kg/m$^3$ are held fixed.}
\label{Tab:NatFreq}
\end{table}

From table \ref{Tab:NatFreq} we can infer that $\omega$ would need to be very large for the term $\mathscr{C}^2$ to be close to one.  For the problems studied herein the frequencies are much lower than the kilo-hertz frequencies needed to see a significant effect on the stresses from the frequency of oscillation.

The exponent $n$ in the generalized neo-Hookean model represents strain hardening or strain softening behavior as given in table \ref{Tab:Behavior}.  To study response of the body, we solve \eqref{NonDimen:Motion} with a prescribed normalized shear displacement of 1 (i.e. displacement equal to the wall thickness) on the inner surface and 0 on the outer surface.  We assume that the values of the parameters $n$ and $b$ conform to the following inequality (see Knowles\cite{knowles1977finite}).

\begin{equation}\label{Knowles}
(2n-1)b>0,
\end{equation}

For this study the parameter $b$ is assumed to positive.  From equation \eqref{Knowles} we note that values of $n$ equal to or less that 0.5 violate this inequality.  For this reason we restrict our study to values of $n$ greater than 0.50.

\begin{table}[H]
\centering
\begin{tabularx}{1\textwidth}{C|C} 
\toprule
\multicolumn{2}{c}{\textbf{Range of values for exponential parameter n}} \\
\midrule
Loss of Ellipticity  & n $\leq$ 0.5 \\
\hline
Strain Softening & 0.5 < n < 1.0 \\
\hline
Neo-Hookean & n = 1.0 \\
\hline
Strain Hardening & n > 1.0 \\ 
\bottomrule
\end{tabularx}
\caption{$T_{rz}$ vs. $\kappa$ behavior for values of the exponent n.}
\label{Tab:Behavior}
\end{table}

% =============================================================
% Static Analysis
% =============================================================

\section{Static analysis}

We consider static longitudinal shear deformation governed by the equations of equilibrium given below, 

% ==================================================
% Quasi-static motion
% ==================================================

\begin{equation}\label{Quasi:Motion}
0 = \frac{\partial}{\partial R^*}(R^*\Phi^*(f^*)'),
\end{equation}

\noindent
For the classical neo-Hookean case(n=1), this equation can be solved analytically to obtain the following displacement 

% ==================================================
% Solution
% ==================================================

\begin{equation}
f^* = C_0\ln(R^*)+C_1.
\end{equation}

\noindent
Applying the boundary conditions

% ==================================================
% Boundary conditions
% ==================================================

\begin{gather}
    f^*(R^*_i) = 1 \textrm{, \hspace{0.5cm}} f^*(R^*_o) = 0,
\end{gather}

\noindent
integration constants $C_0$ and $C_1$ can be determined as follows

% ==================================================
% Integration constants
% ==================================================

\begin{gather}
    C_0=\frac{1}{\ln\left(\frac{R^*_i}{R^*_o}\right)} \textrm{, \hspace{0.5cm}}  C_1=\frac{\ln\left(\frac{1}{R^*_o}\right)}{\ln\left(\frac{R^*_i}{R^*_o}\right)}.
\end{gather}

\noindent Shear stress can now be calculated as

% ==================================================
% Shear stress
% ==================================================

\begin{equation}\label{Quasi:Soln}
T^*_{rz} = \frac{C_o}{R^*}.
\end{equation}

\begin{figure}[H]
% =================================================
% Stresses, neo-hookean
% =================================================
\begin{subfigure}{.5\linewidth}
\centering
\includegraphics[width=\textwidth]{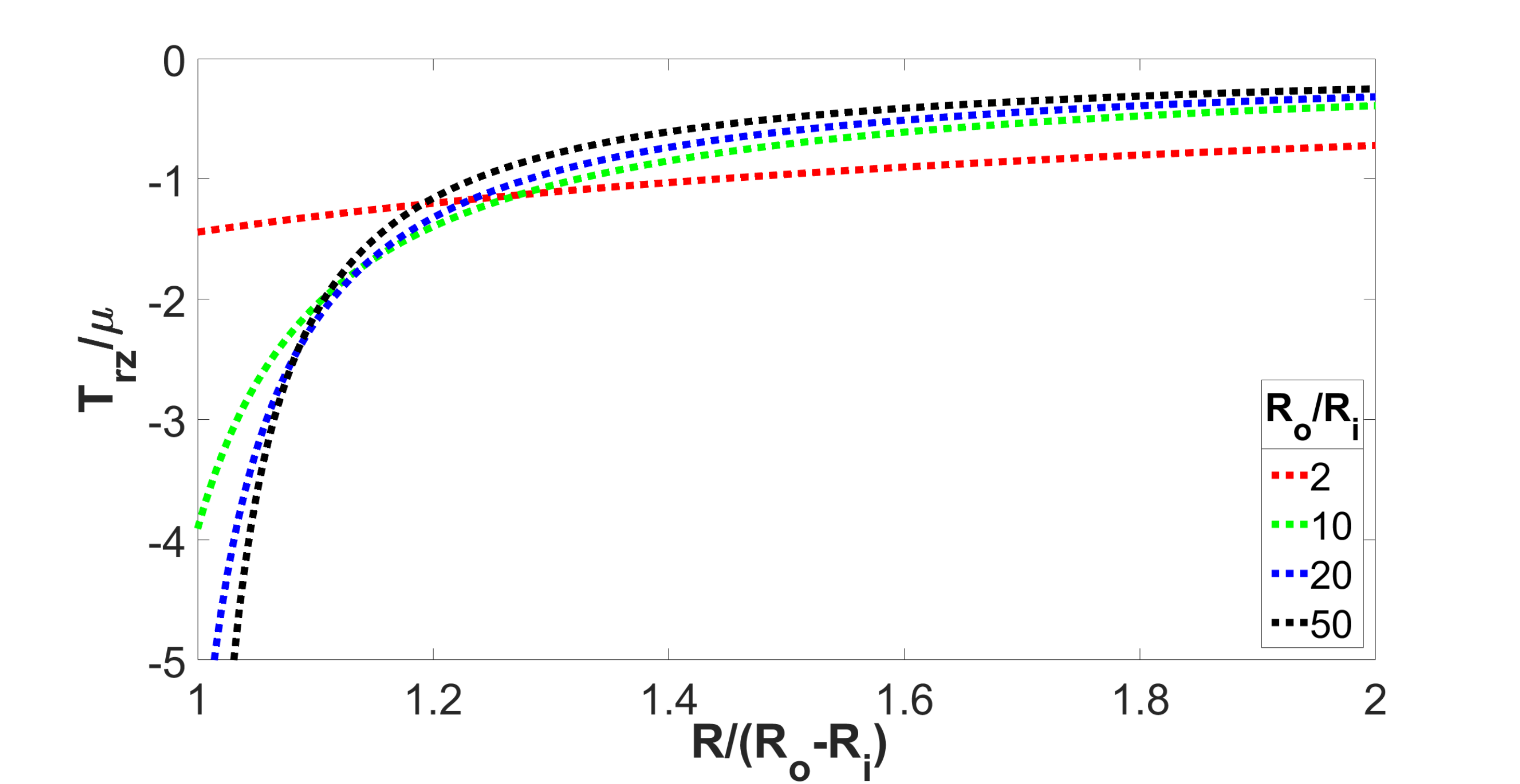}
\caption{}
\label{NH:A}
\end{subfigure} %\\[1ex]
% =================================================
% Displacements, neo-hookean
% =================================================
\begin{subfigure}{.5\linewidth}
\centering
\includegraphics[width=\textwidth]{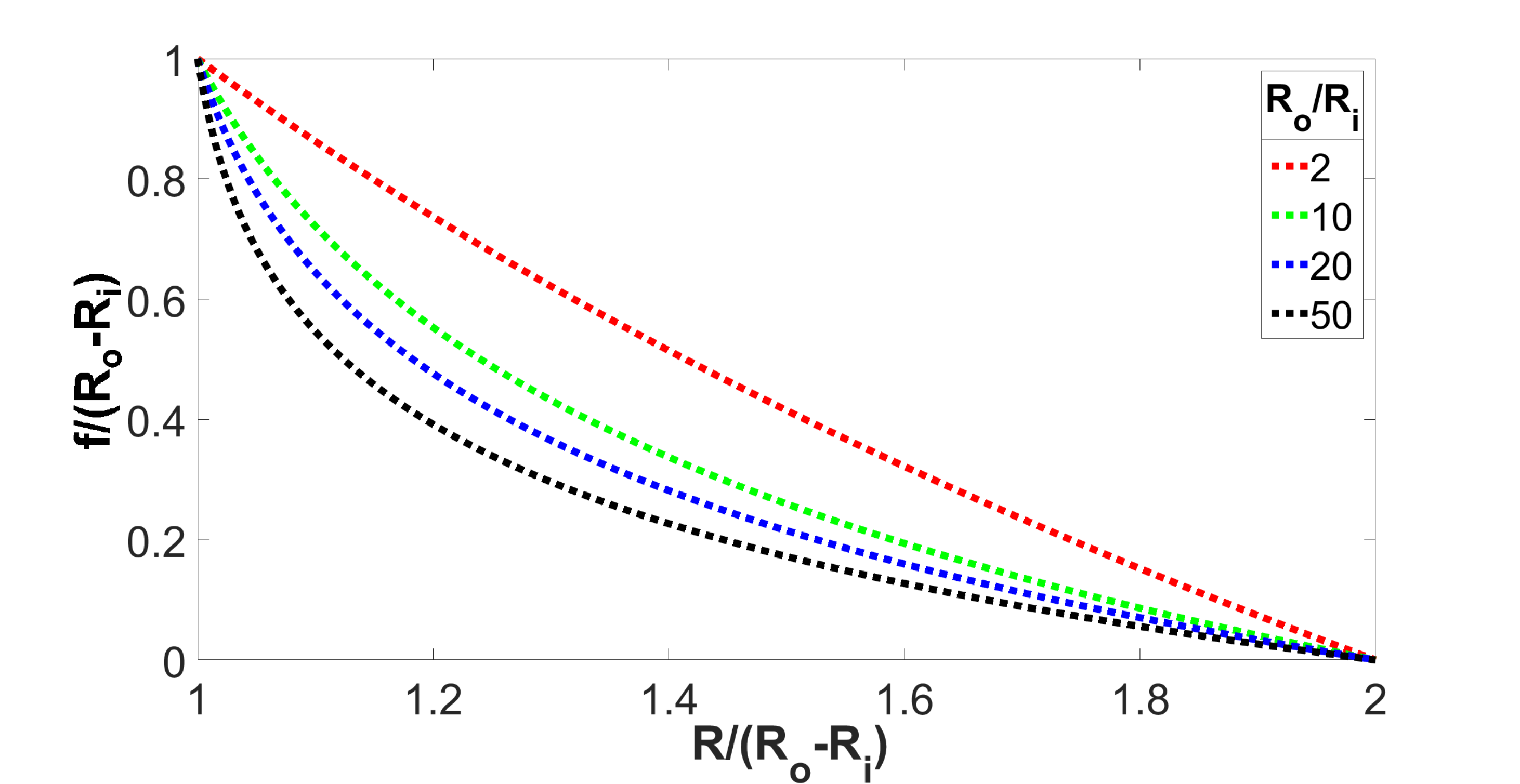}
\caption{}
\label{NH:B}
\end{subfigure}%\\[1ex]
% =================================================
% Caption for entire figure
% =================================================
\caption{\subref{NH:A}.) The shear stress is plotted for different thickness ratios (TR), \subref{NH:B}.) the displacements for the same (TR) are plotted.}
\label{NH:Total}
%\vspace{-20pt}
% =================================================
\end{figure}

\noindent
It can be observed (see figure \ref{NH:Total}) that the shear stress is only a function of the normalized radius (and the boundary conditions) for a cylinder made of neo-Hookean material.  We define the following parameter, $TR$=$R_o/R_i$\footnote{The term $TR$ stand for \emph{thickness ratio}} for the following plots.  Plotting the solution, \eqref{Quasi:Soln}, for increasing (TR) we begin to see stress boundary layers forming.

% =================================================
% =================================================

 In the case of strain softening or strain hardening the displacements are dependent on the values of the parameter $n$ and $b$ and the full ODE needs to be solved.  It follows from \eqref{Quasi:Motion} that,

\begin{equation}\label{NonDim:Static}
(f^*)'' = \frac{-(f^*)'[n+b(f^*)'^2]}{R^*(n+b(f^*)'^2[2n-1])}.
\end{equation}

% ==========================================================
% Numerical Method
% ==========================================================

\subsection{Numerical Method}

We drop asterisk from the non-dimensional quantities for convenience. We approximate the spatial derivatives using second order finite difference formulae,

\begin{equation}
\begin{split}
\frac{\partial f}{\partial R}\bigg\rvert_{R_n} & \approx \frac{f_{n+1}-f_{n-1}}{2\delta r} \\
\frac{\partial^2 f}{\partial R^2}\bigg\rvert_{R_n} & \approx \frac{f_{n+1}-2f_{n}+f_{n-1}}{\delta r^2}
\end{split}
\end{equation}

\noindent Using the boundary conditions, 

\begin{gather}
f({R_i}) = 1\textrm{, \hspace{0.5cm}} f({R_o}) = 0.
\end{gather}

\noindent We now change the differential equation into a difference equation,

\begin{equation}\label{Disc:01}
\begin{split}
f_n & = \frac{f_{n+1}+f_{n-1}}{2}+\Gamma(f) \\
\Gamma(f) & = \frac{dr^2[n\frac{f_{n+1}-f_{n-1}}{2dr}b(\frac{f_{n+1}-f_{n-1}}{2dr})^3]}{2\tilde{R}(n+b(2n-1)(\frac{f_{n+1}-f_{n-1}}{2dr})^2)}.
\end{split}
\end{equation}

In the interval ($R_i\leq R\leq R_o$), the values for $f_n$ are approximated by equation \eqref{Disc:01}.  The problem is solved iteratively until a tolerance of 10$^{-6}$ is achieved.

\subsection{Boundary layers}

\begin{figure}[H]
% =================================================
% Ri = 1.2 mm, n = 0.55
% =================================================
\begin{subfigure}{.5\linewidth}
\centering
\includegraphics[width=\textwidth]{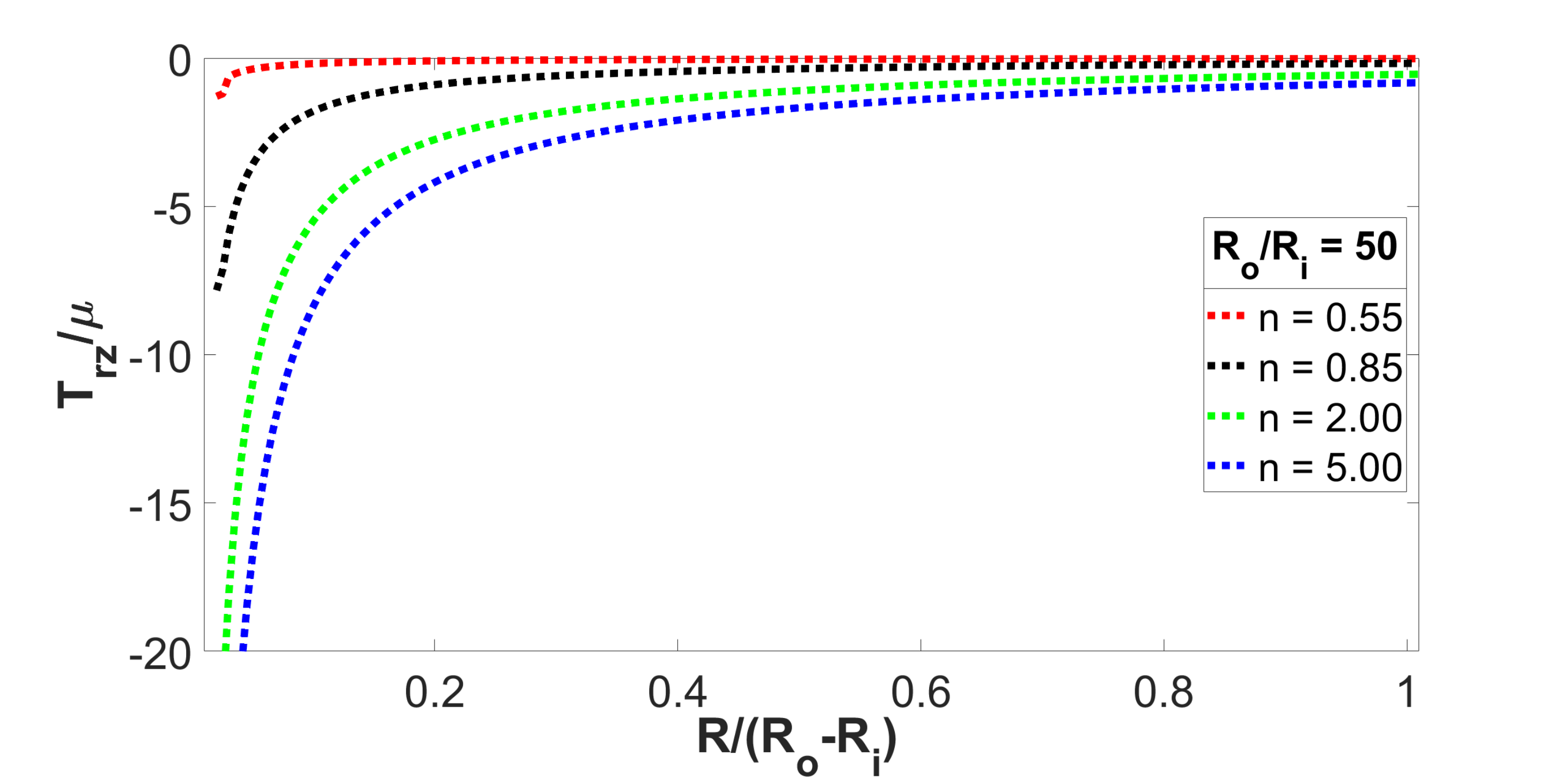}
\caption{}
\label{SD:A}
\end{subfigure} %\\[1ex]
% =================================================
% Ri = 10 mm, n = 0.55
% =================================================
\begin{subfigure}{.5\linewidth}
\centering
\includegraphics[width=\textwidth]{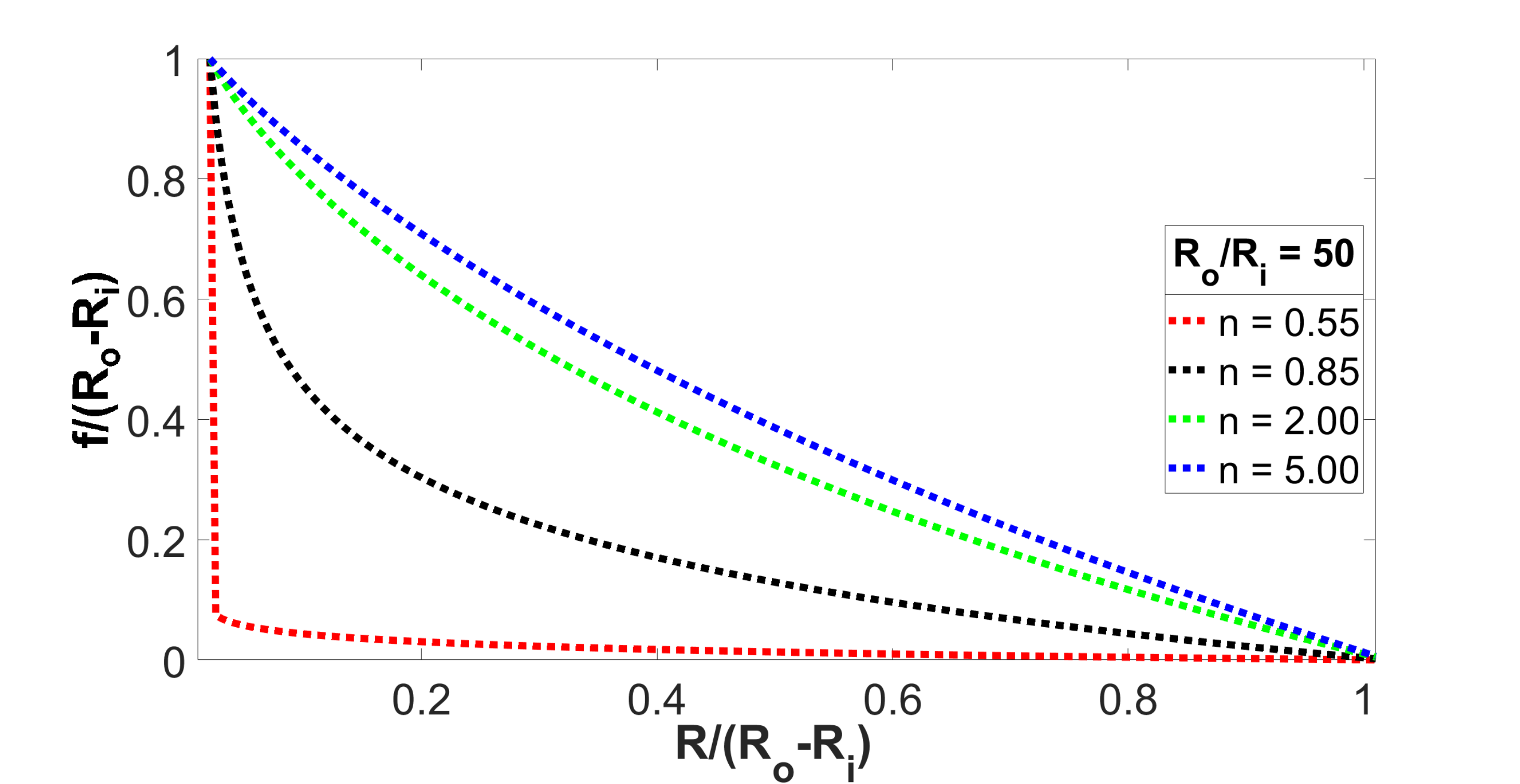}
\caption{}
\label{SD:B}
\end{subfigure}%\\[1ex]
% =================================================
% Caption for entire figure
% =================================================
\caption{\subref{SD:A}.) The shear stress for TR = 50, \subref{SD:B}.) the shear displacement.  For the numerical solution, b = 1.}
\label{SD:Total:03}
%\vspace{-20pt}
% =================================================
\end{figure}

As was noted earlier stress boundary layers could only form with large annular thicknesses (TR) (see fig \ref{NH:A}).  This behavior of the stress can be further seen in shear softening and shear hardening materials(see fig \ref{SD:Total:03}). We take a (TR) value of 50 to represent a very thick annulus and study the effects of varying the power-law parameter $n$.  

% =================================================
% These plots show f(R), Trz and kappa as a function
% of radius.  
% =================================================
% =================================================

\begin{figure}[H]
% =================================================
% displacements
% =================================================
\begin{subfigure}{.49\linewidth}
\centering
\includegraphics[width=\textwidth]{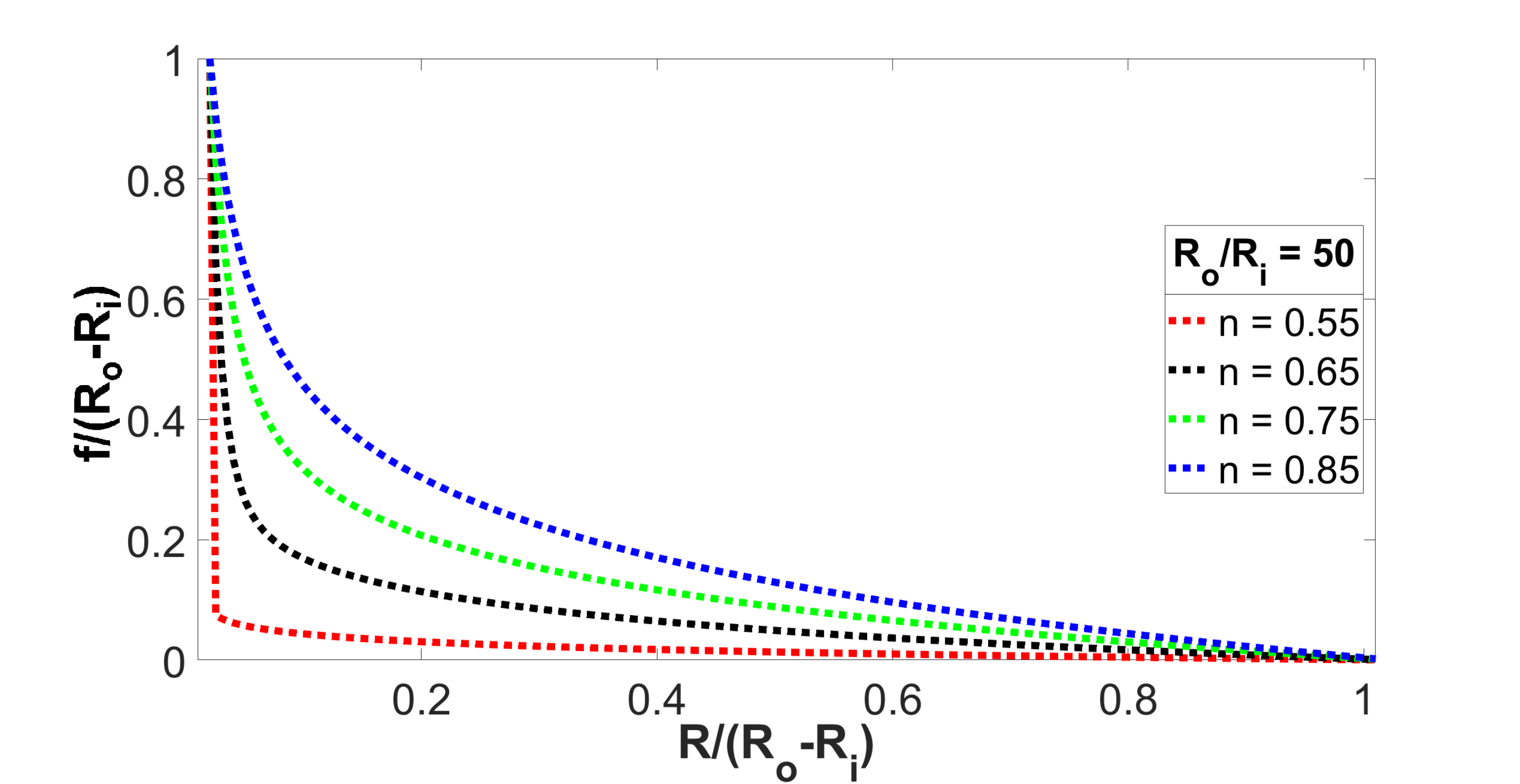}
\caption{}
\label{SD:C}
\end{subfigure} %\\[1ex]
% =================================================
% norm of strain tensor.
% =================================================
\begin{subfigure}{.49\linewidth}
\centering
\includegraphics[width=\textwidth]{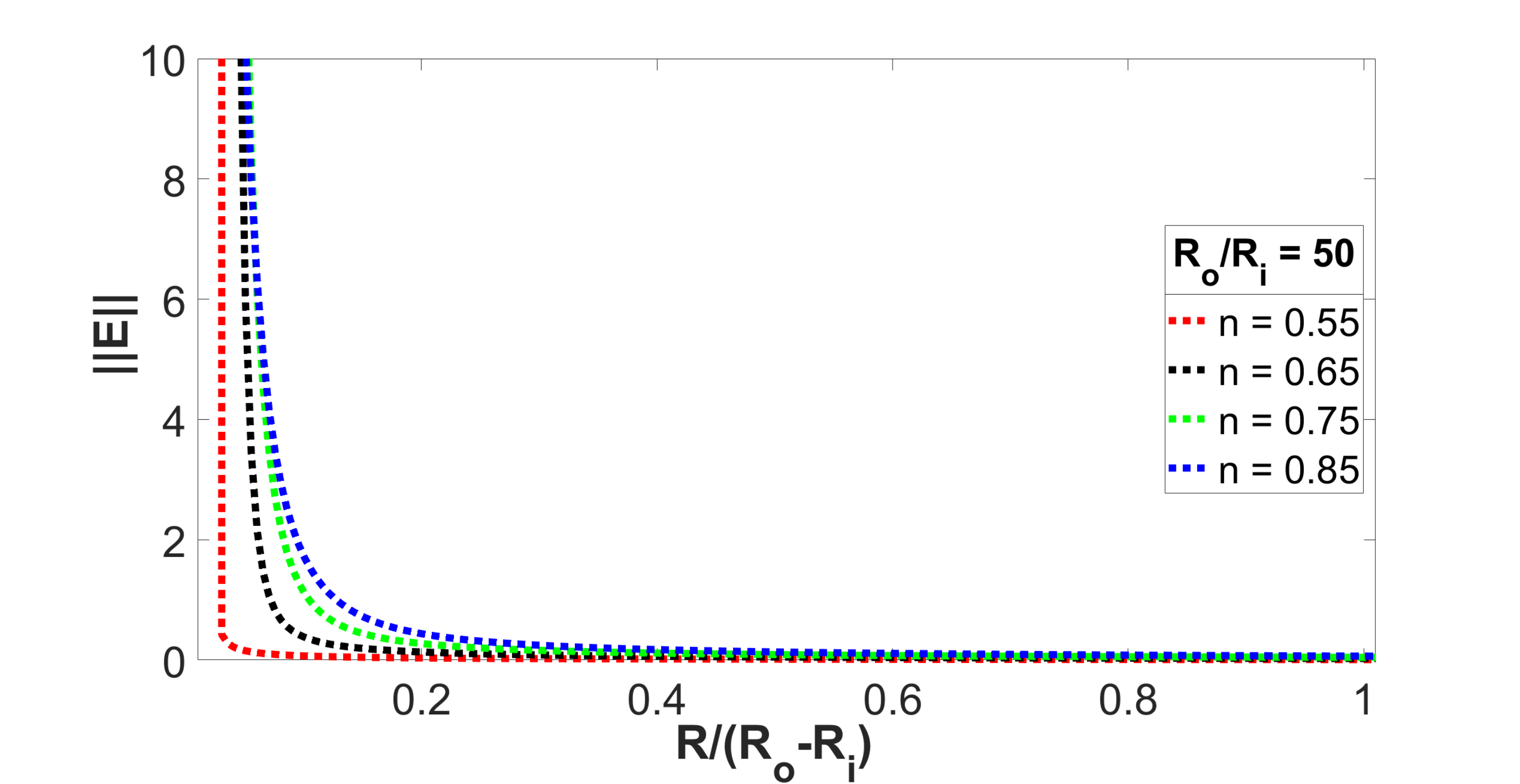}
\caption{}
\label{SD:D}
\end{subfigure}%\\[1ex]
% =================================================
% Caption for entire figure
% =================================================
\caption{In \subref{SD:C}.) the radial distribution of displacements is shown for values of n between 0.55 and 0.85.  Figure \subref{SD:D}.) gives the norm of the strain tensor for values of n ranging from 0.55 to 0.85.}
\label{SD:Total:02}
%\vspace{-20pt}
% =================================================
\end{figure}

In figure \ref{SD:C}, the displacement is plotted as a function of radius for increasing  values of n from n = 0.55 to n = 0.85.  In figure \ref{SD:D} the norm of the strain tensor $\vert\vert\mathbf{E}\vert\vert$ is plotted for the corresponding values of n.  

% =================================================
% These plots show the f(R) and the norm of the 
% strain tensor
% =================================================
% =================================================

A boundary layer type solution for displacements only occurs in the strain softening case when $n$ approaches the limiting value of 0.50.  In figure \ref{SD:D}, the norm of the strain, $\vert\vert\mathbf{E}\vert\vert$,  shows a definite boundary layer solution for the strain softening case.  

From our study of the static neo-Hookean case we observe that the stress distribution is entirely dependent on the cylinder geometry with stress boundary layers appearing only in very thick walled cylinders.  When hardening ($n$ > 1) or softening ($n$ < 1) is taken into account, we note that strain boundary layers formed for softening materials when the value of n approached the limiting value of 0.50.

% Single layer Solutions
\section{Deformation of an infinitely long cylindrical annulus due to periodic longitudinal shearing}\label{Sect:Inf01}

\label{Sect:Inf02}

We now consider the effects of inertia.  For a infinitely long cylindrical annulus, we assume the motion to be of the form \eqref{Motion_Zero}.  Boundary and initial conditions imposed in this problem are

\begin{subequations}
\begin{gather}
    f(R_i,t)=A \cos(\omega t) \textrm{, \hspace{0.5cm}} f(R_o,t)=0 \\
f(R,0)= f_{static}\vert_{f(R_i,0)=A} \textrm{, \hspace{0.5cm}} \frac{\partial f}{\partial t}(R,0)=0
\end{gather}
\end{subequations}

\noindent
The non-dimensional form of the boundary and initial conditions can be obtained using the appropriate length and time scales introduced in the previous sections(see eq.\eqref{non_dim_variables} ). $A^* = A/(R_o-R_i)$, is taken as 1 in solving the time dependent equations. Equation \eqref{NonDimen:Motion} is solved numerically using the method described in section \ref{Num:Dynamic}. 

% 
% ==================================
% Numerical Methods
% ==================================
\subsection{Numerical Method}\label{Num:Dynamic}

\begin{figure}[H]%[!]
    \centering
    \includegraphics[width=.5\textwidth]{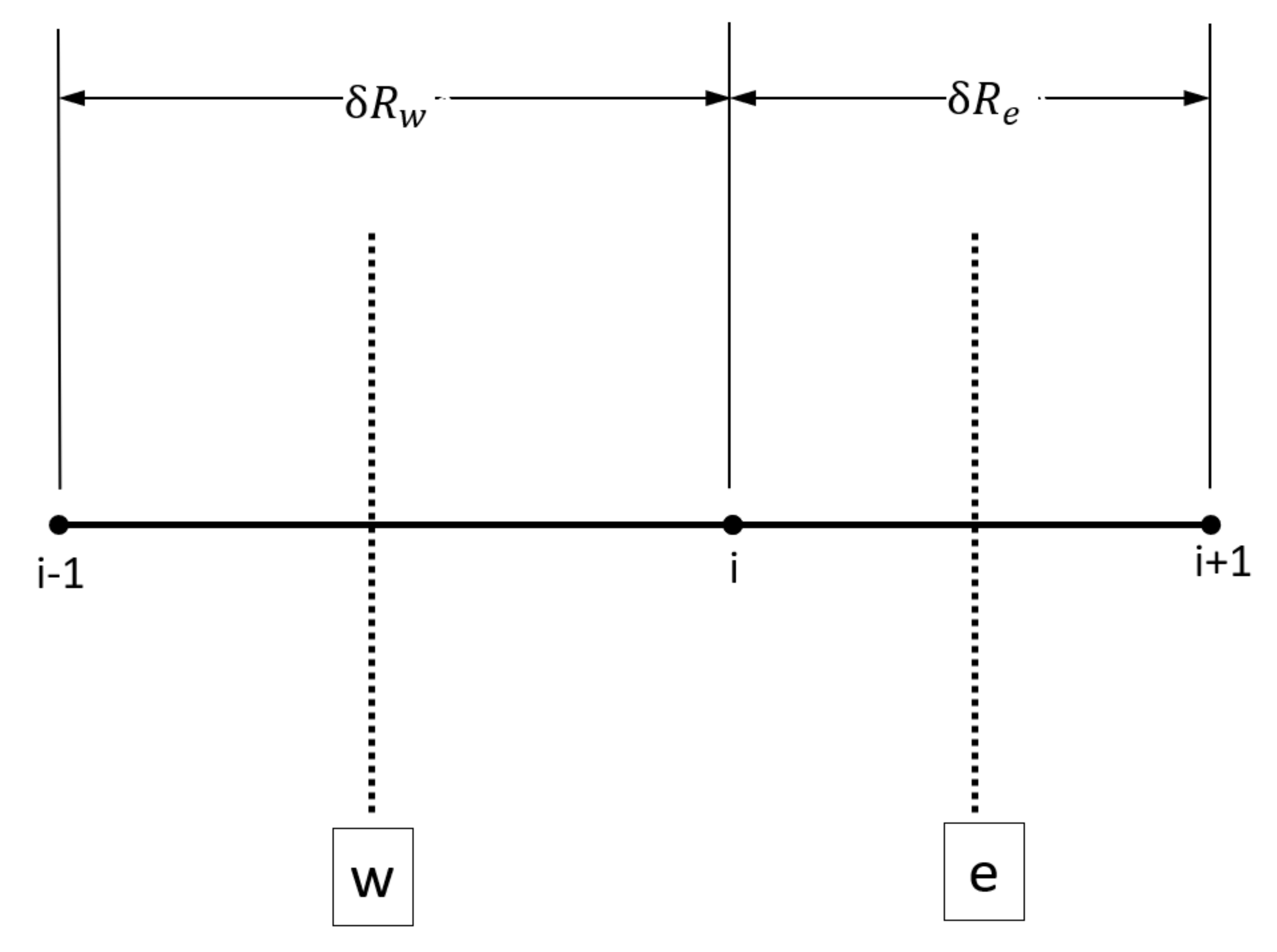}
    \caption{Computational stencil used in the finite volume method}
    \label{comp_stencil_FV_method}
\end{figure}

We used the finite volume method with implicit time stepping for the numerical solution of the equilibrium equations. The second order finite difference formula is used for the spatial derivatives and the first order backward difference formula is used for the time derivatives. Thickness of the cylindrical wall is divided into a number of intervals by specifying the number of nodes/mesh points. For a composite cylinder, the number of mesh points within each layer is specified. A finite volume containing the $i^{th}$ node is shown in the figure \ref{comp_stencil_FV_method}.  The finite volume is bounded by the east(e) and west(w) interfaces which occur at the mid points between the nodes i,i+1 and, i,i-1.

Integrating this equation over the finite volume at a time instant t gives

\begin{gather}
    \int^{R_e}_{R_w} \frac{\partial}{\partial R}\left(\omega^2_L R\Phi\frac{\partial f}{\partial R} \right)dR = \omega^2 \frac{\partial^2 }{\partial {t}^2}\left(\int^{R_e}_{R_w} RfdR\right)\\
    J_e-J_w = \omega^2\frac{(R^2_e-R^2_w)}{2}\frac{\partial^2 f_i}{\partial t^2}
\end{gather}

where $J_e$, $J_w$ are the fluxes at the east and west boundaries of the control volume

\begin{gather}
J\bigg\rvert_{\alpha} = \left(\omega^2_L R \Phi  \frac{\partial f}{\partial R}\right)\Bigg\rvert_{\alpha} \textrm{ \hspace{0.2cm} },\alpha=e,w
\end{gather}

Spatial derivatives at the east and west interfaces are approximated using the second order finite difference formula

\begin{subequations}
\begin{gather}
    \frac{\partial f}{\partial R}\bigg\rvert_w = \frac{f_{i}-f_{i-1}}{\delta R_w} \\
    \frac{\partial f}{\partial R}\bigg\rvert_e = \frac{f_{i+1}-f_{i}}{\delta R_e}.
\end{gather}
\end{subequations}

\noindent
The second time derivative of $f$ is approximated using the first order backward difference formula at the $(n+1)^{th}$ time step

\begin{gather}
    \frac{\partial^2 f_i}{\partial t^2}\bigg\rvert_{n+1} = \frac{f^{n+1}_i-2f^n_i+f^{n-1}_i}{\delta t^2}.
\end{gather}

The fluxes at the east and west interfaces are evaluated at the $(n+1)^{th}$ time step. This results in a system of difference equations at the $(n+1)^{th}$ time step having the form,

\begin{gather}
    A(F^{n+1})F^{n+1} = B(F^{n},F^{n-1}),
\end{gather}

\noindent
where $F$ is the vector containing values of $f$ at the nodes in the domain. This nonlinear system is solved using the Picard iteration scheme with an appropriate acceleration factor (0.1 to 0.2). Initial conditions are used to specify the values of $f$ in the first and second time steps. For the composite case, material properties at the interface between the inner and outer layers are defined using a smooth step function within a span of 0.4\% of the thickness of the cylinder. Simulations are run for one time period of deformation.

% ==================================
% Periodic Motion
% ==================================

\subsection{Periodic Motion}

\begin{figure}[H]
% =================================================
% R_i = 6.0 mm
% =================================================
\begin{subfigure}{.49\linewidth}
\centering
\includegraphics[width=\textwidth]{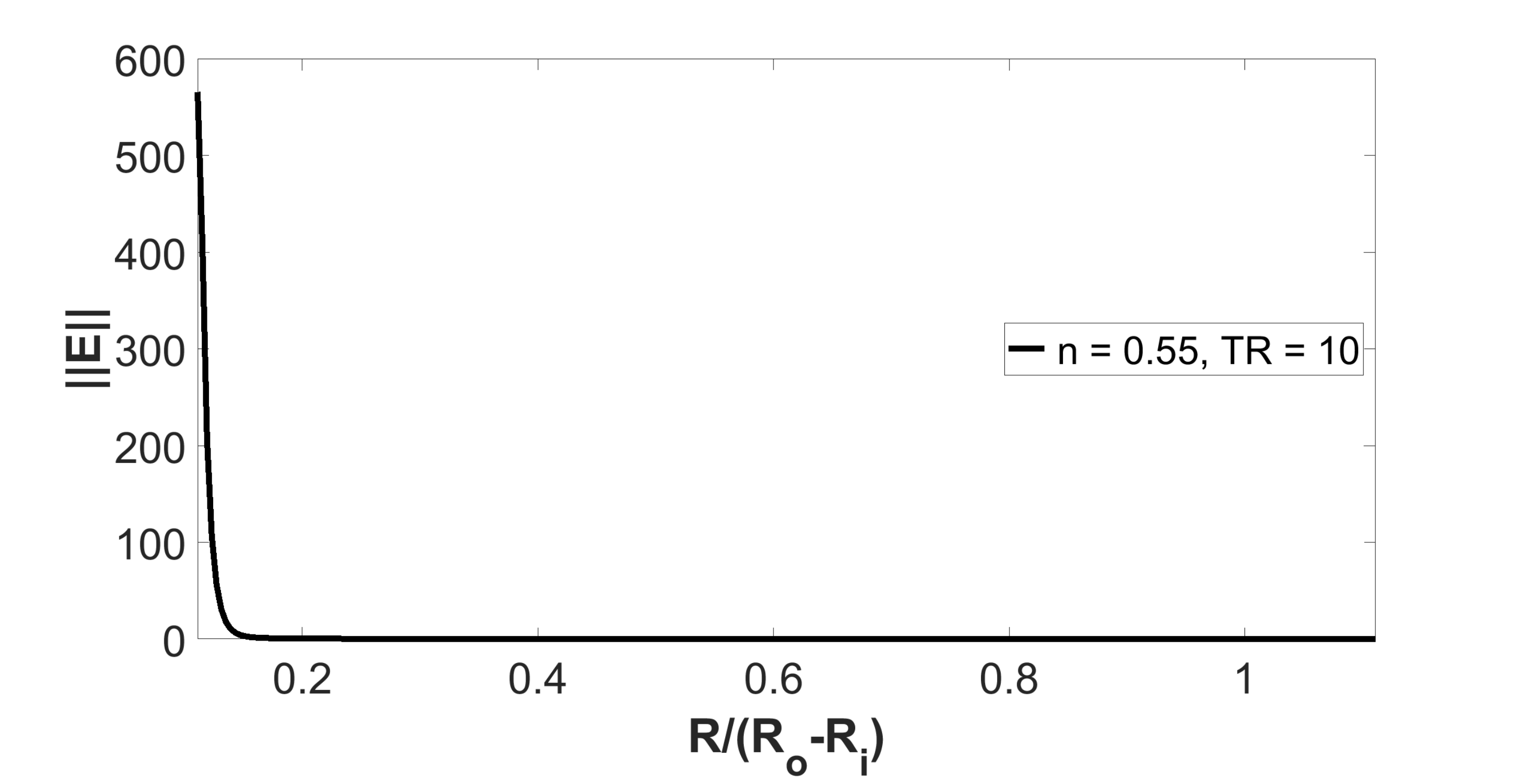}
\caption{$\vert\vert\mathbf{E}\vert\vert$, Thickness ratio 10}
\label{Norm:TR10}
\end{subfigure} %\\[1ex]
% \end{figure}
% \begin{figure}[H]\ContinuedFloat
% =================================================
% R_i 10 mm
% =================================================
\begin{subfigure}{.5\linewidth}
\centering
\includegraphics[width=\textwidth]{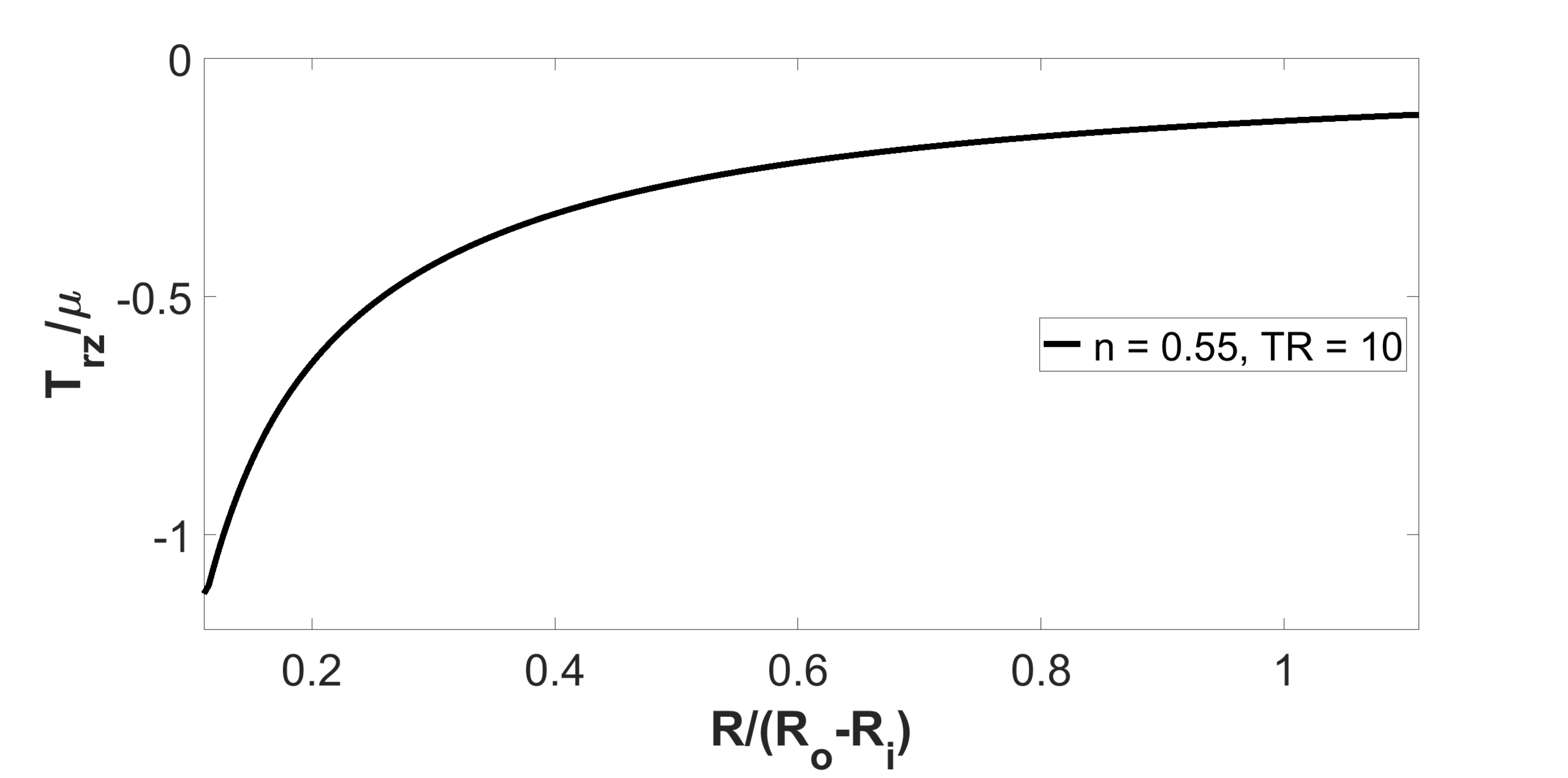}
\caption{$T_{rz}$, Thickness ratio 10}
\label{Trz:TR10}
\end{subfigure}%\\[1ex]
% % =================================================
% % Stress vs Strain
% % =================================================
% \begin{subfigure}{\linewidth}
% \centering
% \includegraphics[width=.5\textwidth]{Kappa_01}
% \caption{$R_i$ = 6.0 mm, $R_o$ = 12 mm}
% \label{Stress:Kappa}
% \end{subfigure}
% =================================================
% Caption for entire figure
% =================================================
\caption{The norm of the strain tensor $\vert\vert\mathbf{E}\vert\vert$ is shown for TR = 10 and n = 0.55 in \subref{Norm:TR10}.) and the shear stress on the interior of the annulus is shown \subref{Trz:TR10}.) at $t^*=2\pi$} 
\label{Stress:Function}
%\vspace{-20pt}
% =================================================
\end{figure}

\begin{figure}[H]
% =================================================
% R_i = 6.0 mm
% =================================================
\begin{subfigure}{.49\linewidth}
\centering
\caption{n = 0.55}
\includegraphics[width=\textwidth]{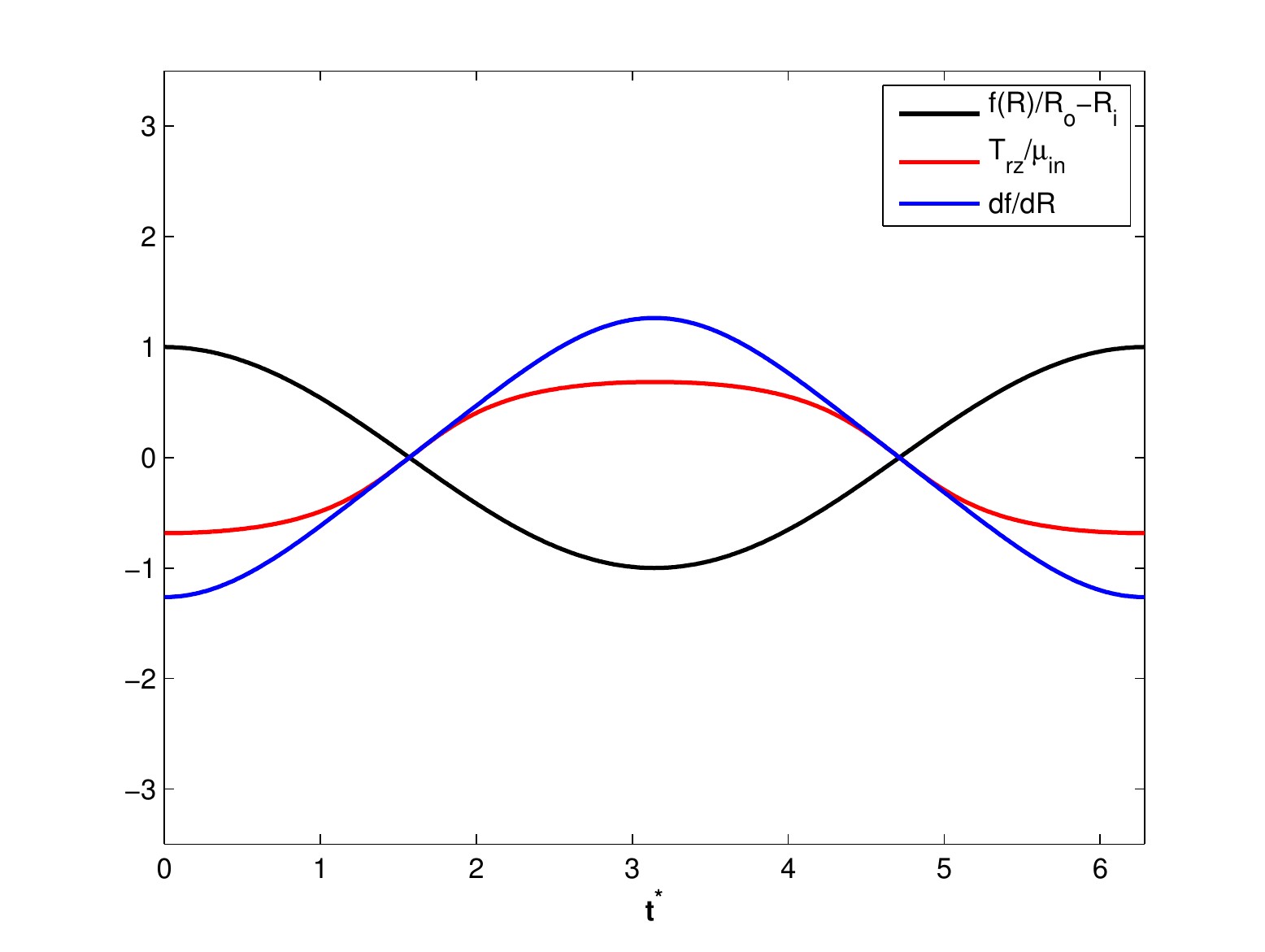}
\label{DSS:N055}
\end{subfigure} %\\[1ex]
% =================================================
% Stress vs Strain
% =================================================
\begin{subfigure}{.5\linewidth}
\centering
\caption{n = 5.00}
\includegraphics[width=\textwidth]{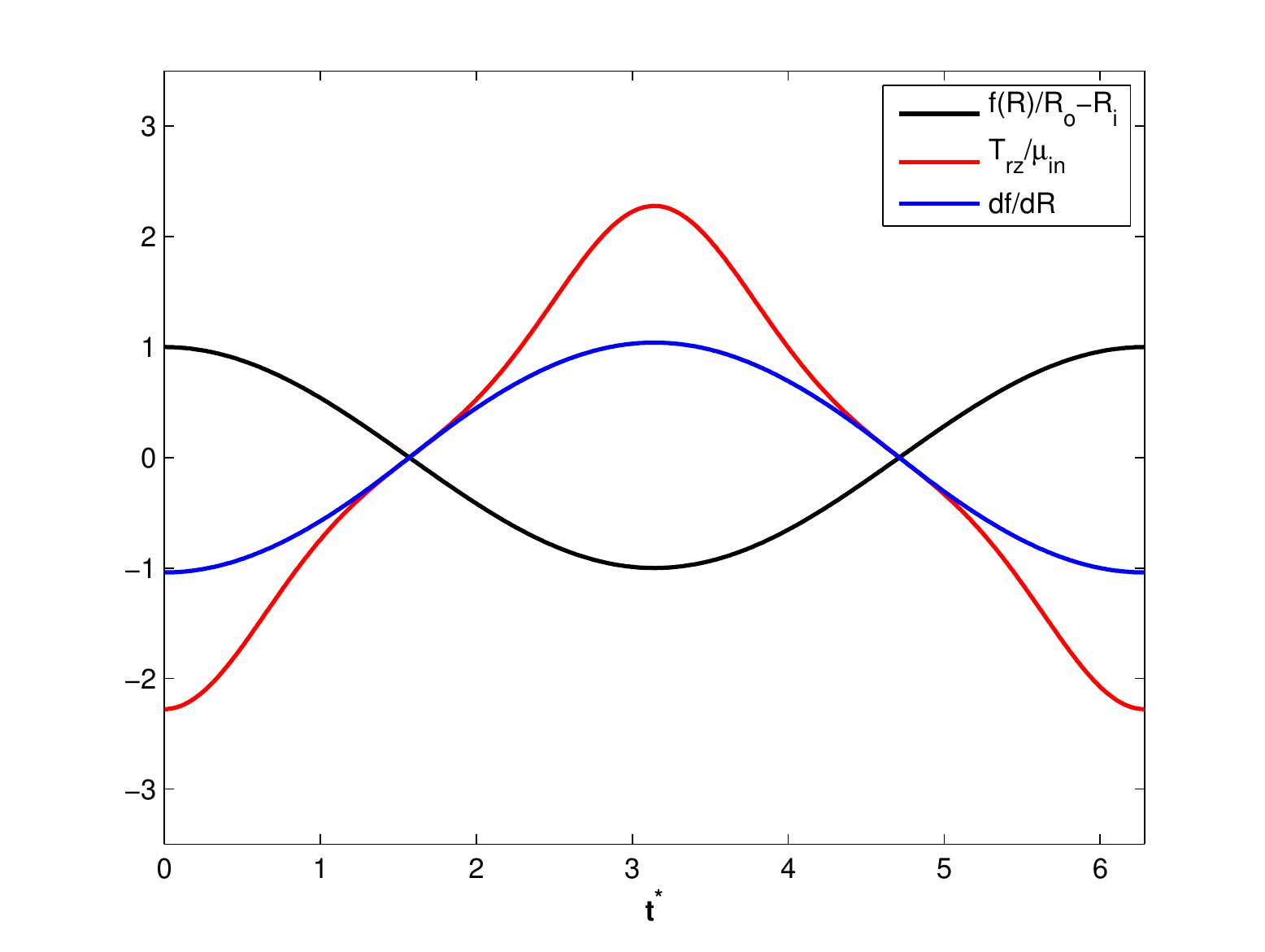}
\label{DSS:N500}
\end{subfigure}
% =================================================
% Caption for entire figure
% =================================================
\caption{These plots show the temporal distribution of the motion $f(R^*)$, the shear strain $\kappa$ and the normalized shear stress $T_{rz}/\mu$ at the inner surface for n = 0.55 and n = 5.00. $R_i$}
\label{DSS:All}
%\vspace{-20pt}
% =================================================
\end{figure}

In plot \ref{Norm:TR10}, the $\vert\vert\mathbf{E}\vert\vert$ is shown for a thickness ratio of 10 and n of 0.55 at $t^*=2\pi$.  This clearly shows that the boundary layer develops in terms of strain not in terms of stress.  For a smaller TR of 1.2, the displacement, shear strain and shear stress over one time period are shown in figure, \ref{DSS:All}.   

% =================================================
% These plots show the distribution of Stress T_rz
% as a function of time for different values of n.
% Additionally the shear stress vs. shear strain
% is shown.
% =================================================
% =================================================

% =================================================
% These plots show f(R), \kappa and T_rz on one 
% plot for each value of n.
% =================================================
% =================================================

% =================================================
% Shear Stress
% =================================================

\section{Deformation of a finite length cylindrical annulus due to periodic longitudinal shearing}\label{Sect:06}

We used ABAQUS finite element software to study the periodic longitudinal shear deformation of a finite length cylindrical annulus. The normalized shear stress is studied over a period of 2$\pi$ for two different inner radii. Two different cases of length to thickness ratio (LTR) of 5:1 and 10:1 are considered, to see what influence the length of the annulus has on the shear stresses.

% =================================================
% geometry used in abaqus
% =================================================
\begin{figure}[H] %{r}{0.5\textwidth}
\centering
\vspace{-20pt}
\includegraphics[width=.3\textwidth]{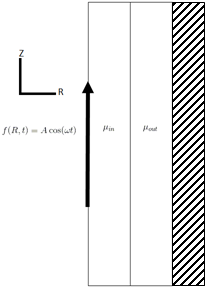}
% =================================================
% Caption for entire figure
% =================================================
\caption{A representation of the geometry used in the finite element simulations.  The two layers are comprised of a power-law neo-Hookean material with differing values for the small strain modulus $\mu$.  A periodic displacement of $f(R,t)=A\cos(\omega t)$ is prescribed on the inner boundary.  To analyze a single layer we set $\mu_{in}$ and $\mu_{out}$ equal to one another.}
\label{axisymmetric_geometry}
% \vspace{-10pt}
\end{figure}
% =================================================

The Abaqus subroutine UHYPER is used for the simulations. The following stored energy is used in FE simulations for easier convergence.

\begin{gather}
    W = \kappa(J-1)^2+\frac{\mu}{2b}\left[1+\frac{b}{n}(\overline{I}_1-3)\right]^{n}
\end{gather}

where $J=det(\textbf{F})$ and $\overline{I}_1=J^{-2/3}I_1$.  The axisymmetric geometry used in FE simulations is shown in fig \ref{axisymmetric_geometry}. 

A boundary value problem that is the closest representation to the infinite case and that is experimentally feasible is studied.  For the FE analysis, axial symmetry is used to reduce the problem to 2D. Finite element geometry consists of a cylindrical annulus represented using a rectangular domain. For the composite case(see section \ref{Sect:07}), the annulus is further divided into inner and outer layers.  Motion is studied for $n=0.55,1,5$ and $b=1$. The ratio $\kappa/\mu$ is taken as 5000 to approximate incompressibility.  Due to near incompressibility, reduced integration of hybrid elements(CAX4RH) is used for inner and outer layers.  The outer surface of the cylindrical annulus is fixed in the radial and axial directions. The inner surface is sheared sinusoidally at 1Hz with an amplitude equal to the wall thickness of the cylinder. Traction free boundary condition is applied on the top and bottom surfaces of the cylinder.  Lateral movement(i.e., movement in R-direction) of the top and the bottom surfaces beyond the rightmost surface of the domain is not allowed during the deformation. Mesh size is chosen such that the displacements converged to within 0.2\%. The solutions to the finite cases studied are compared to the infinite solutions obtained in sections \ref{Sect:Inf01}. All comparisons between the infinite length and finite length annulus are given in terms of the shear stress at the inner surface where stresses are the highest. The relative difference in the shear stress is defined as

\begin{gather}
    \Delta T_{rel} = \frac{T_{fin}-T_{inf}}{T_{inf}}
\end{gather}

\begin{figure}[H]
% =================================================
% Infinite and finite solutions
% =================================================
\begin{subfigure}{.49\linewidth}
\centering
\caption{Infinite and Finite solutions.}
\includegraphics[width=\textwidth]{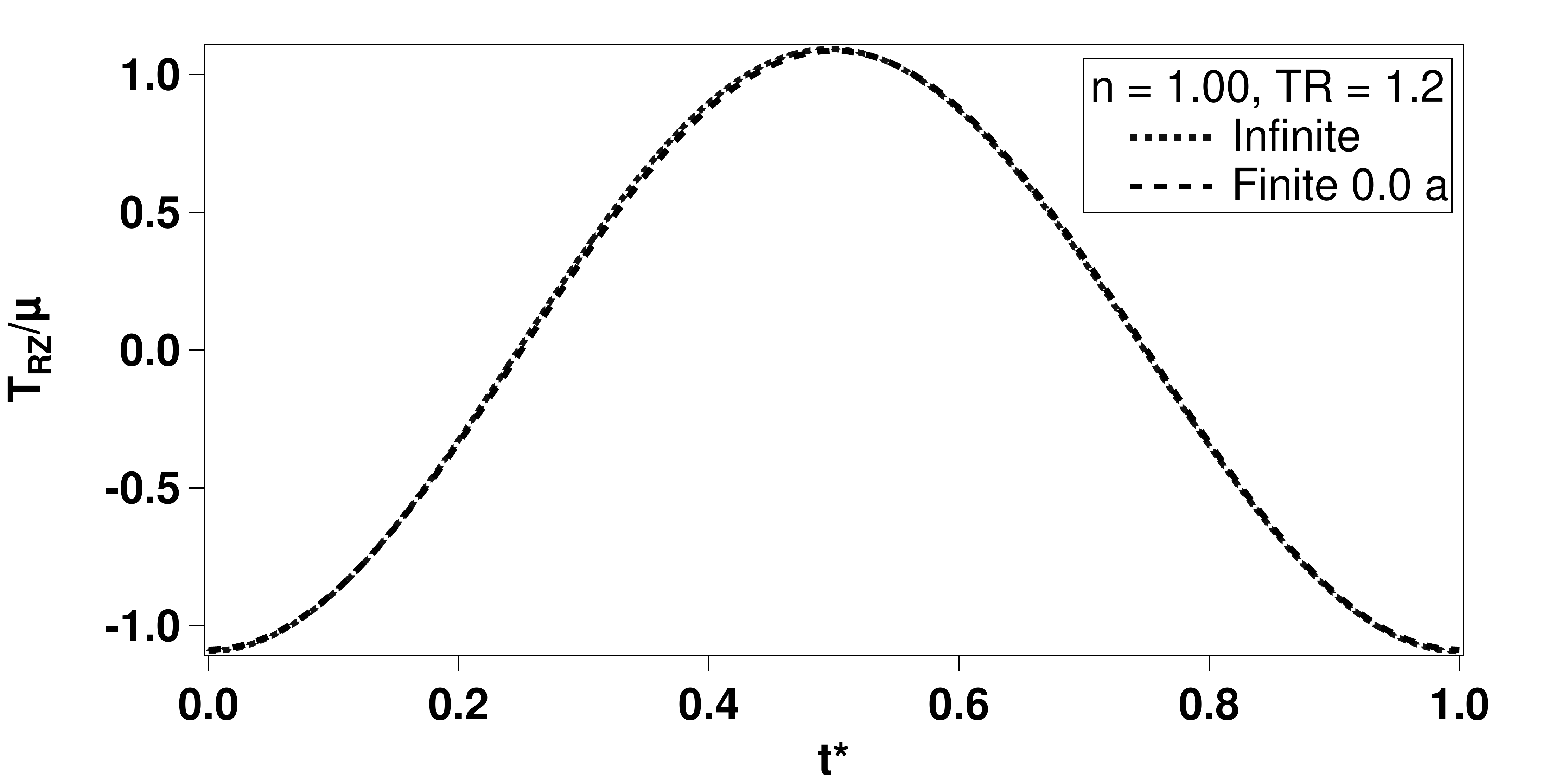}
\label{Fin:R10:N100:S110}
\end{subfigure}
% =================================================
% b = 1, n = 1.0
% =================================================
\begin{subfigure}{.5\linewidth}
\centering
\caption{b = 1, n = 1.0}
\includegraphics[width=\textwidth]{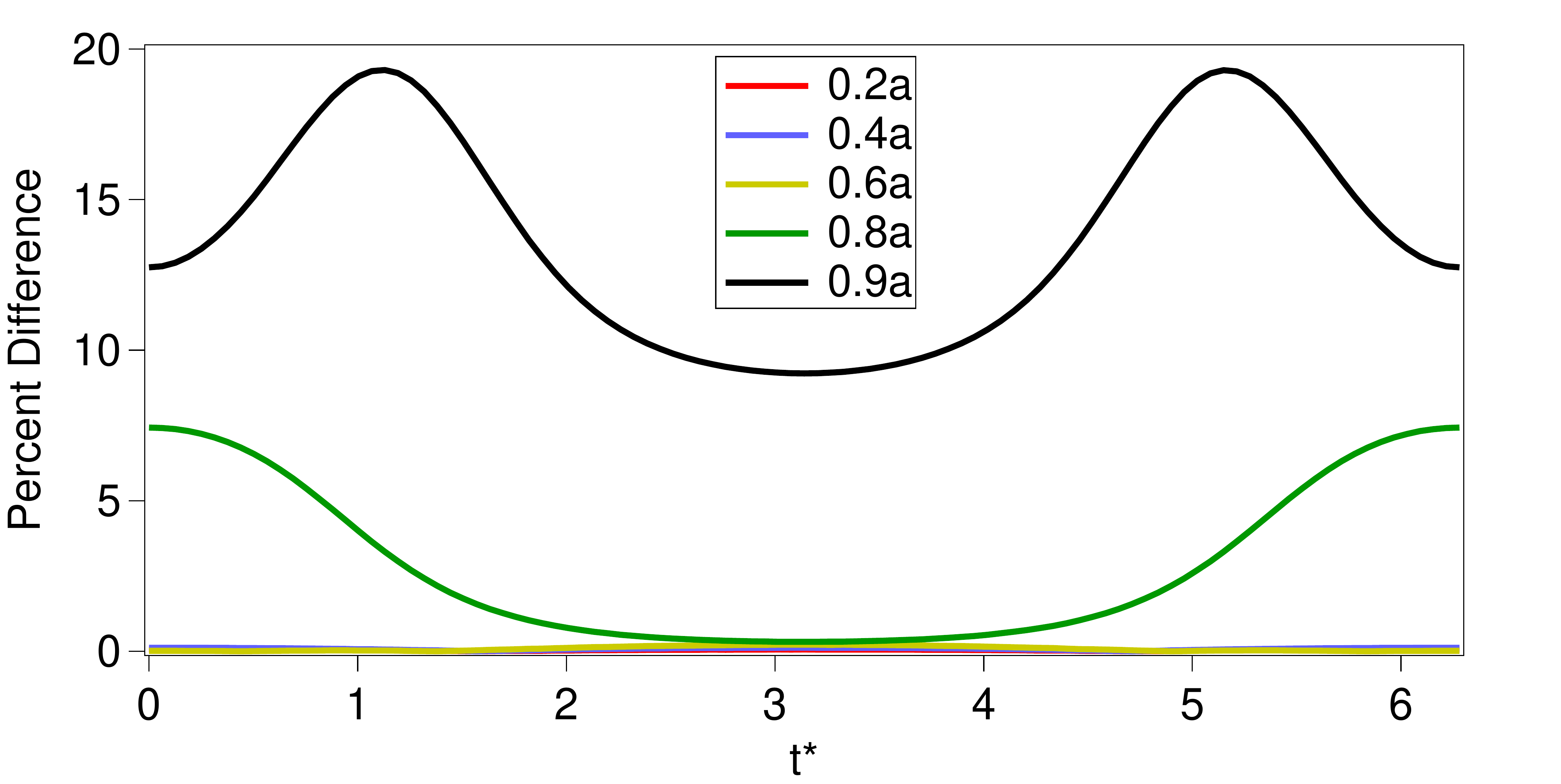}
\label{Err:S01}
\end{subfigure}\\[1ex]
% =================================================
% b = 1, n = 5.0
% =================================================
\begin{subfigure}{.49\linewidth}
\centering
\caption{b = 1, n = 5.0}
\includegraphics[width=\textwidth]{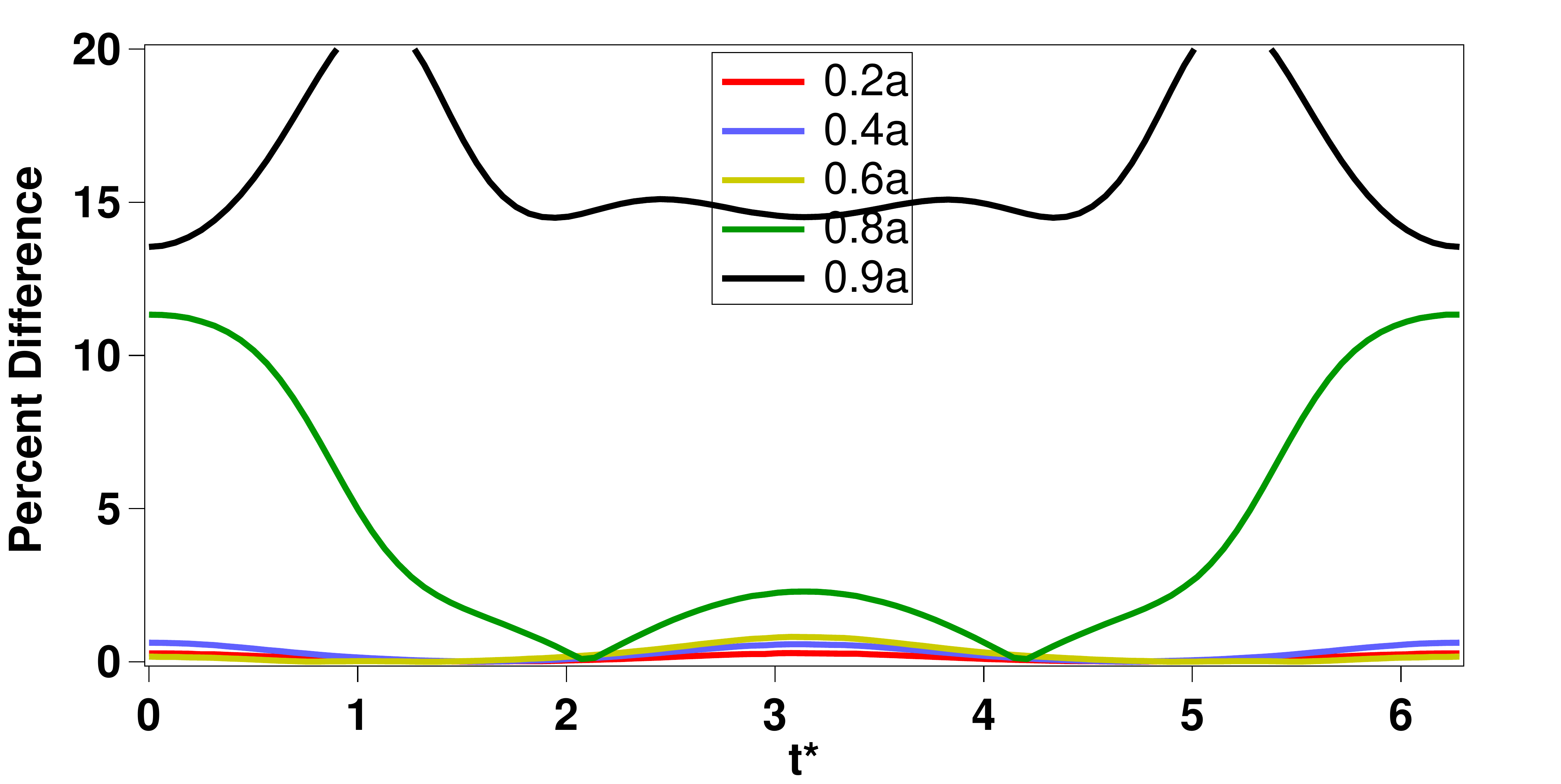}
\label{Err:S02}
\end{subfigure}
% =================================================
% b = 1, n = 0.55
% =================================================
\begin{subfigure}{.5\linewidth}
\centering
\caption{b = 1, n = 0.55}
\includegraphics[width=\textwidth]{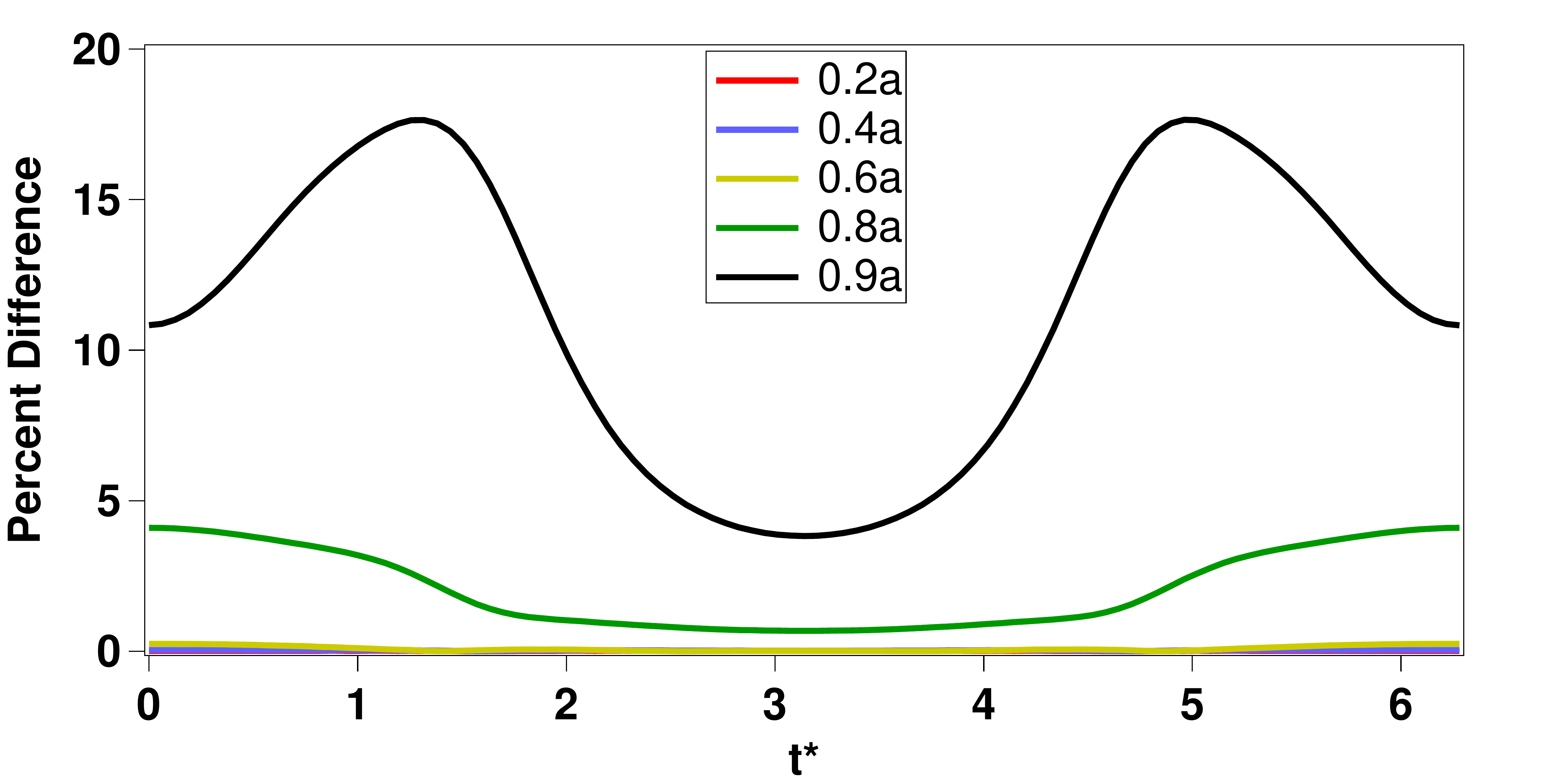}
\label{Err:S03}
\end{subfigure}
% =================================================
% Caption for entire figure
% =================================================
\caption{In figure \subref{Fin:R10:N100:S110}.) the stress is given as a function of time for the infinite solution and finite solution in the middle of the annulus.  In figures  \subref{Err:S01}.), \subref{Err:S02}.), and \subref{Err:S03}.), the percent difference in comparison to the middle of the annulus is reported.  The percent difference is reported for the length to thickness ratio (LTR) of 10:1 with TR = 1.2.}
\label{Fin:N055}
%\vspace{-20pt}
% =================================================
\end{figure}

In figure \ref{Fin:R10:N100:S110}, the infinite solution is compared to the finite solution at the middle of the annulus for LTR 10:1 homogeneous body.  It can be seen that the solutions match very well. In figure \ref{Err:S01}, the length from the middle of the annulus to either end is $a$(i.e., half-length of the cylindrical annulus); thus the total length is L=2$a$.  The number on the legend represents how far away from the center the stress is being taken, i.e., 0.4a represents 40\% of the half-length from the center of the annulus and 0.9a would be 90\% of the half-length or very close to the free end.  The region of validity of the infinite solution corresponds to a length of 2 times the annular thickness (2t) away from the ends of the cylinder.    

% =================================================
% These plots show the comparison of the infinite
% and finite solutions at the the middle of the 
% annulus.  The percent difference is reported for
% LTR = 10:1, TR = 1.2
% =================================================
% =================================================

% =================================================
% Figures from FEA analysis for finite composite.
% =================================================
\begin{figure}[H]
% =================================================
% n = 1, t = \pi
% =================================================
\begin{subfigure}{.49\linewidth}
\centering
\includegraphics[width=0.5\textwidth]{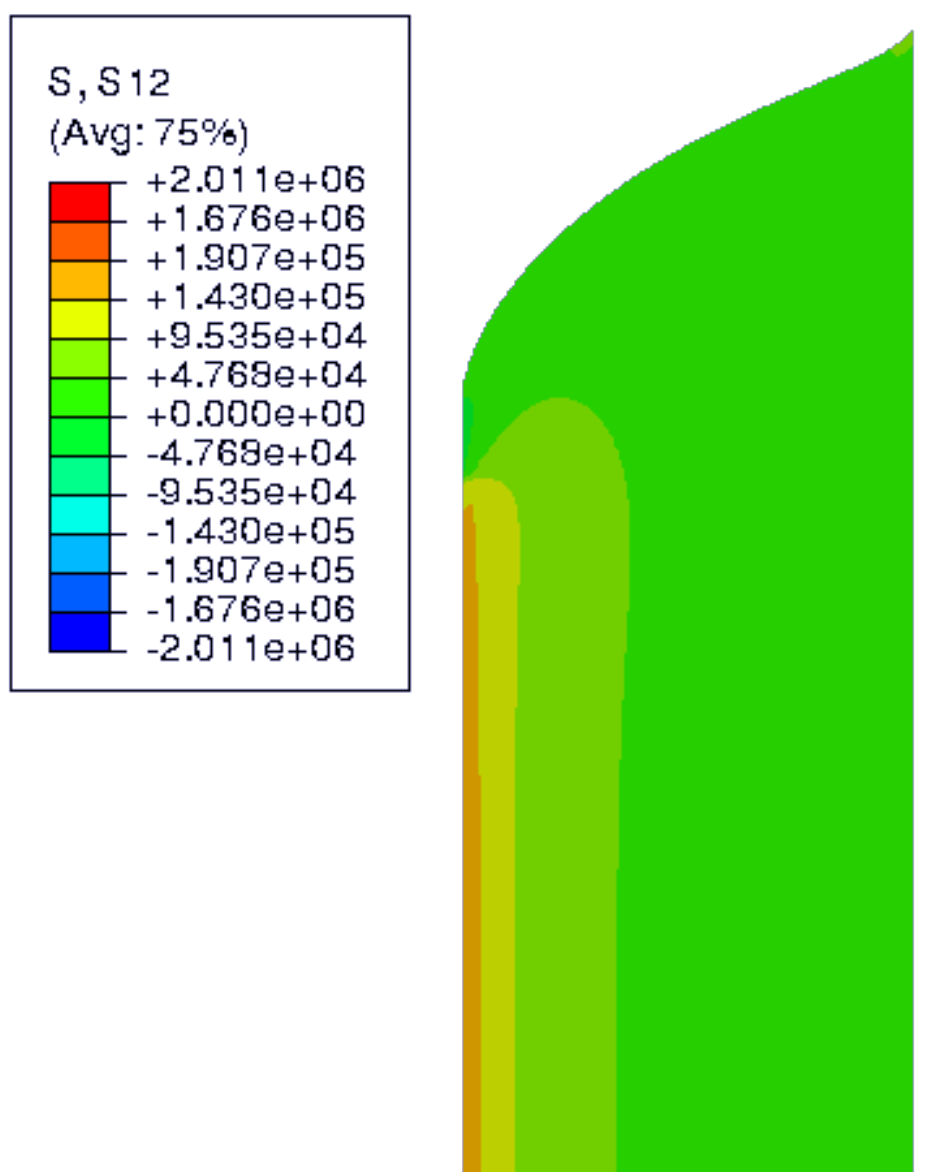}
\caption{}
\label{asymmetry_at_05_s}
\end{subfigure} %\\[1ex]
% =================================================
% n = 1, t = 2\pi
% =================================================
\begin{subfigure}{.49\linewidth}
\centering
\includegraphics[width=0.5\textwidth]{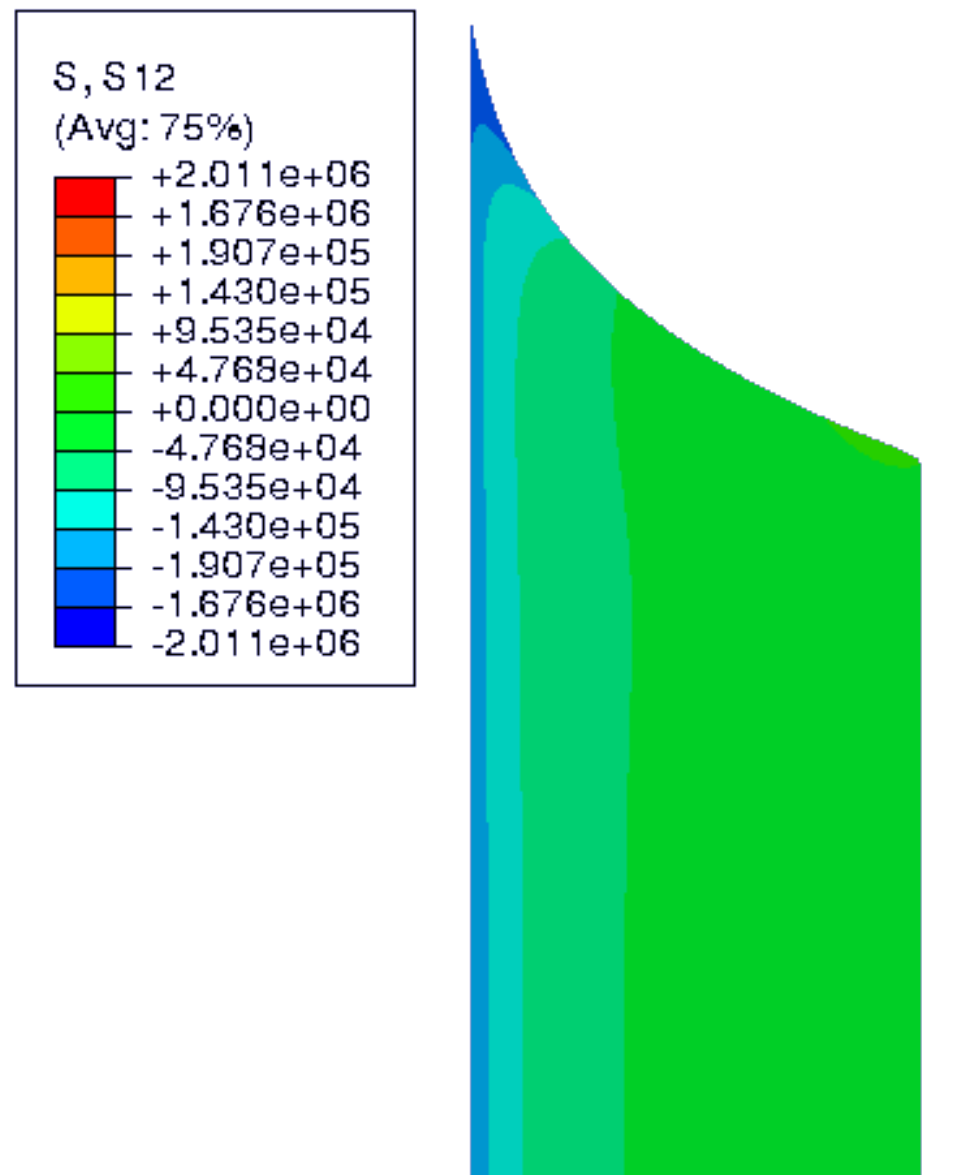}
\caption{}
\label{asymmetry_at_1s}
\end{subfigure}%\\[1ex]
% =================================================
% Caption for entire figure
% =================================================
\caption{In \subref{asymmetry_at_05_s}.) stress distribution near the edges for n=1 at $t^*=\pi$ is shown.  In \subref{asymmetry_at_1s}.) stress distribution near the edges for n=1 at $t^*=2\pi$ is shown. }
\label{stress_asymmetry_in_2d}
%\vspace{-20pt}
% =================================================
\end{figure}

In figure \ref{stress_asymmetry_in_2d} we note the differences in the stress distribution near the edges of the finite length annular cylinder at $t^*=\pi$ and $t^*=2\pi$.  The material undergoes compression and tension at the inner surface depending on the magnitude and direction of the motion because the inner surface is constrained to move in the Radial direction. Top surface on the other hand is curved and traction free due to the traction-free boundary condition.  

\color{black}

\begin{figure}[H]
% =================================================
% TR = 1.2 and 10
% =================================================
\begin{subfigure}{.49\linewidth}
\centering
\caption{TR = 1.2 \& 10}
\includegraphics[width=.8\textwidth]{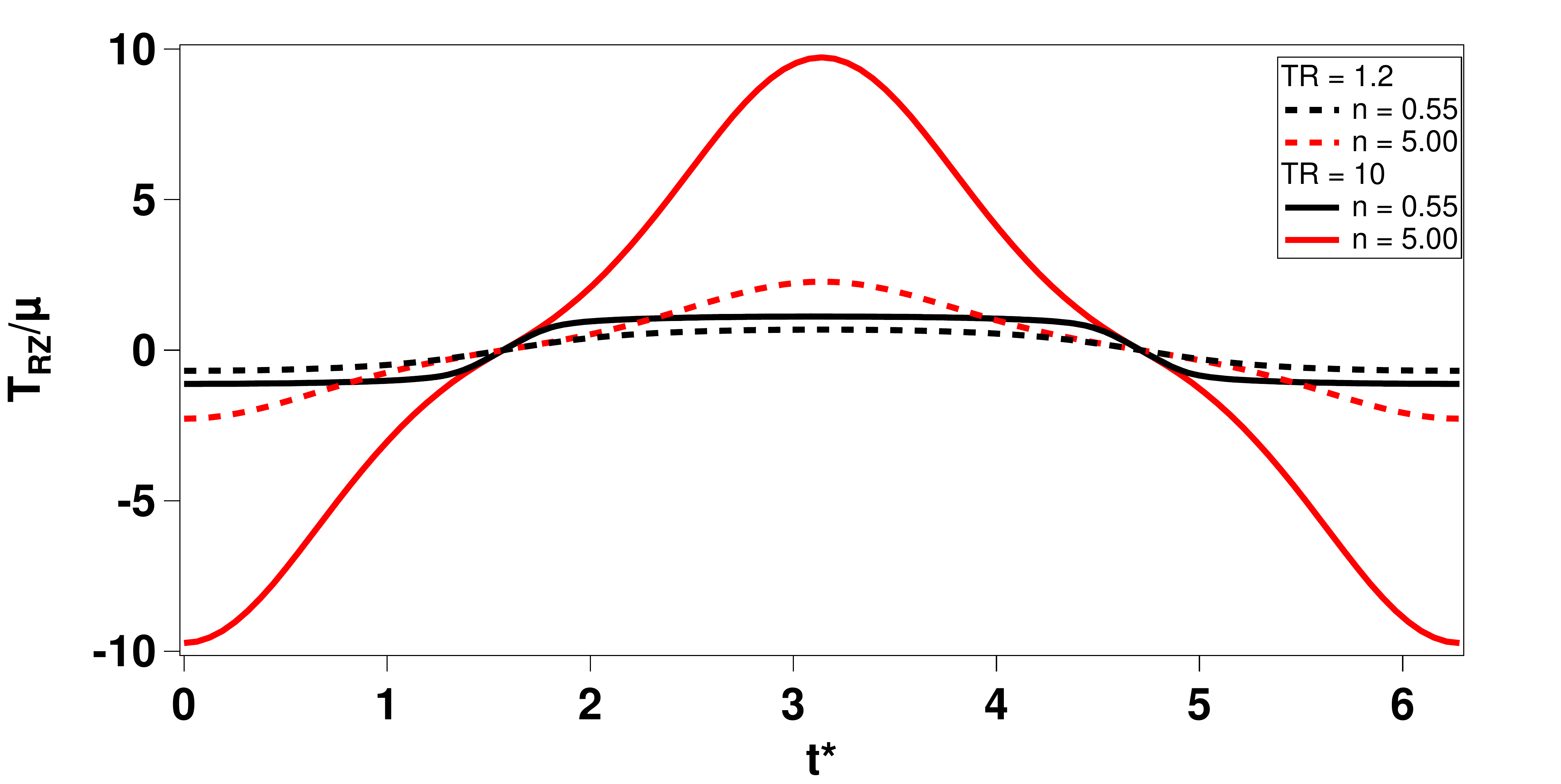}
\label{Comp:Radii}
\end{subfigure}
% =================================================
% b = 1, n = 1.0
% =================================================
\begin{subfigure}{.5\linewidth}
\centering
\caption{b = 1, n = 1.0}
\includegraphics[width=.8\textwidth]{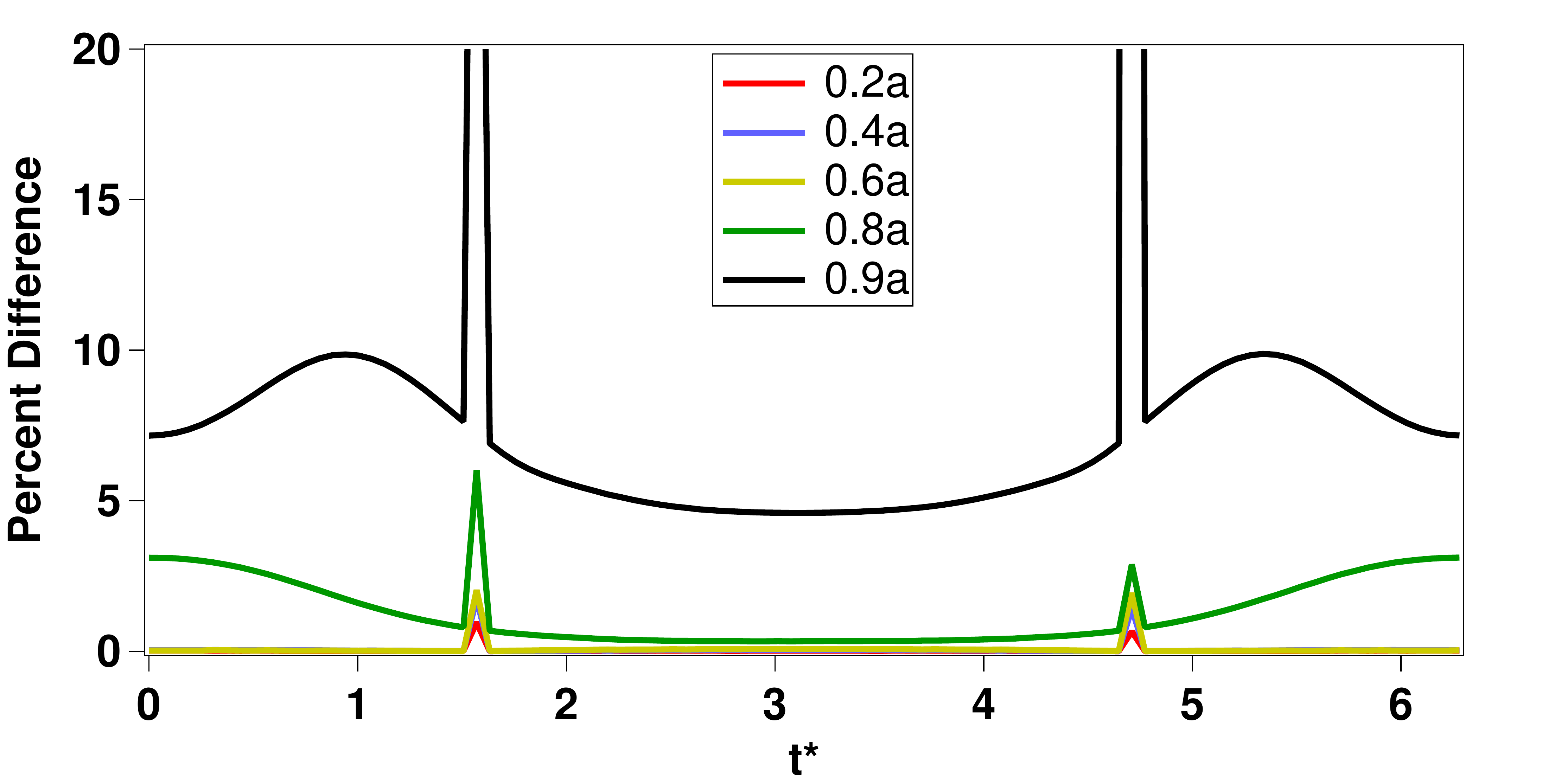}
\label{Err:A01}
\end{subfigure}\\[1ex]
% =================================================
% b = 1, n = 5.0
% =================================================
\begin{subfigure}{.49\linewidth}
\centering
\caption{b = 1, n = 5.0}
\includegraphics[width=.8\textwidth]{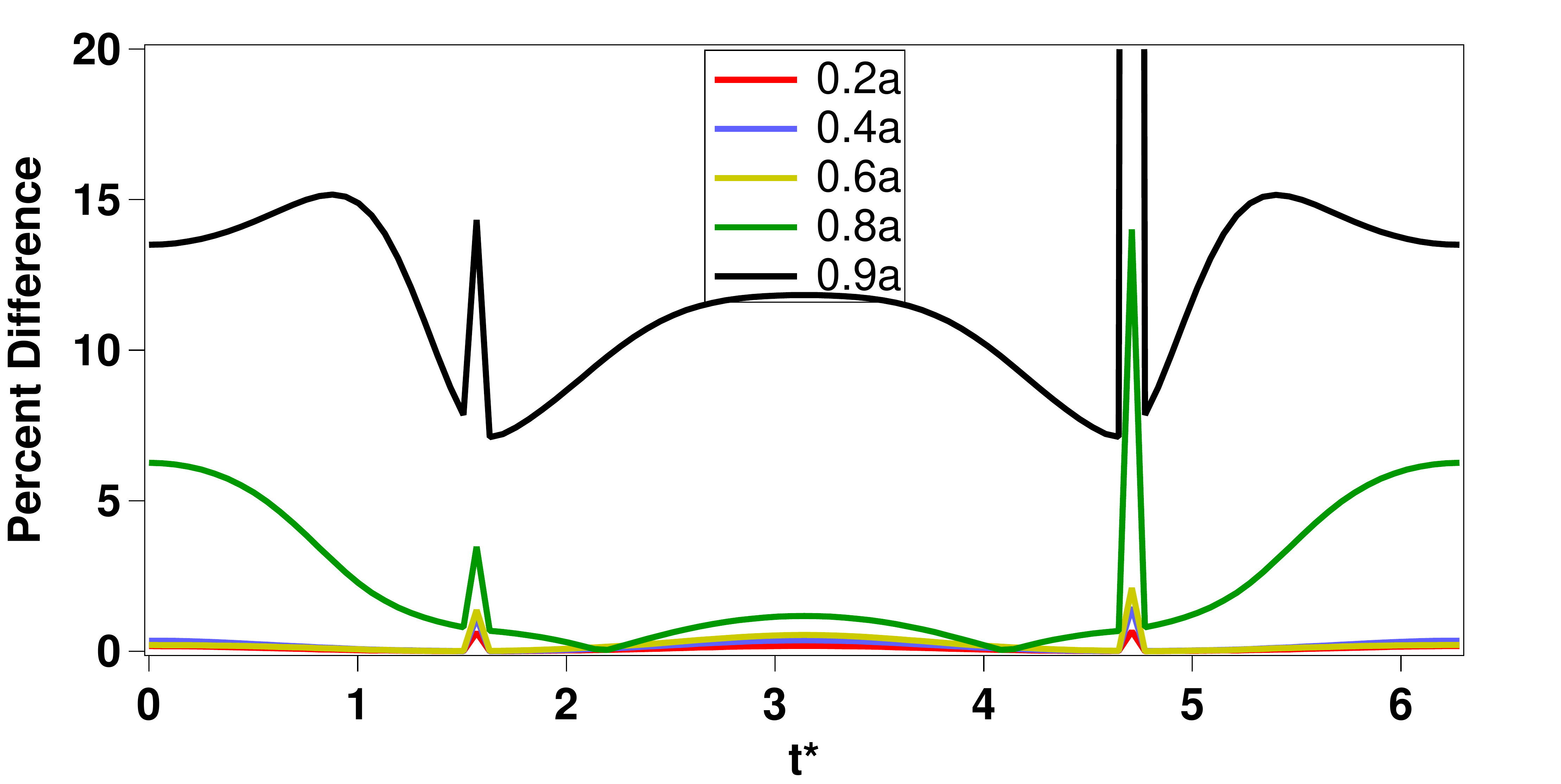}
\label{Err:A02}
\end{subfigure}
% =================================================
% b = 1, n = 0.55
% =================================================
\begin{subfigure}{.5\linewidth}
\centering
\caption{b = 1, n = 0.55}
\includegraphics[width=.8\textwidth]{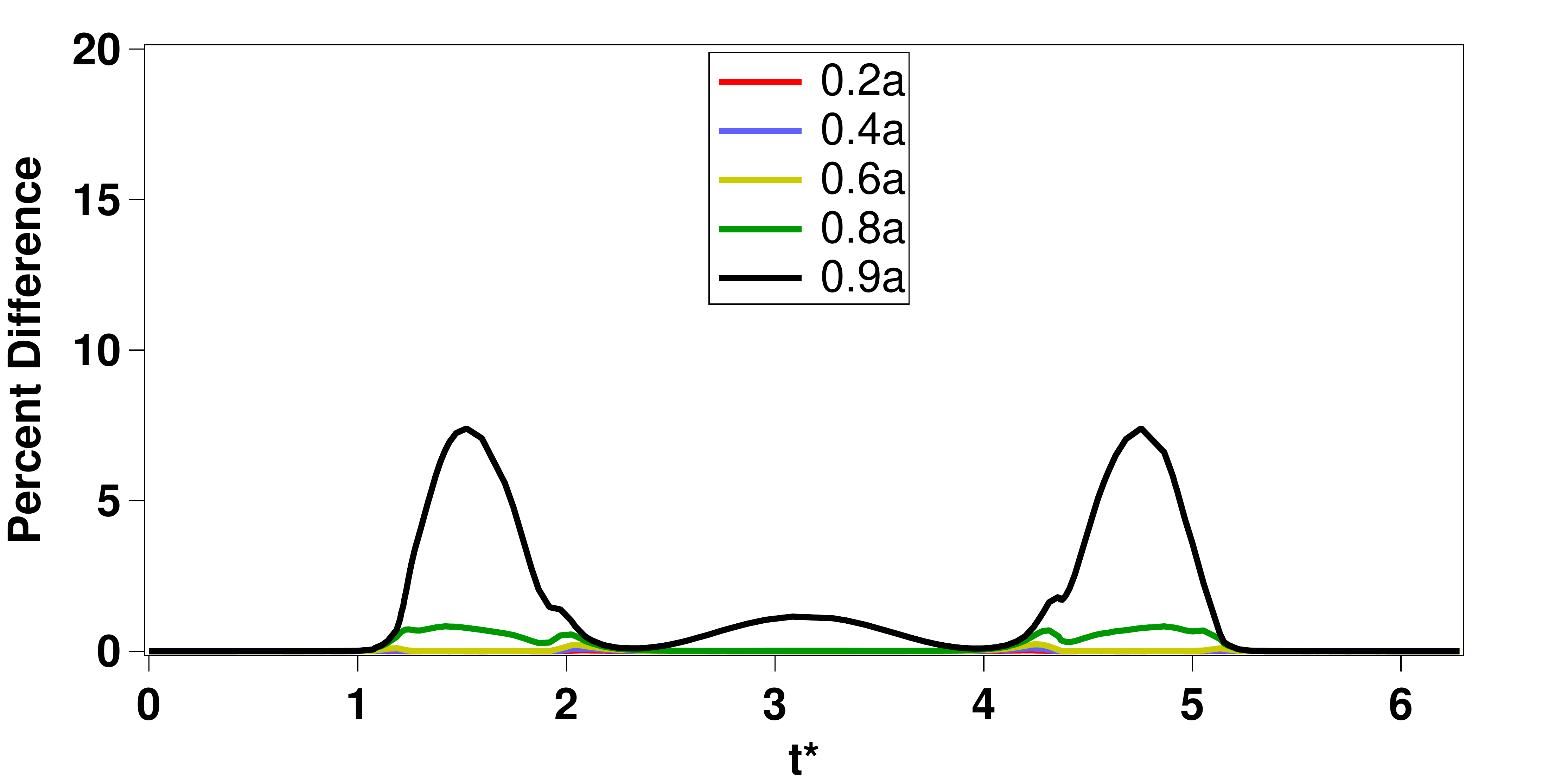}
\label{Err:A03}
\end{subfigure}
% =================================================
% Caption for entire figure
% =================================================
\caption{In figure \subref{Comp:Radii}.) the shear stress as a function of time is compared for TR's of 1.2 and 10.  In figure \subref{Err:A01}.) the percent difference in comparison to the middle of the annulus is reported.  The percent difference is reported for the length to thickness ratio (LTR) of 10:1 with TR = 10.}
\label{Final:R12}
%\vspace{-20pt}
% =================================================
\end{figure}

In a region that is two times the thickness (2t) from the cylinder ends the solution to the infinite annular cylinder and the solutions to the finite annular cylinder do not match.  This is shown in figures \ref{Err:A01}, \ref{Err:A02}, and \ref{Err:A03}.  Beyond this (2t) distance the solutions to the finite annular cylinder and the infinite annular cylinder are within a range of 0.25\% to 2\% of one another.  This corresponds to a region that is 60\% of the length for a cylinder with a (LTR) of 10:1.  For a cylinder with an (LTR) of 5:1 this corresponds to a region that is 20\% of the cylinder length.  As the cylinder gets longer the region of applicability becomes larger.

The spikes in the data are associated with transition points when the stress was alternating from positive to negative or in the other way.   The error was highest for n = 5.00 while for the thick-walled cylinder (TR = 10) the error was small and only approached its highest error at the transition points.  In the case of strain softening, n = 0.55, the stress varies little with increased strain over a significant period of the motion.  As a result, the error is small for these cases.  In figure \ref{Comp:Radii} we note that the stress on the inner surface is many times larger for the thicker annulus (TR = 10) as is to be expected for both the hardening and softening cases.
 
\begin{figure}[H]
% =================================================
% R_i 10 mm
% =================================================
\begin{subfigure}{.5\linewidth}
\centering
\caption{Located 20\% of annular length from center.}
\includegraphics[width=.8\textwidth]{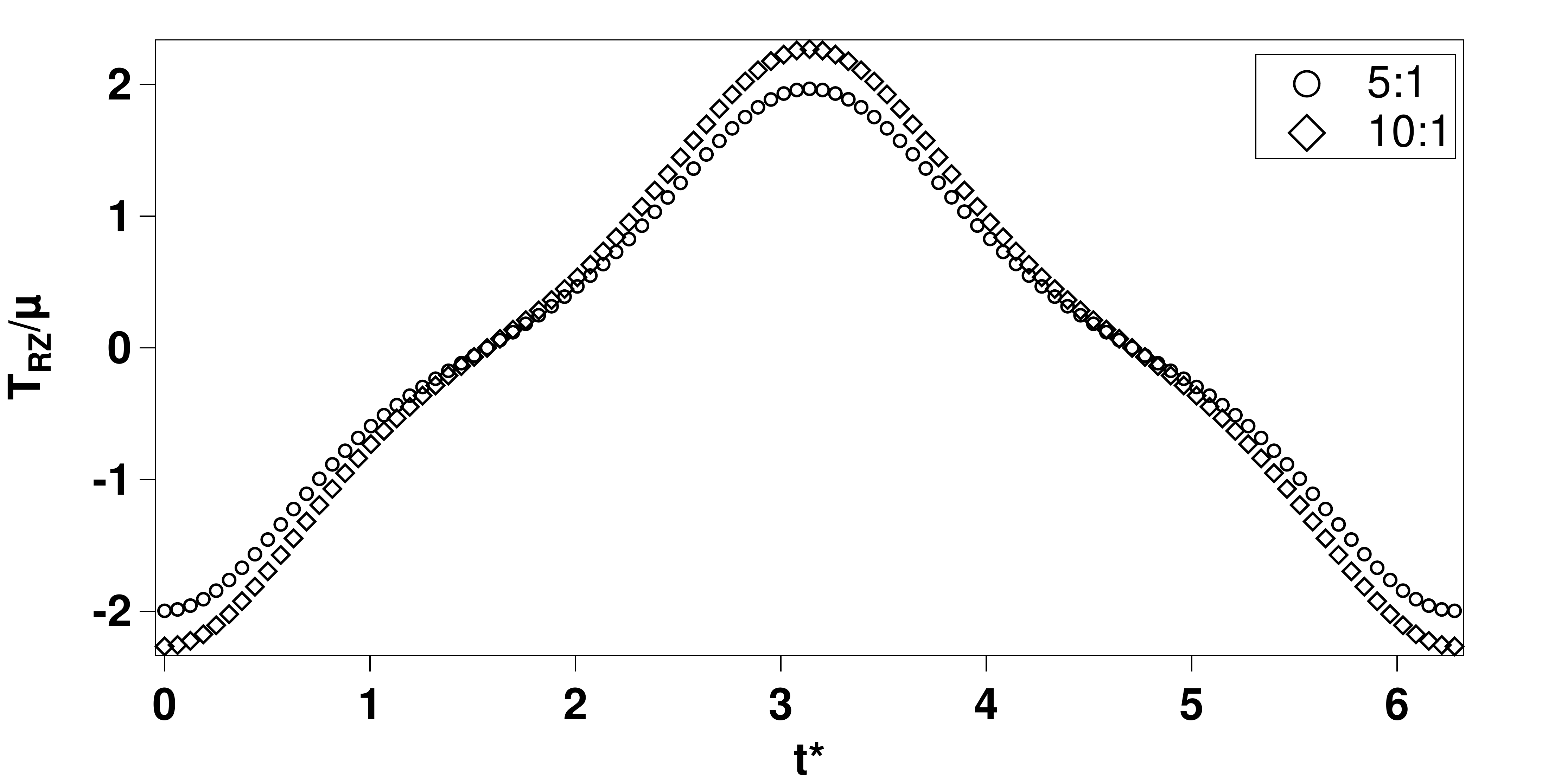}
\label{Comp:01}
\end{subfigure}
% =================================================
% Stress vs Strain
% =================================================
\begin{subfigure}{.5\linewidth}
\centering
\caption{Located twices the thickness (2t) from the cylinder ends.}
\includegraphics[width=.8\textwidth]{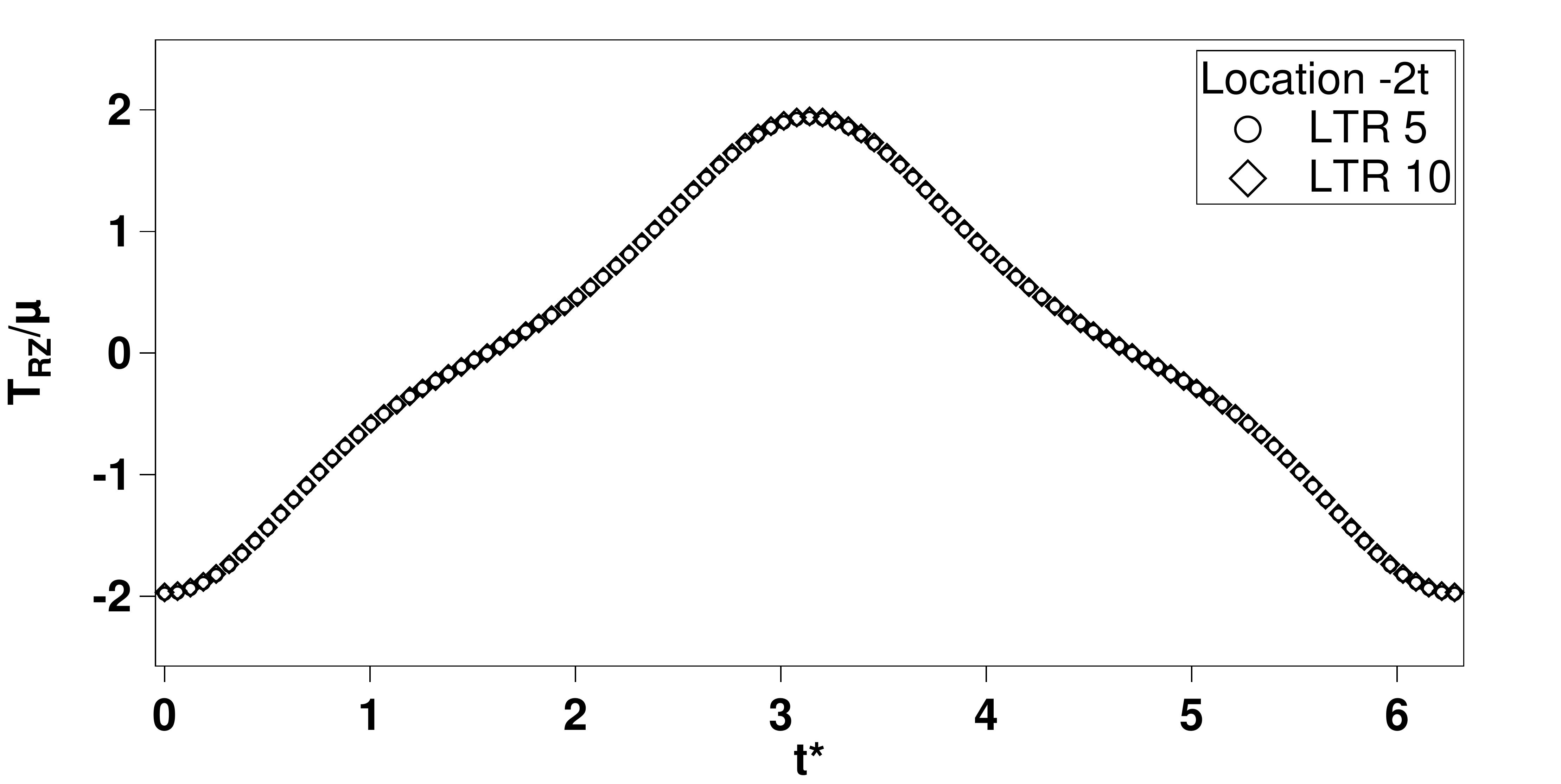}
\label{Comp:02}
\end{subfigure}\\[1ex]
% =================================================
% R_i 10 mm
% =================================================
\begin{subfigure}{\linewidth}
\centering
\caption{b = 1, n = 5.0}
\includegraphics[width=.4\textwidth]{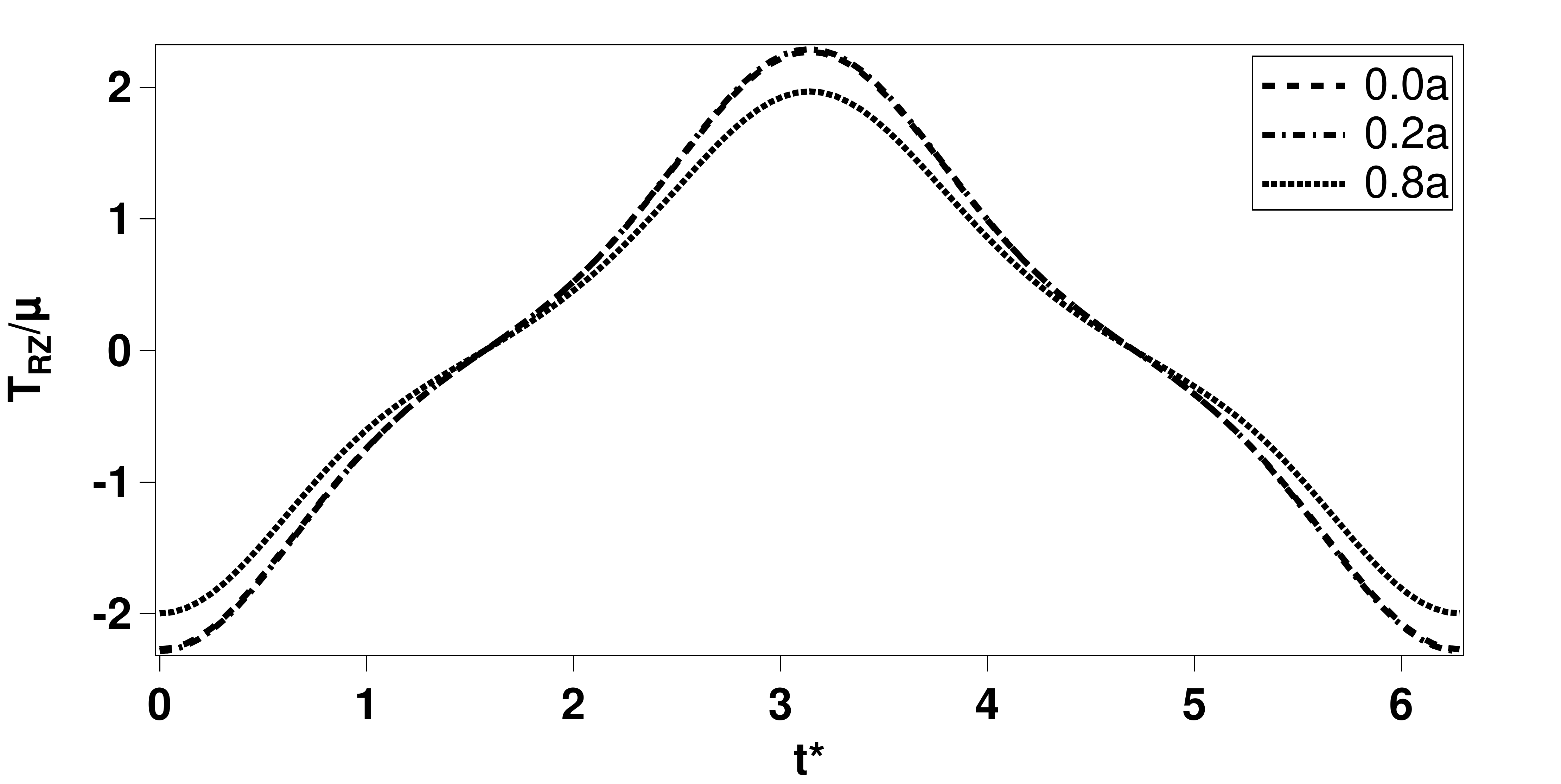}
\label{Err:E}
\end{subfigure}
% =================================================
% Caption for entire figure
% =================================================
\caption{In figure \subref{Comp:01}.) The normalized stress ($T_{rz}/\mu$) at a distance 20\% of the length from the middle of the annulus is plotted for (LTR's) of 5:1 and 10:1.  Figure \subref{Comp:02}.) is plotted against at a distance 2t away from the cylider edges.  Figure \subref{Err:E}.) shows $T_{rz}/\mu$ as a function of $t^*$ for (LTR) = 5:1 for three difference points on the annulus.}
\label{Final:R1}
%\vspace{-20pt}
% =================================================
\end{figure}

Figure \ref{Comp:01} compares the solution to the finite case for (LTR's) of 5:1 and 10:1 at a distance of 0.2a from the annular center.  The finite solution does have a dependence on LTR giving difference solutions 20\% from the middle of the respective lengths.  Figure \ref{Comp:02} shows the solution to the finite case for (LTR's) of 5:1 and 10:1 at a distance 2t from the annular edges.  At this distance that is 2t away from the edges for LTR 5:1 and LTR 10:1 the solutions match well.  For the middle of the cylinder between a distance of 2t from each edge the solution to the infinite case and the solution to the finite case match to within 2\% of one another.  Figure \ref{Err:E} shows solution to the finite case for a distance of 0.0a, 0.2a and 0.8a from the middle of the annulus for LTR 5:1.  

\color{black}

% =================================================
% These plots show the stress for the infinite
% solution for n = 0.55 and n = 5.0.  The solution
% is compared for TR = 1.2 and 10  and the percent 
% difference compared to the middle of the annulus 
% and the infinite solution.  The percent 
% difference is reported for
% LTR = 10:1, TR = 10
% =================================================
% =================================================

% =================================================
% These plots show f(R), \kappa and T_rz on one 
% plot for each value of n.
% =================================================
% =================================================

% +++++++++++++++++++++++++++++++++++++++++++++++++

For all the cases studied, the error between the finite and infinite solutions was relatively low (< 2.0\%) for strain hardening and softening materials when the location in question is more than two times the thickness of the annulus away from the edges of the sample.

% \clearpage

% Composite layer solutions

\section{Deformation of a composite two layered infinitely long cylindrical annulus due to periodic longitudinal shearing}\label{Sect:Inf03}

\par
\begin{figure}[H]
% =================================================
% R_i 1.2 mm
% =================================================
\begin{subfigure}{.49\linewidth}
\centering
\caption{$n$ = 1,5 \& $\mu_{in}$:$\mu_{out}$ = 10:1}
\includegraphics[width=\textwidth]{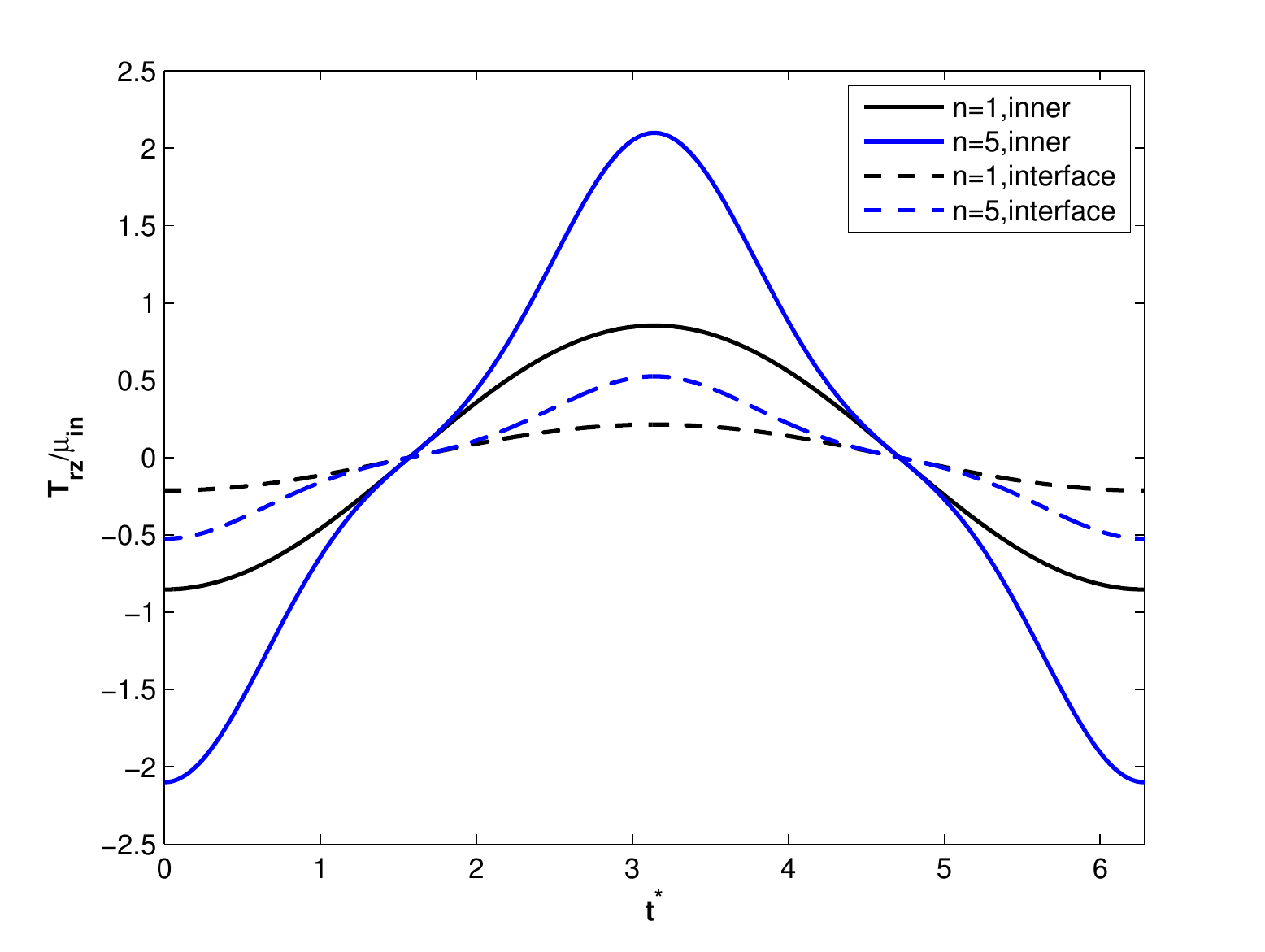}
\label{Ri_1_2_stress_modulus_ratio_10_1}
\end{subfigure}
% =================================================
% Stress vs Strain
% =================================================
\begin{subfigure}{.5\linewidth}
\centering
\caption{$n$ = 1,5 \& $\mu_{in}$:$\mu_{out}$=1:10}
\includegraphics[width=\textwidth]{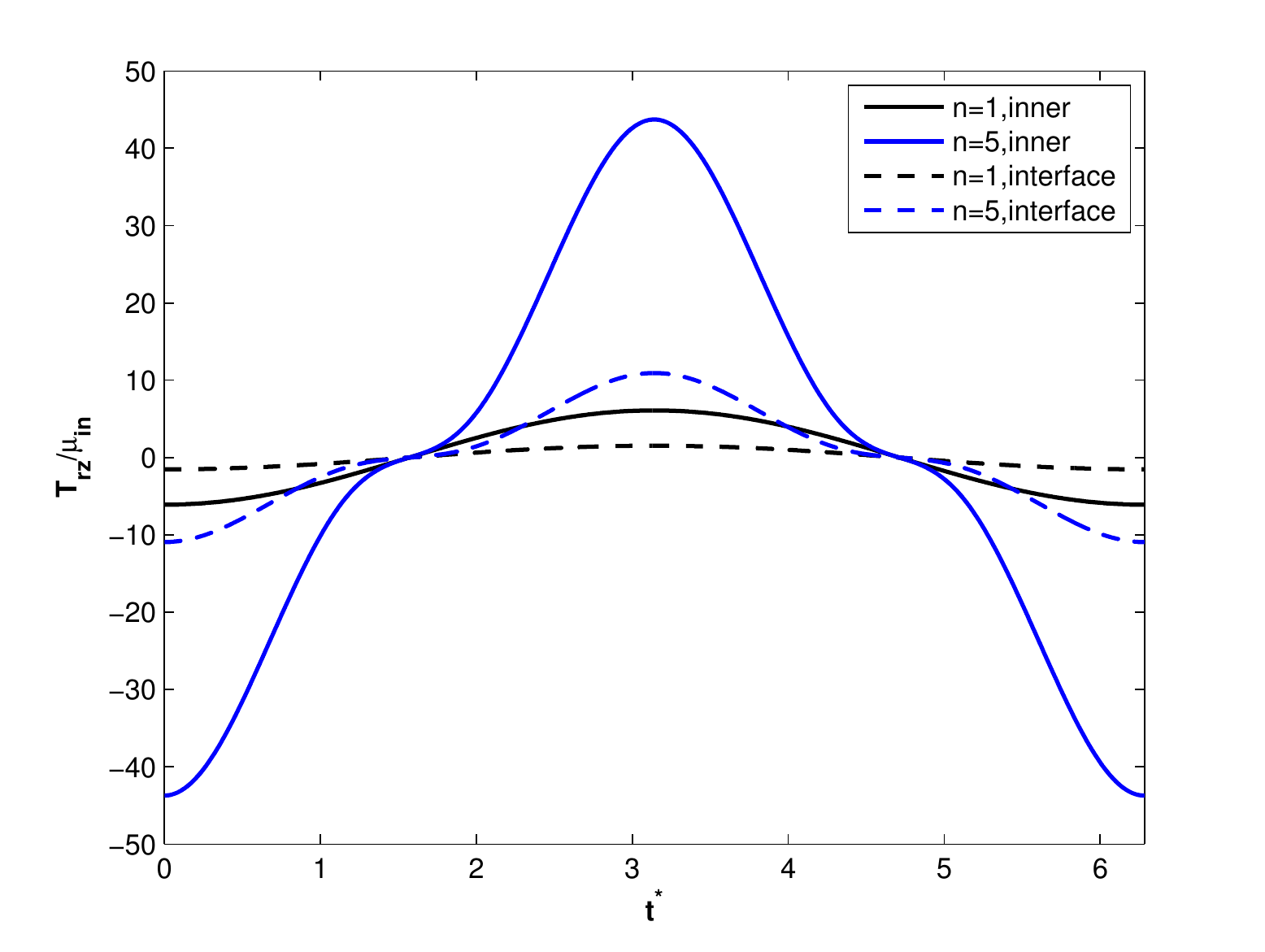}
\label{Ri_1_2_stress_modulus_ratio_1_10}
\end{subfigure}\\[1ex]
% =================================================
% R_i 1.2 mm
% =================================================
\begin{subfigure}{.49\linewidth}
\centering
\caption{$n$ = 0.55 \& $\mu_{in}$:$\mu_{out}$ = 10:1}
\includegraphics[width=\textwidth]{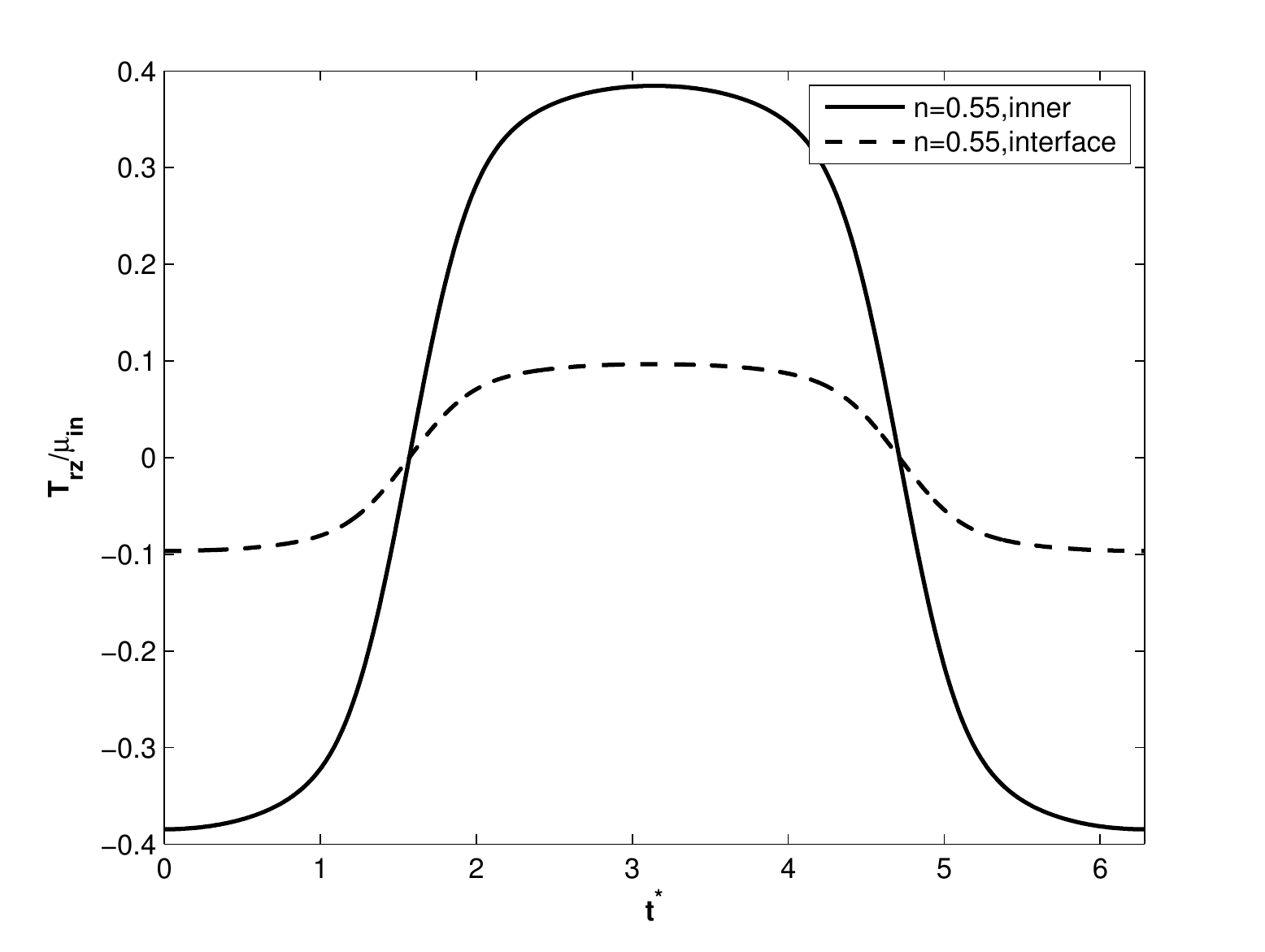}
\label{Ri_1_2_stress_modulus_ratio_10_1_n_055}
\end{subfigure}
% =================================================
% Stress vs Strain
% =================================================
\begin{subfigure}{.5\linewidth}
\centering
\caption{$n$ = 0.55 \& $\mu_{in}$:$\mu_{out}$ = 1:10}
\includegraphics[width=\textwidth]{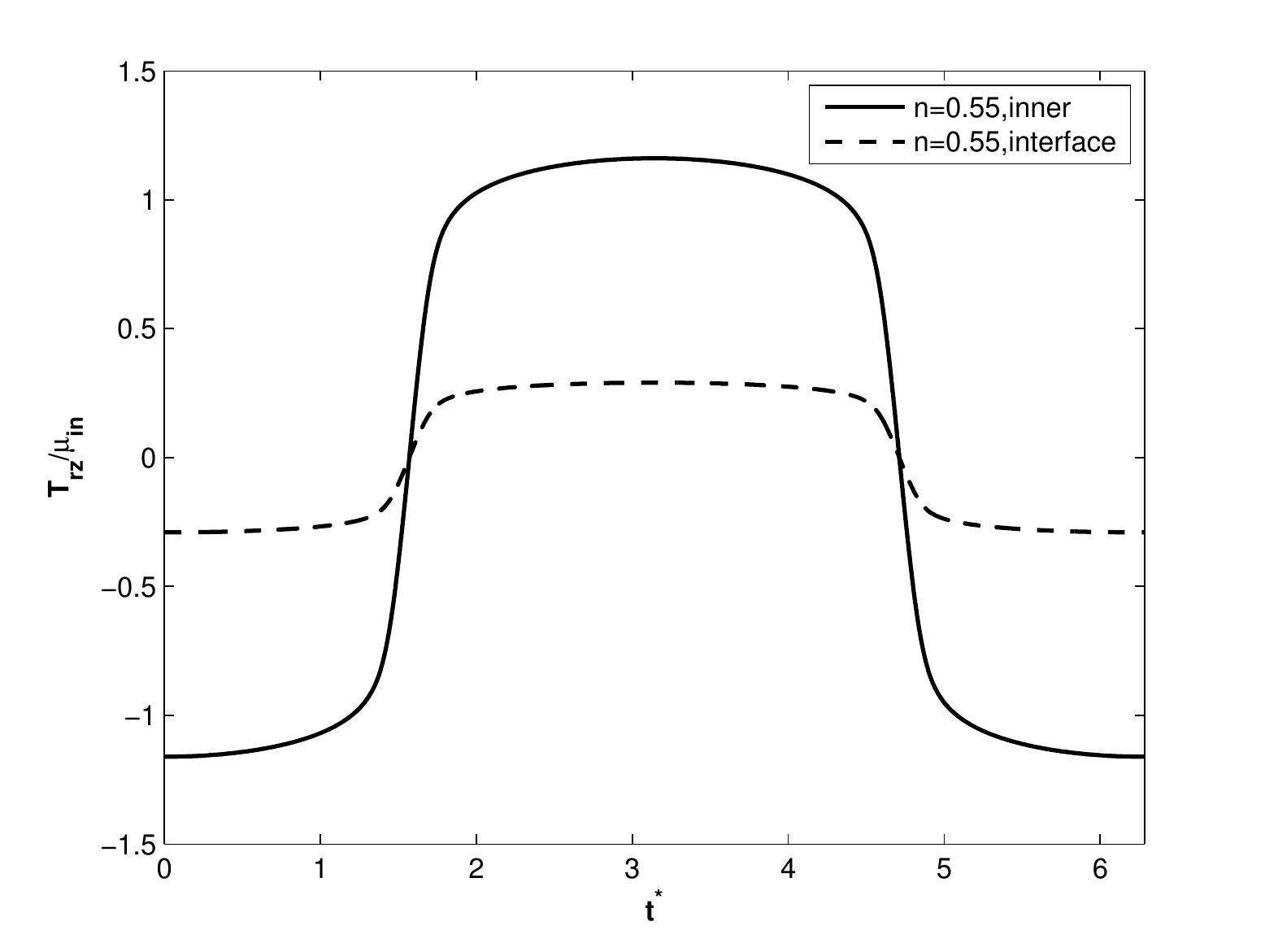}
\label{Ri_1_2_stress_modulus_ratio_1_10_n_055}
\end{subfigure}
% =================================================
% Caption for entire figure
% =================================================
\caption{Figures \subref{Ri_1_2_stress_modulus_ratio_10_1}.) through \subref{Ri_1_2_stress_modulus_ratio_1_10_n_055} show the normalized shear stress at the inner surface and the interface between inner and outer layers as a function of normalized time.}
\label{Ri_1_2_stress_infinite}
%\vspace{-20pt}
% =================================================
\end{figure}

 For dynamic simulations, we vary the material parameters $\mu$, $n$ and $b$ to alter the inertia term $\omega_L$(i.e., $\mu$), and the stiffening or softening characteristics of the layer(i.e., $n$ and $b$). Firstly, $\mu_{in}$ is held fixed at 50 kPa matching the single-layered case and $\mu_{out}$ depends on the stiffness ratio (SR) $\mu_{in}$:$\mu_{out}$.  As an example if $\mu_{in}$:$\mu_{out}$ = 10:1 then the inner stiffness is ten times as large as the outer stiffness; thus the outer layer stiffness would be 5 kPa.  If (SR = 1:10) the outer layer is ten times as stiff as the inner layer and the outer layer stiffness is 500 kPa.  Results in these cases are used for comparison with the results in finite length cylindrical annulus. Figure \ref{Ri_1_2_stress_infinite} shows the variation in shear stress at the inner surface and the interface of the inner and outer layers for an annulus with inner and outer radius 1.2 mm and 12.0 mm respectively.  Studies were performed for thin-walled annuli (not shown), and similar results were obtained.

It can be observed that a stiffer outer layer produces lower stresses compared to a softer outer layer.  Similar to the case of a single-layered cylinder, thick-walled cylinder produces considerable variation in shear stress across the wall.  

\par
\begin{figure}[H]
% =================================================
% R_i 1.2 mm
% =================================================
\begin{subfigure}{.49\linewidth}
\centering
\caption{$n$ = 1,5, at $t^*=2\pi$}
\includegraphics[width=\textwidth]{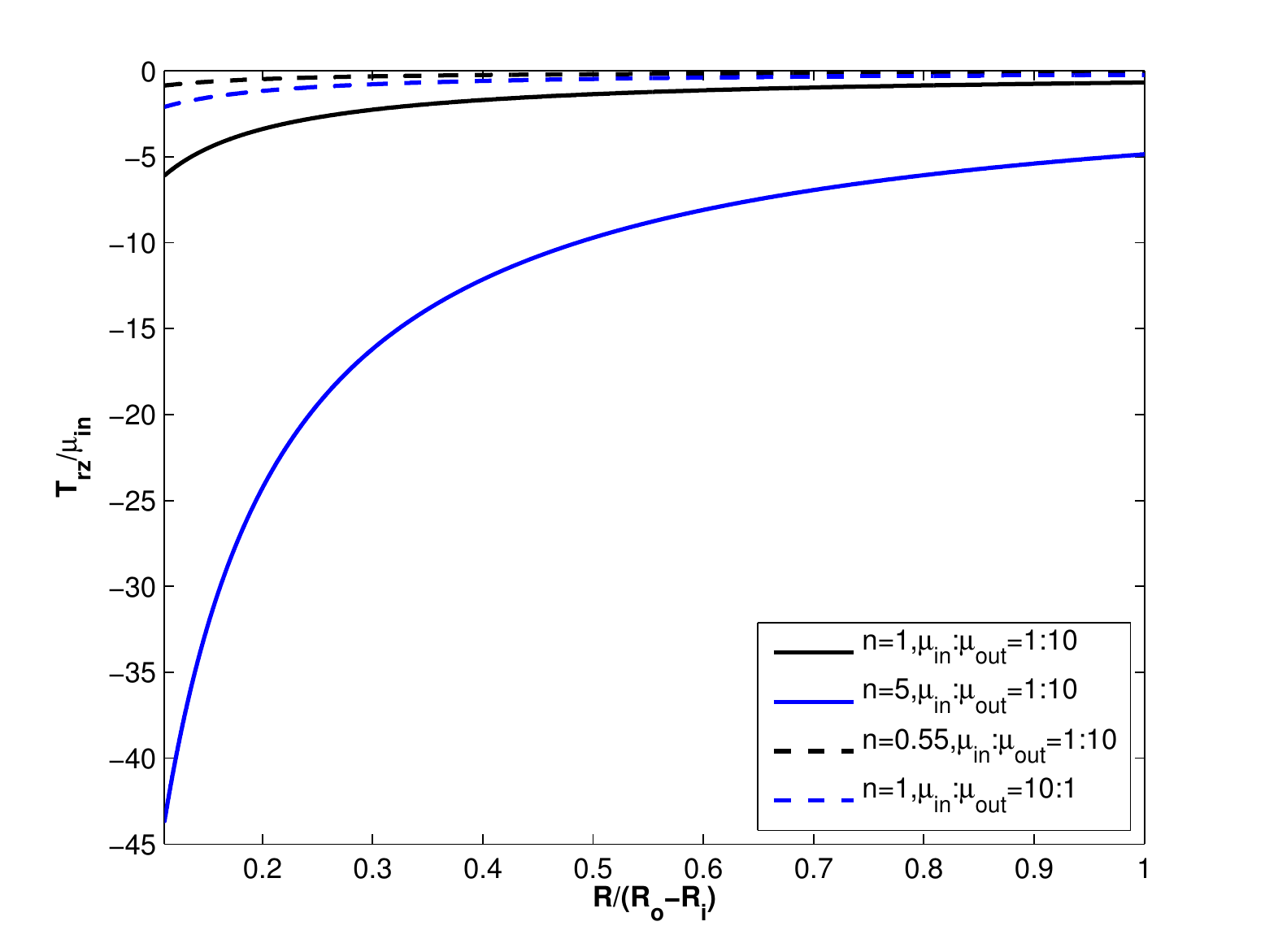}
\label{Ri_1_2_stress_1s}
\end{subfigure}
% =================================================
% R_i 1.2 mm
% =================================================
\begin{subfigure}{.49\linewidth}
\centering
\caption{$n$ = 0.55, at $t^*=2\pi$}
\includegraphics[width=\textwidth]{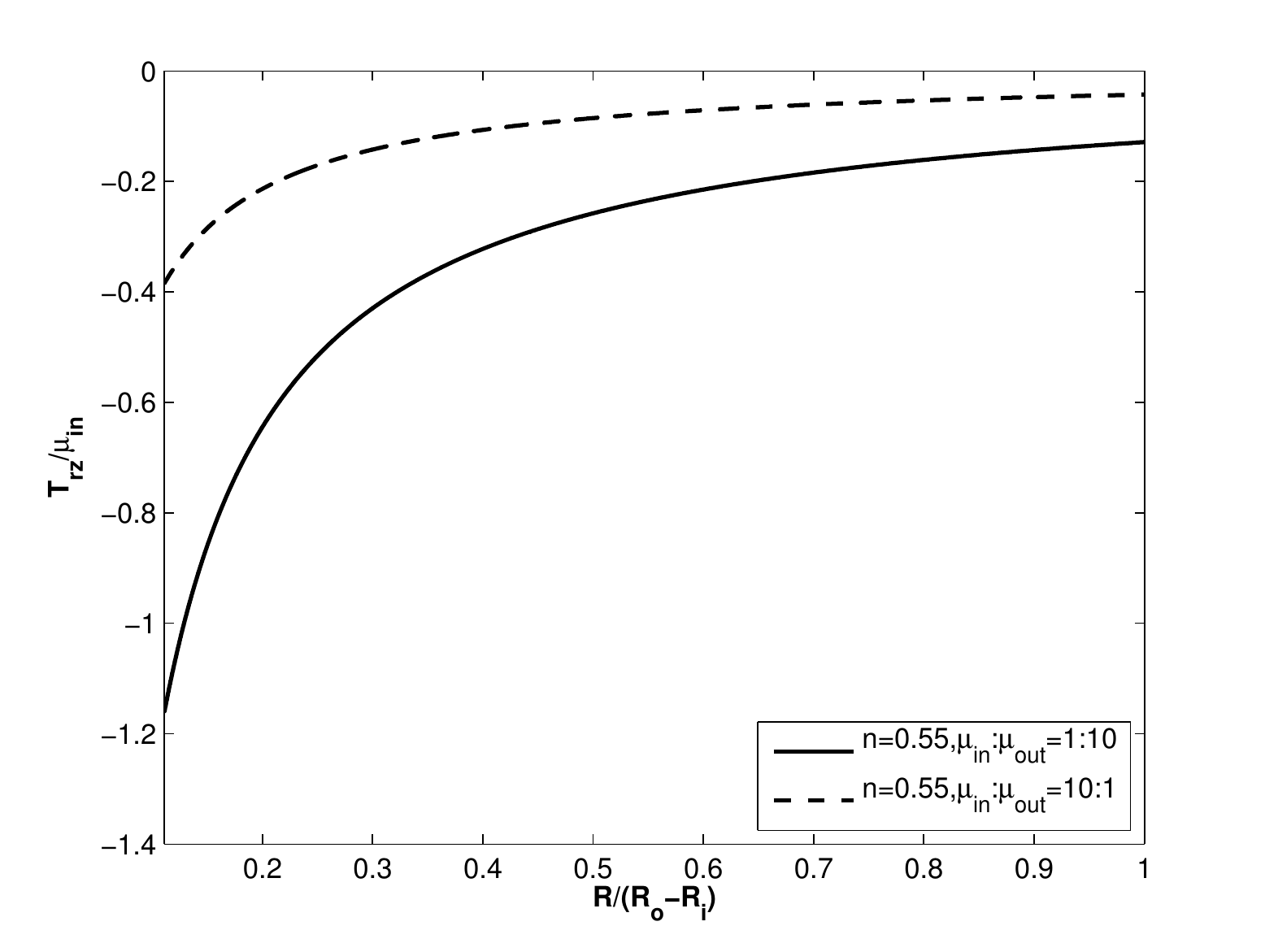}
\label{Ri_1_2_stress_1s_n_055}
\end{subfigure}\\[1ex]
% =================================================
% Stress vs Strain
% =================================================
\begin{subfigure}{.5\linewidth}
\centering
\caption{$n$ = 1,5, at $t^*=2\pi$}
\includegraphics[width=\textwidth]{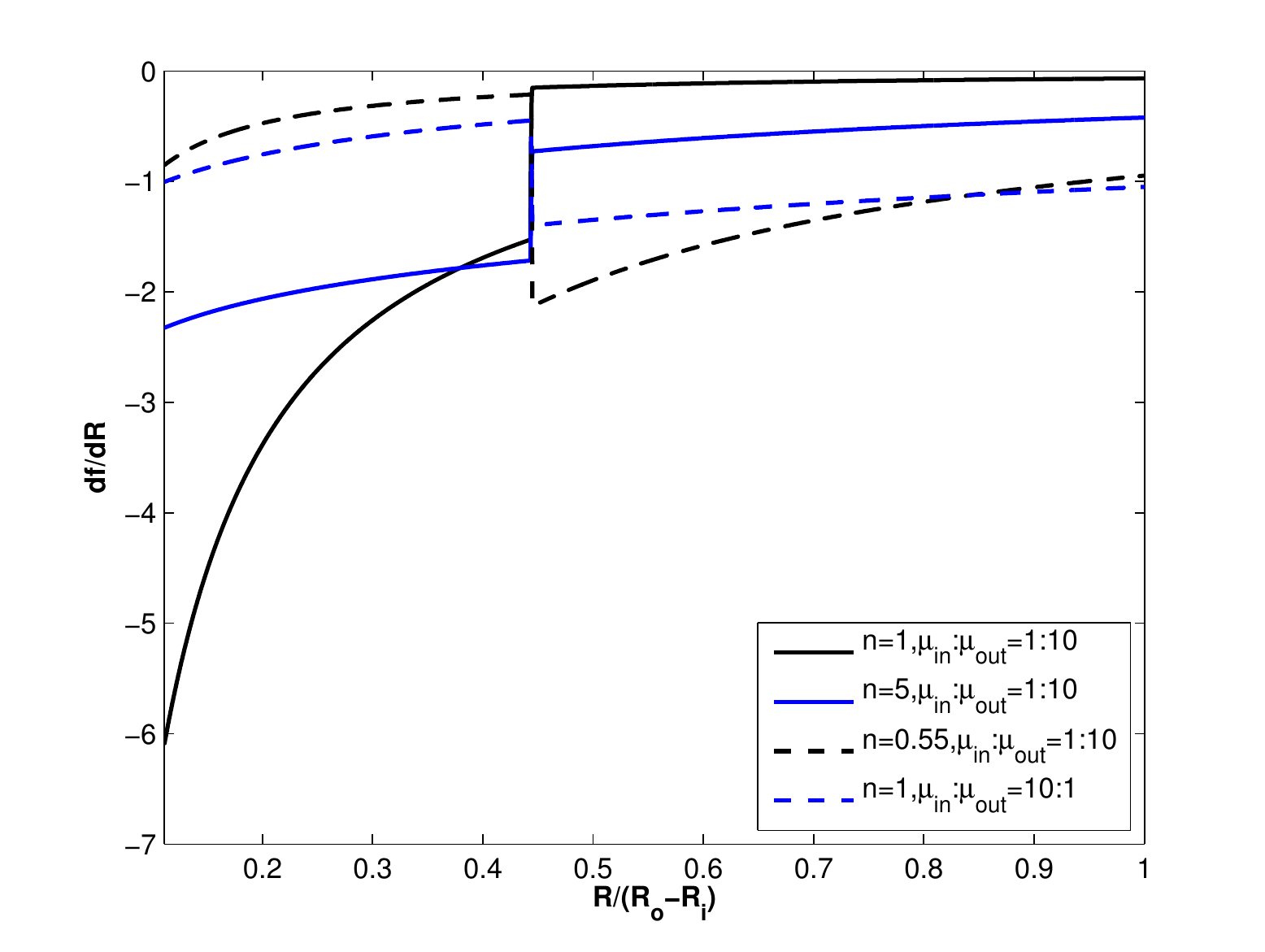}
\label{Ri_1_2_strain_1s}
\end{subfigure}
% =================================================
% Stress vs Strain
% =================================================
\begin{subfigure}{.5\linewidth}
\centering
\caption{$n$ = 0.55, at $t^*=2\pi$}
\includegraphics[width=\textwidth]{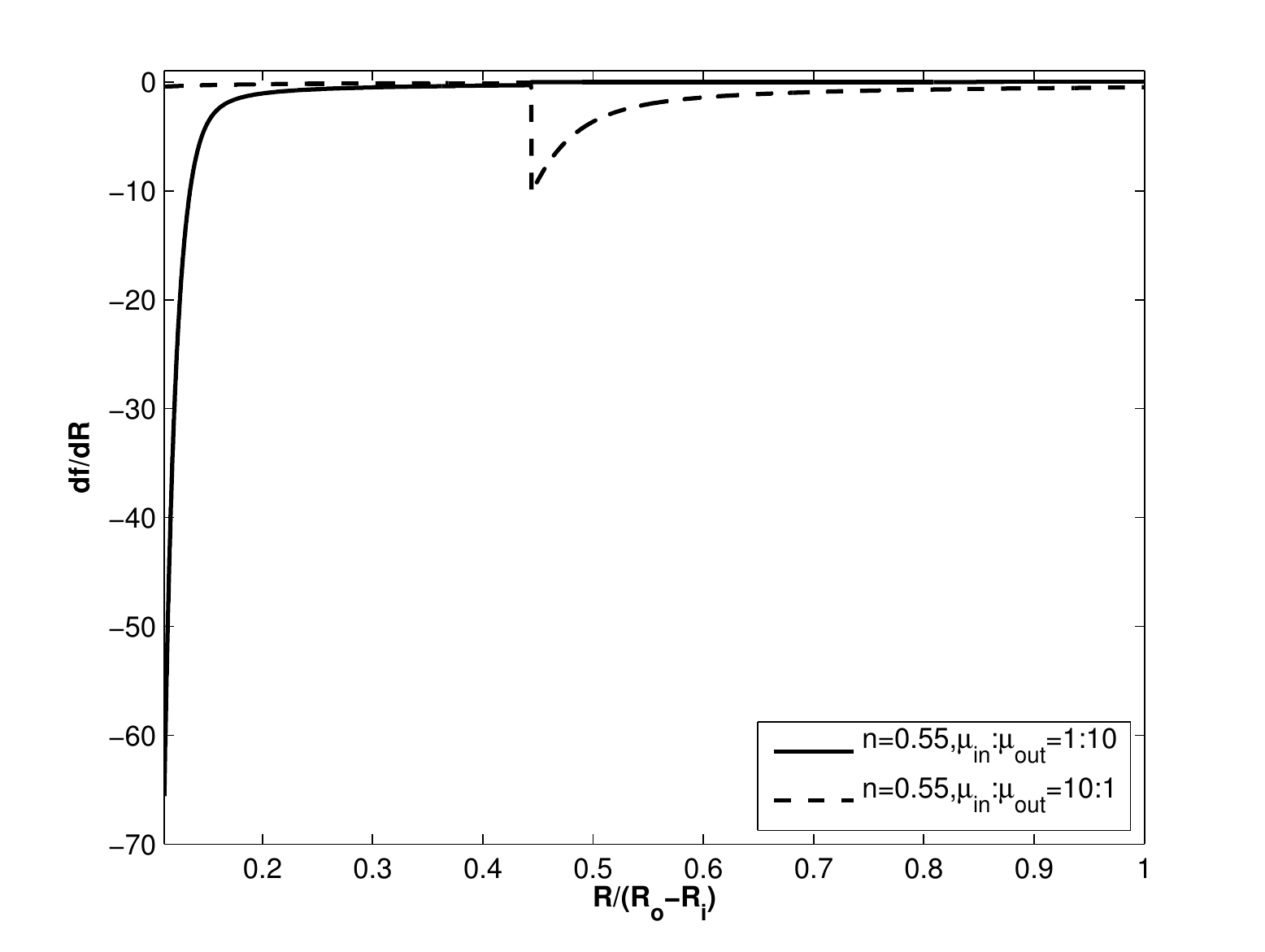}
\label{Ri_1_2_strain_1s_n_055}
\end{subfigure}
% =================================================
% Caption for entire figure
% =================================================
\caption{Figures \subref{Ri_1_2_stress_1s}.) and \subref{Ri_1_2_stress_1s_n_055}.) show the normalized shear stress across the wall thickness at $t^*=2\pi$.  Figures \subref{Ri_1_2_strain_1s}.) and \subref{Ri_1_2_strain_1s_n_055}.) shows the shear strain across the wall thickness at $t^*=2\pi$.}
\label{Ri_1_2_stress_strain_1s}
%\vspace{-20pt}
% =================================================
\end{figure}

 Figure \ref{Ri_1_2_stress_strain_1s} shows the variation of shear stress and shear strain across the wall at $\tilde{t}=2\pi$ for an annulus with inner radius 1.2 mm and outer radius of 12.0 mm.  Even though a stiffer outer layer reduces the stress on the inner surface as well as the interface, it increases the shear strain in the outer layer at the interface.

\subsection{On the development of boundary layers in composite cylindrical annulus}

In the previous section, we saw that a softer outer layer having lower $\mu$ produces the most substantial shear strain at the interface of the two layers.  This shear strain can lead to debonding.  In this section, we examine this further by changing the softening and stiffening properties(i.e., $n$ and $b$) of the inner and outer layers. Both the parameters control the rate of softening or stiffening. To simplify the analysis without missing the essential details, the inner layer is considered to be neo-Hookean and outer layer is considered to be softening($n$ = 0.55) and stiffening ($n$ = 5).

\begin{figure}[H]
% =================================================
% R_i 1.2 mm varying b n=0.55
% =================================================
\begin{subfigure}{.49\linewidth}
\centering
\caption{b varied, $\mu$ held constant}
\includegraphics[width=\textwidth]{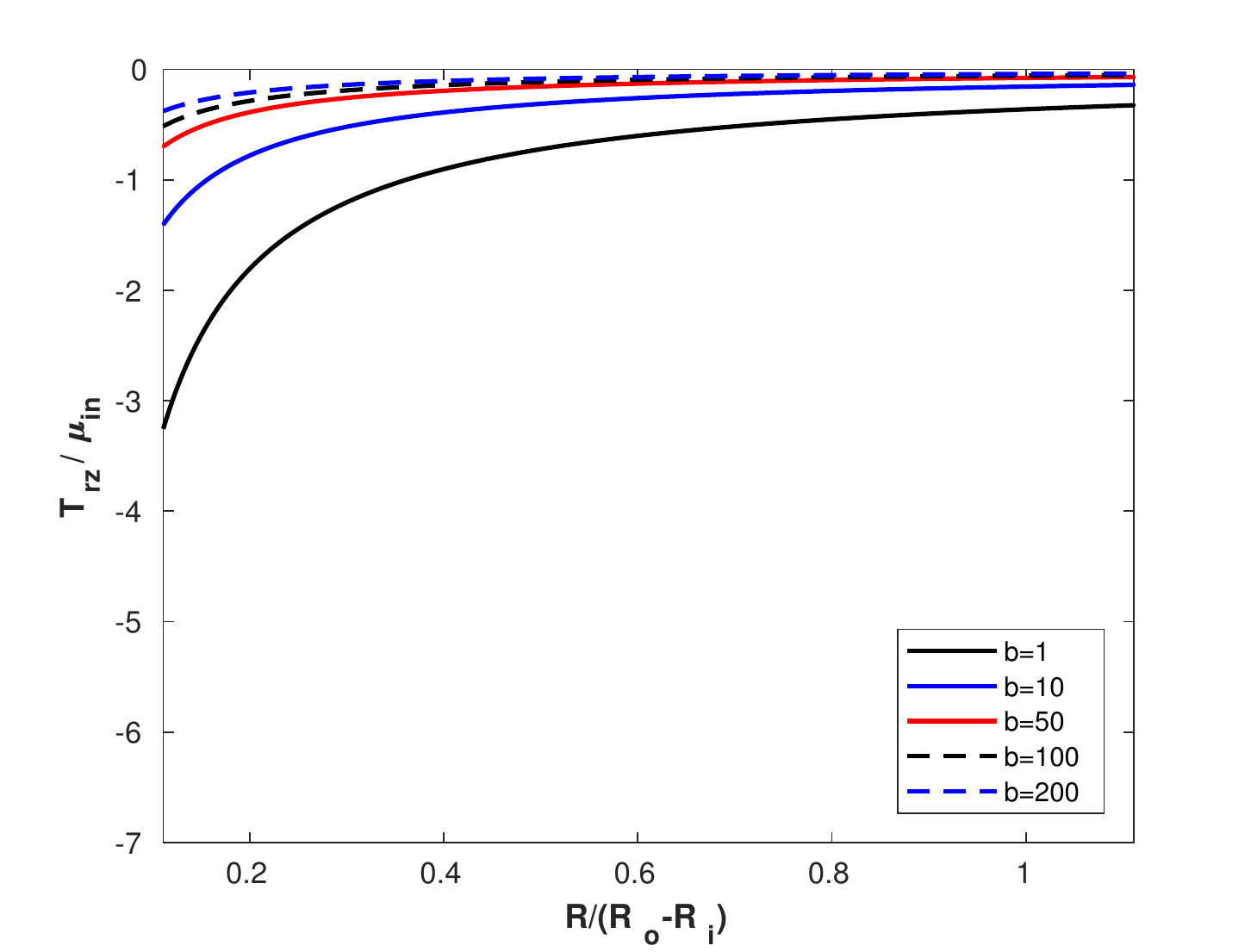}
\label{Ri_1_2_stress_1s_n_055_varying_b}
\end{subfigure}
% =================================================
% R_i 1.2 mm varying mu n=0.55
% =================================================
\begin{subfigure}{.49\linewidth}
\centering
\caption{b held constant, $\mu$ varied}
\includegraphics[width=\textwidth]{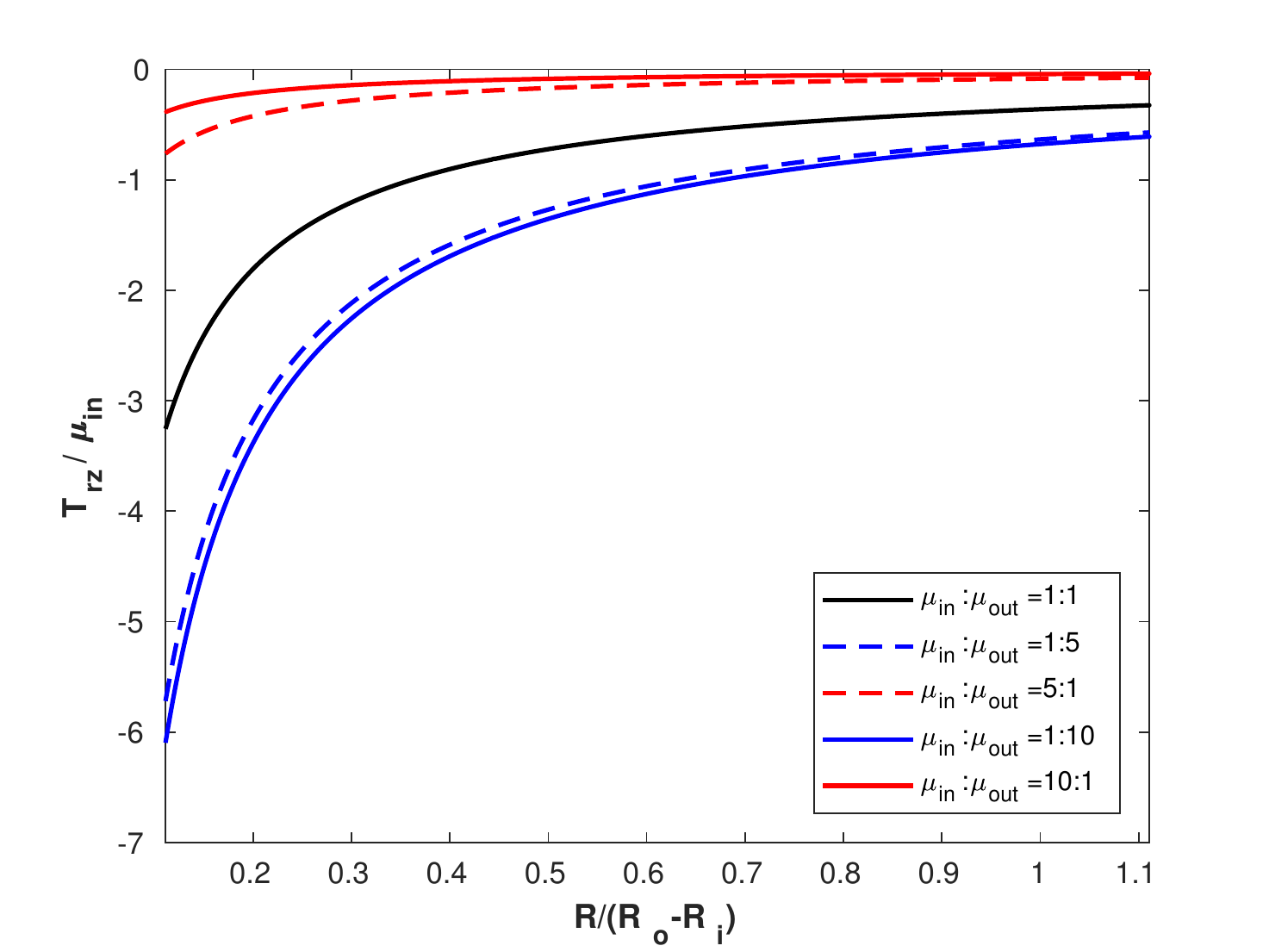}
\label{Ri_1_2_stress_1s_n_055_varying_mu}
\end{subfigure}
% =================================================
% Caption for entire figure
% =================================================
\caption{Normalized stress is plotted as a function of normalized radius for $n$ = 0.55.  The parameter $b$ is varied in figure \subref{Ri_1_2_stress_1s_n_055_varying_b}.) while the parameter $\mu$ is held constant at 50 kPa.  In figure \subref{Ri_1_2_stress_1s_n_055_varying_mu}.) the parameter $\mu$ is varied while $b$ is held constant at 1.}
\label{Ri_1_2_stress_strain_1s_n_055_varying_b}
%\vspace{-20pt}
% =================================================
\end{figure}

\begin{figure}[H]
% =================================================
% Stress vs Strain
% =================================================
\begin{subfigure}{.5\linewidth}
\centering
\caption{b varied, $\mu$ held constant}
\includegraphics[width=\textwidth]{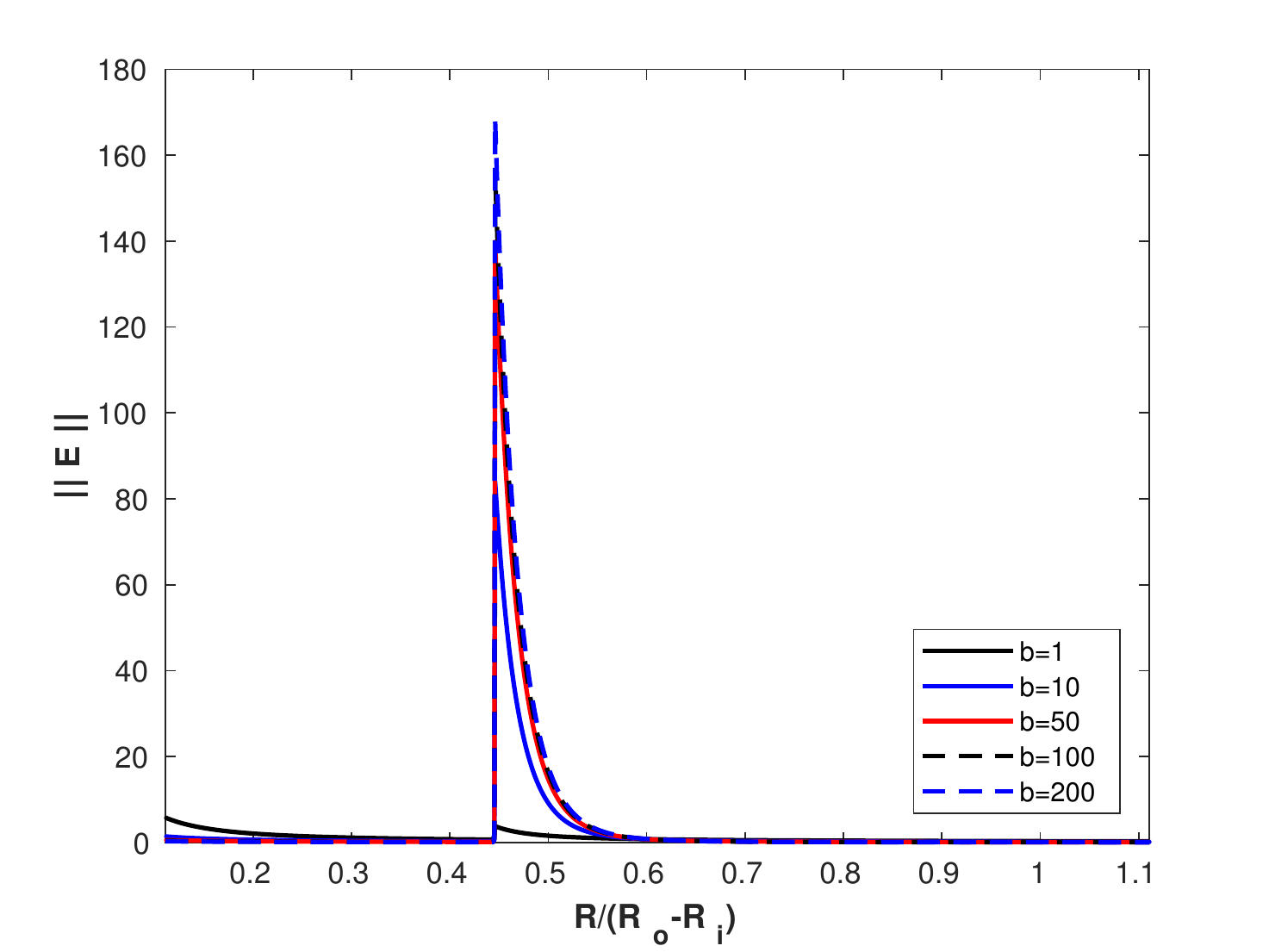}
\label{Ri_1_2_strain_1s_n_055_varying_b}
\end{subfigure}%\\[1ex]
% =================================================
% Stress vs Strain
% =================================================
\begin{subfigure}{.5\linewidth}
\centering
\caption{b held constant, $\mu$ varied}
\includegraphics[width=\textwidth]{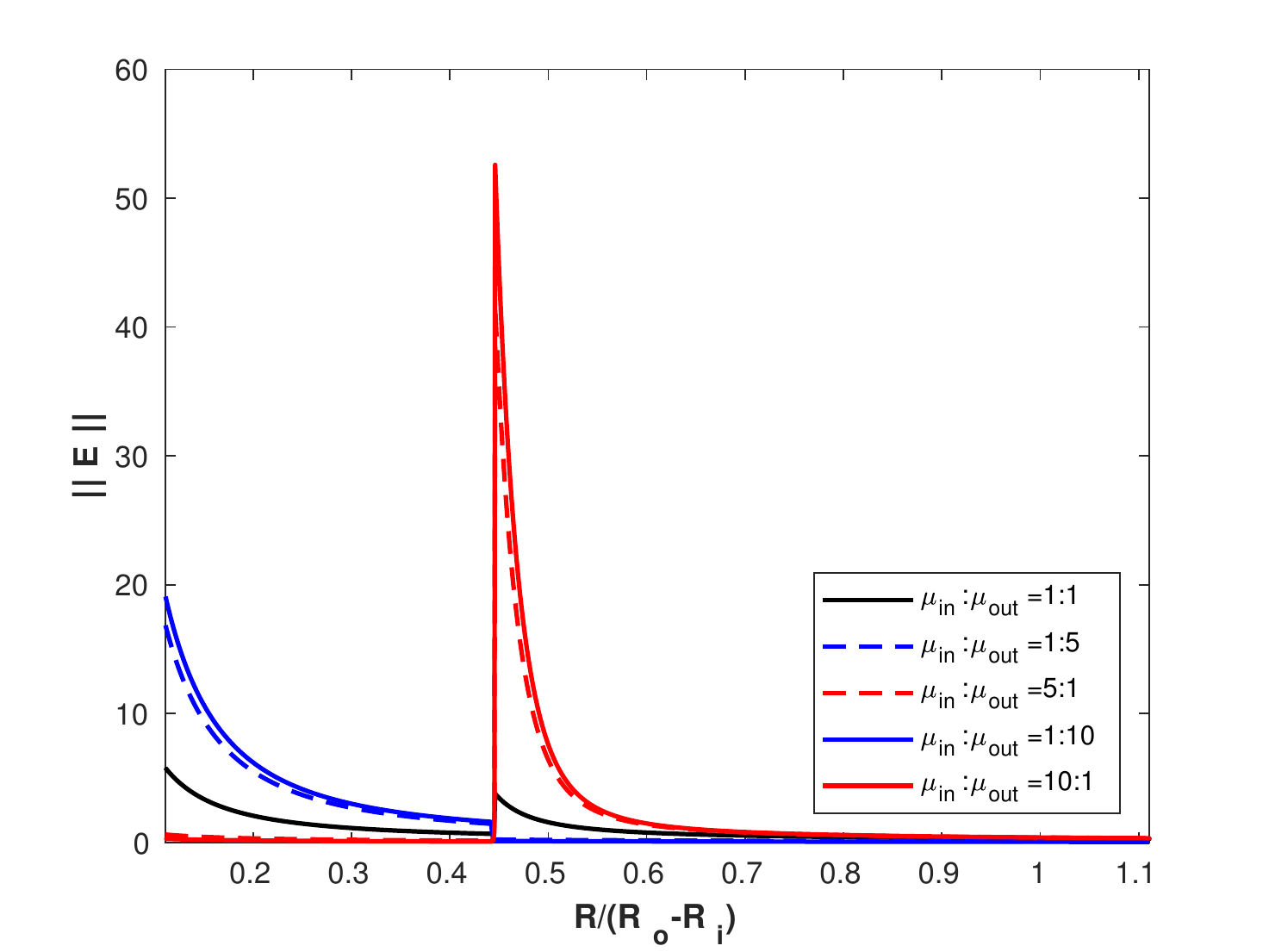}
\label{Ri_1_2_strain_1s_n_055_varying_mu}
\end{subfigure}
% =================================================
% Caption for entire figure
% =================================================
\caption{Normalized strain is plotted as a function of normalized radius for $n$ = 0.55.  The parameter $b$ is varied in figure \subref{Ri_1_2_strain_1s_n_055_varying_b}.) while the parameter $\mu$ is held constant at 50 kPa.  In figure \subref{Ri_1_2_strain_1s_n_055_varying_mu}.) the parameter $\mu$ is varied while $b$ is held constant at 1.}
\label{Ri_1_2_stress_strain_1s_n_055_varying_mu}
%\vspace{-20pt}
% =================================================
\end{figure}

\begin{figure}[H]
% =================================================
% R_i 1.2 mm,n=0.55, displacement
% =================================================
\begin{subfigure}{.49\linewidth}
\centering
\caption{b varied, $\mu$ held constant}
\includegraphics[width=\textwidth]{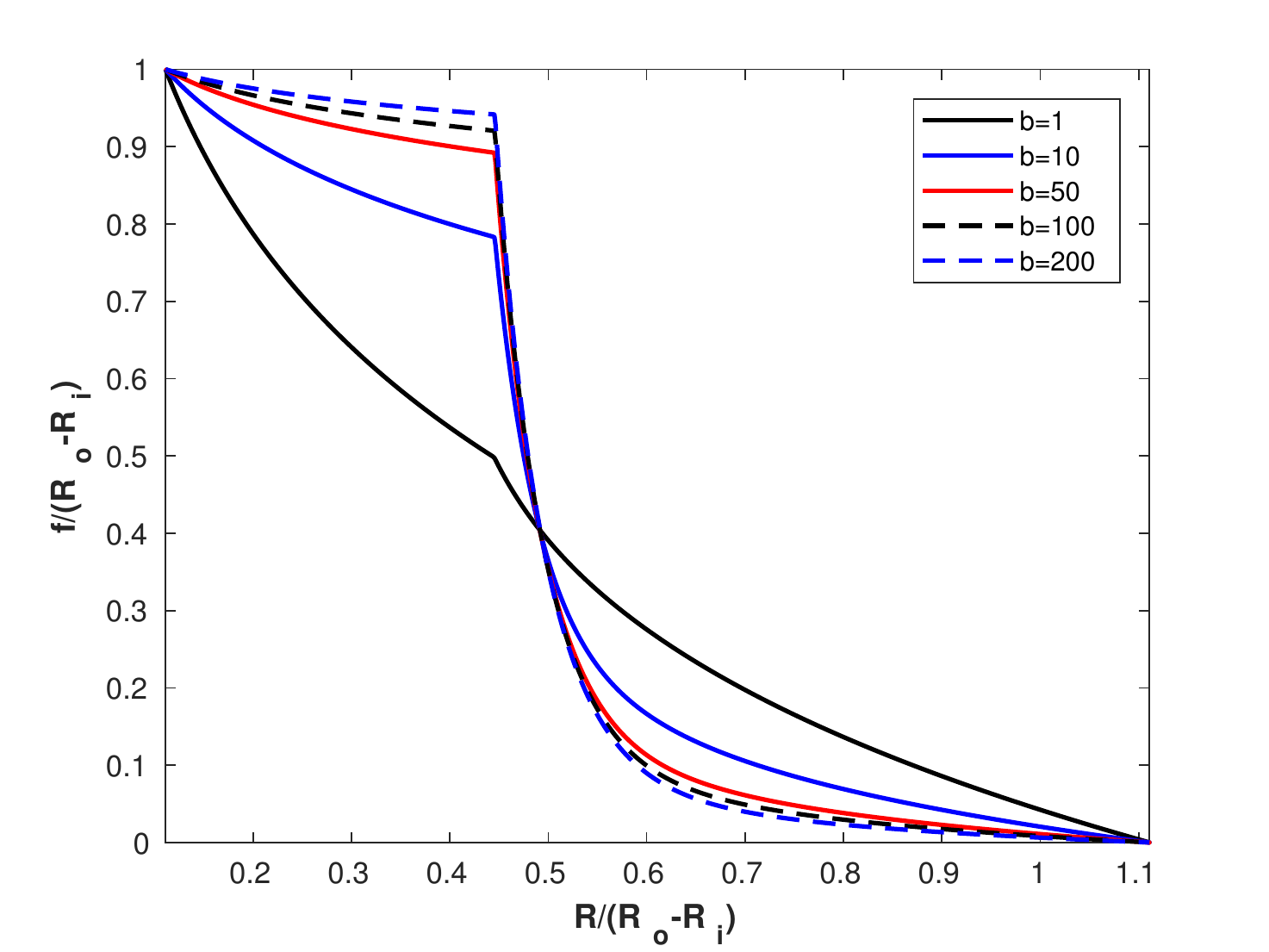}
\label{Ri_1_2_displacement_1s_n_055_varying_b}
\end{subfigure}
% =================================================
% Stress vs Strain
% =================================================
\begin{subfigure}{.5\linewidth}
\centering
\caption{b held constant, $\mu$ varied}
\includegraphics[width=\textwidth]{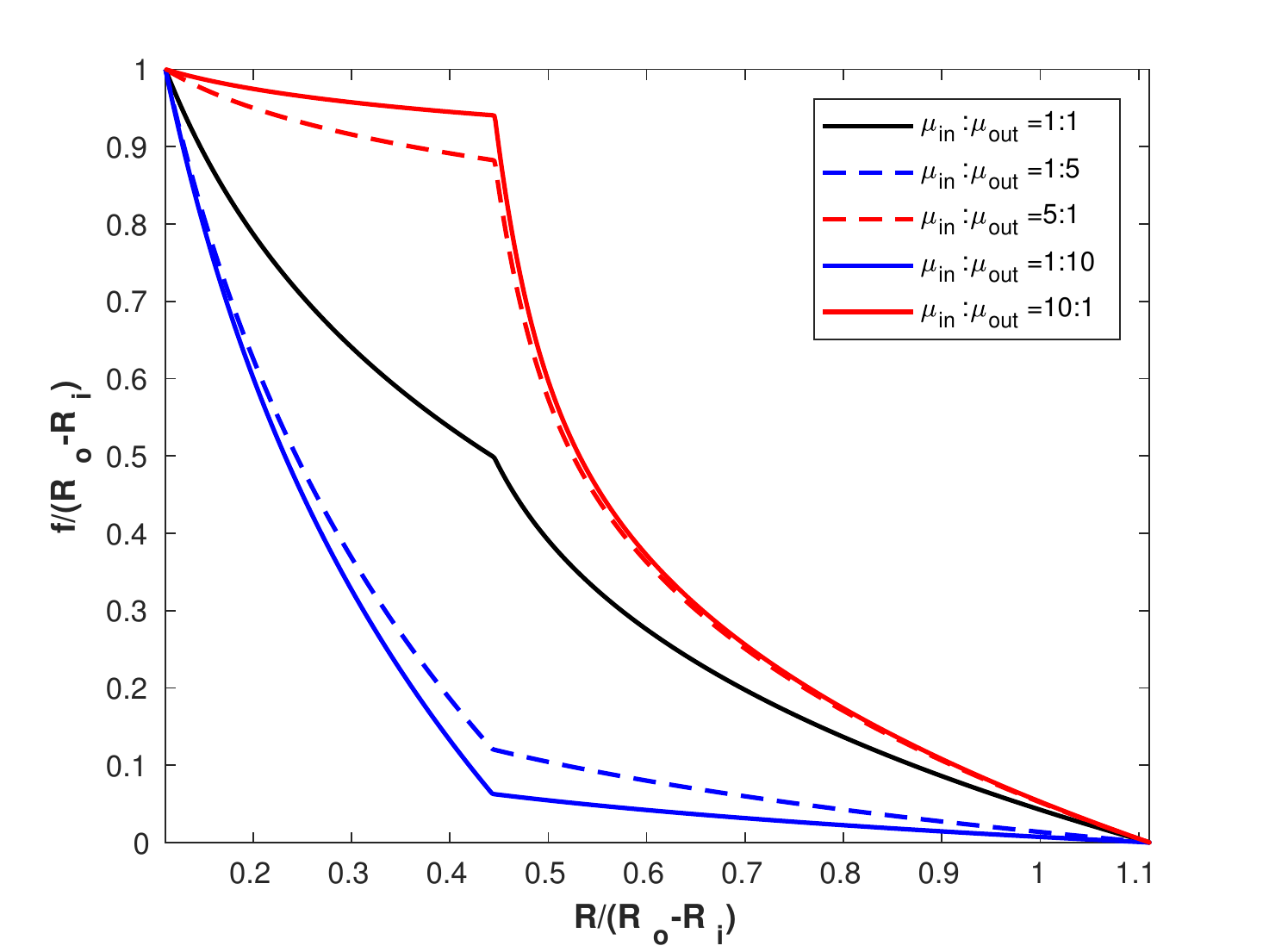}
\label{Ri_1_2_displacement_1s_n_055_varying_mu}
\end{subfigure}
% =================================================
% Caption for entire figure
% =================================================
\caption{Normalized displacement is plotted as a function of normalized radius for $n$ = 0.55.  The parameter $b$ is varied in figure \subref{Ri_1_2_displacement_1s_n_055_varying_b}.) while the parameter $\mu$ is held constant at 50 kPa.  In figure \subref{Ri_1_2_displacement_1s_n_055_varying_mu}.) the parameter $\mu$ is varied while $b$ is held constant at 1.}
\label{Ri_1_2_displacement_1s_n_055}
%\vspace{-20pt}
% =================================================
\end{figure}

Figures \ref{Ri_1_2_stress_strain_1s_n_055_varying_b}, \ref{Ri_1_2_stress_strain_1s_n_055_varying_mu} and \ref{Ri_1_2_displacement_1s_n_055} show the normalized stress, strain and displacement as a function of normalized radius respectively for varying $b$ and stiffness ratios ($\mu_{in}$:$\mu_{out}$) for $n$ = 0.55 at $t^*=2\pi$.  We can observe the formation of a boundary layer near the interface of the layers with increasing $b$. Stress distribution is controlled purely by geometry of the cylinder as discussed earlier. Strain, on the other hand, depends on material parameters in addition to geometry. Similar appearance of boundary layers at the interface can be seen in figures \ref{Ri_1_2_stress_1s_n_055_varying_mu} and \ref{Ri_1_2_strain_1s_n_055_varying_mu} in which $\mu$ is varied for an outer layer with $n=0.55$. Displacement plots in figure \ref{Ri_1_2_displacement_1s_n_055} shows the development of boundary layers near the interface.

% =================================================
% Stresses
% =================================================

% =================================================
% Strains
% =================================================

% =================================================
% Displacements
% =================================================

\section{Deformation of a two-layered finite length cylindrical annulus due to periodic longitudinal shearing}\label{Sect:07}

\begin{figure}[H]
% =================================================
% R_i 10 mm
% =================================================
\begin{subfigure}{.5\linewidth}
\centering
\caption{TR = 1.2, $R_o=12$ mm}
\includegraphics[width=.9\textwidth]{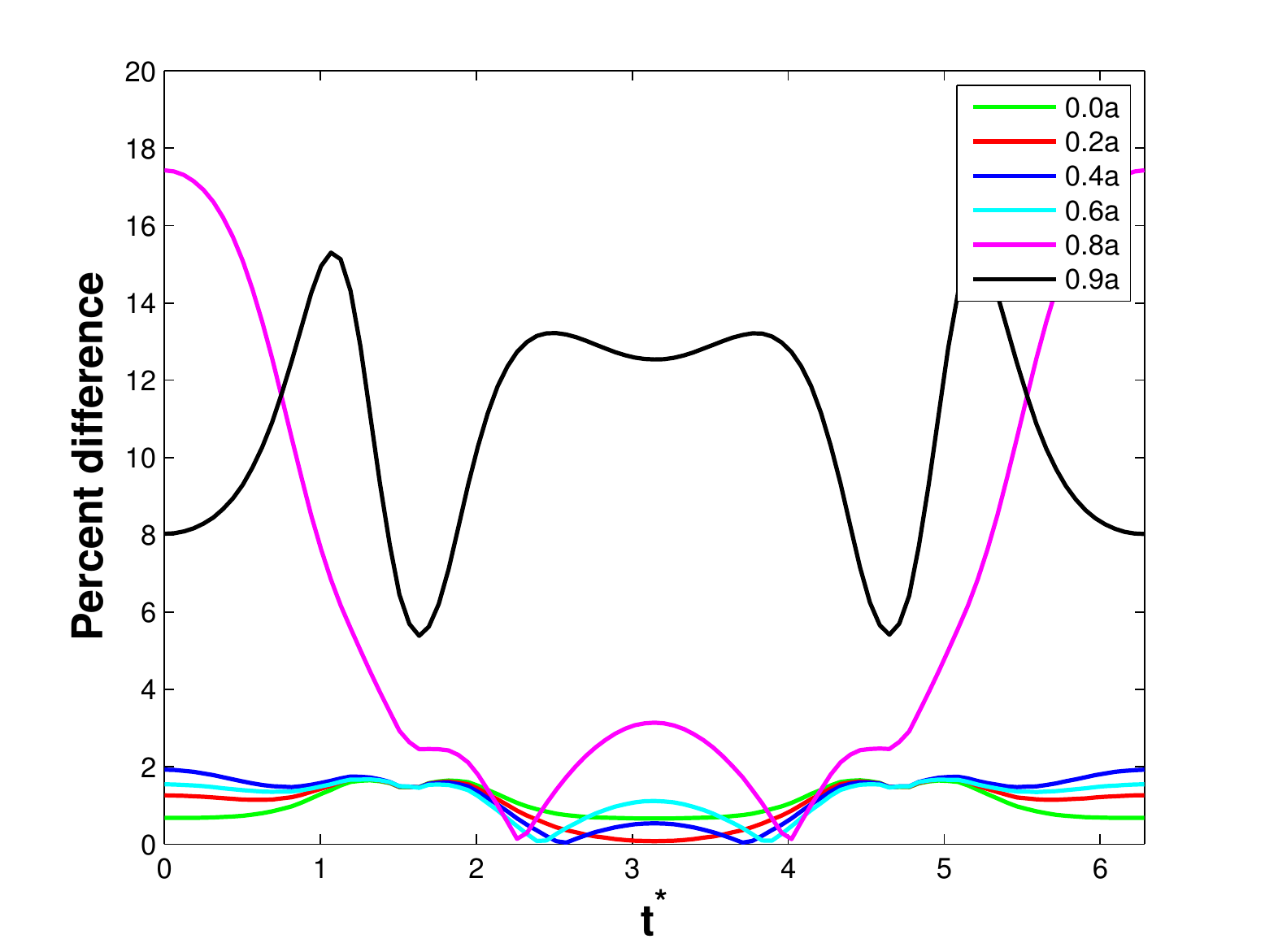}
\label{Ri_10_n_5_1_10_time}
\end{subfigure}
% =================================================
% R_i 1.2 mm
% =================================================
\begin{subfigure}{.49\linewidth}
\centering
\caption{TR = 10, $R_o = 12$ mm}
\includegraphics[width=.9\textwidth]{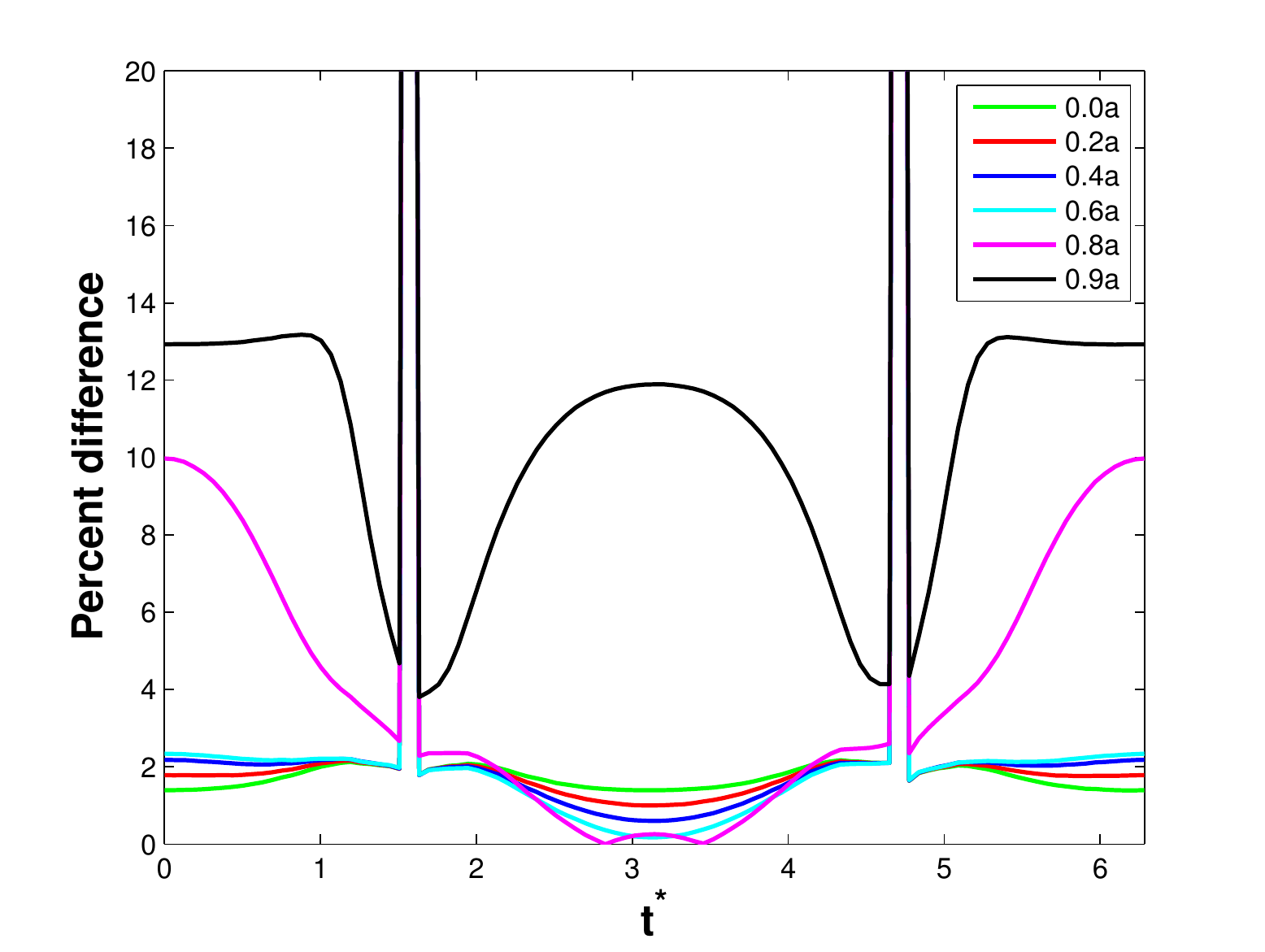}
\label{Ri_1_2_n_5_1_10_time}
\end{subfigure}\\[1ex]
% =================================================
% Stress vs Strain
% =================================================
\begin{subfigure}{.5\linewidth}
\centering
\caption{TR = 1.2, $R_o=12$ mm}
\includegraphics[width=.9\textwidth]{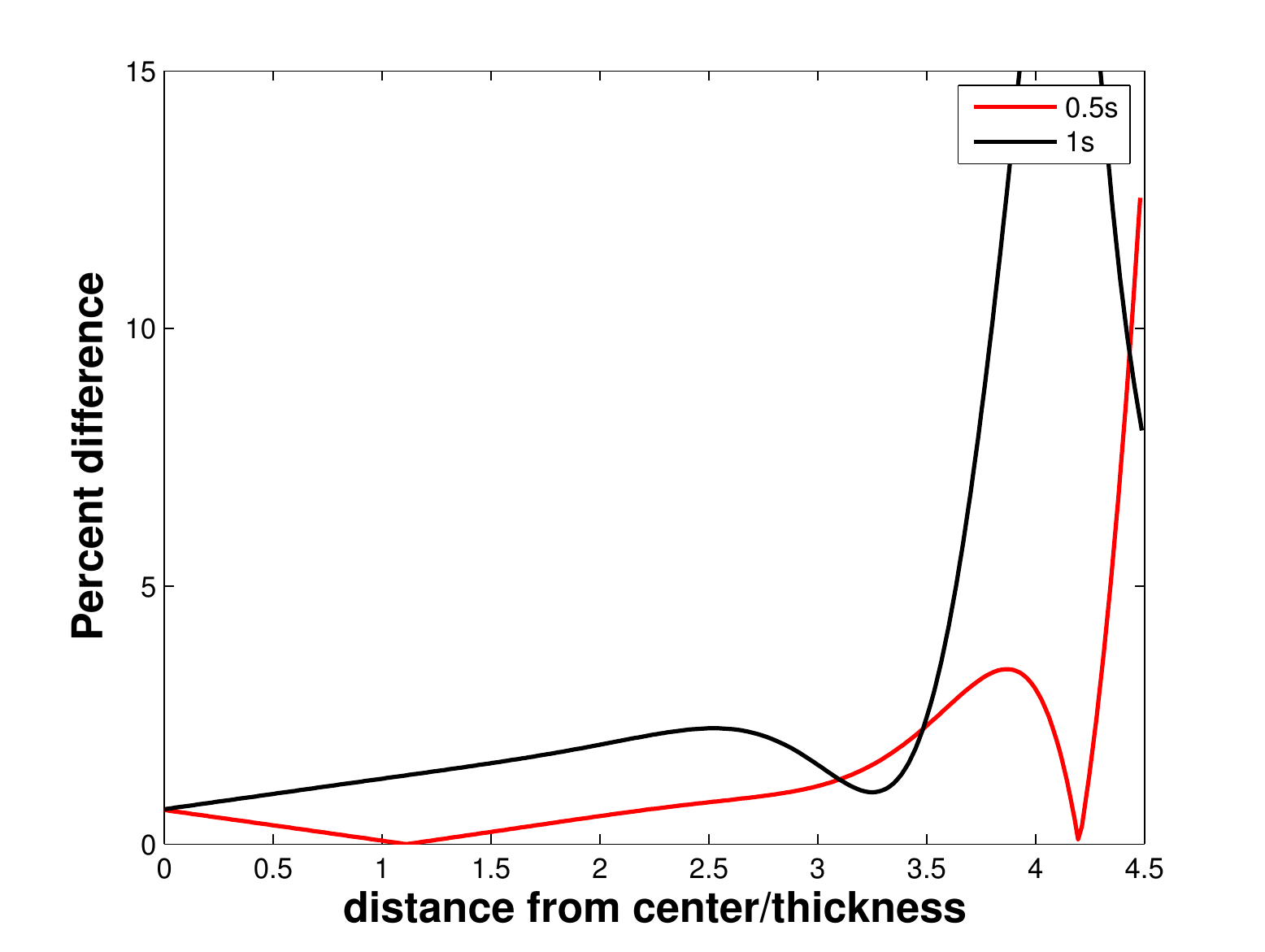}
\label{Ri_10_n_5_1_10_along_z}
\end{subfigure}
% =================================================
% Stress vs Strain
% =================================================
\begin{subfigure}{.5\linewidth}
\centering
\caption{TR = 10, $R_o = 12$ mm}
\includegraphics[width=.9\textwidth]{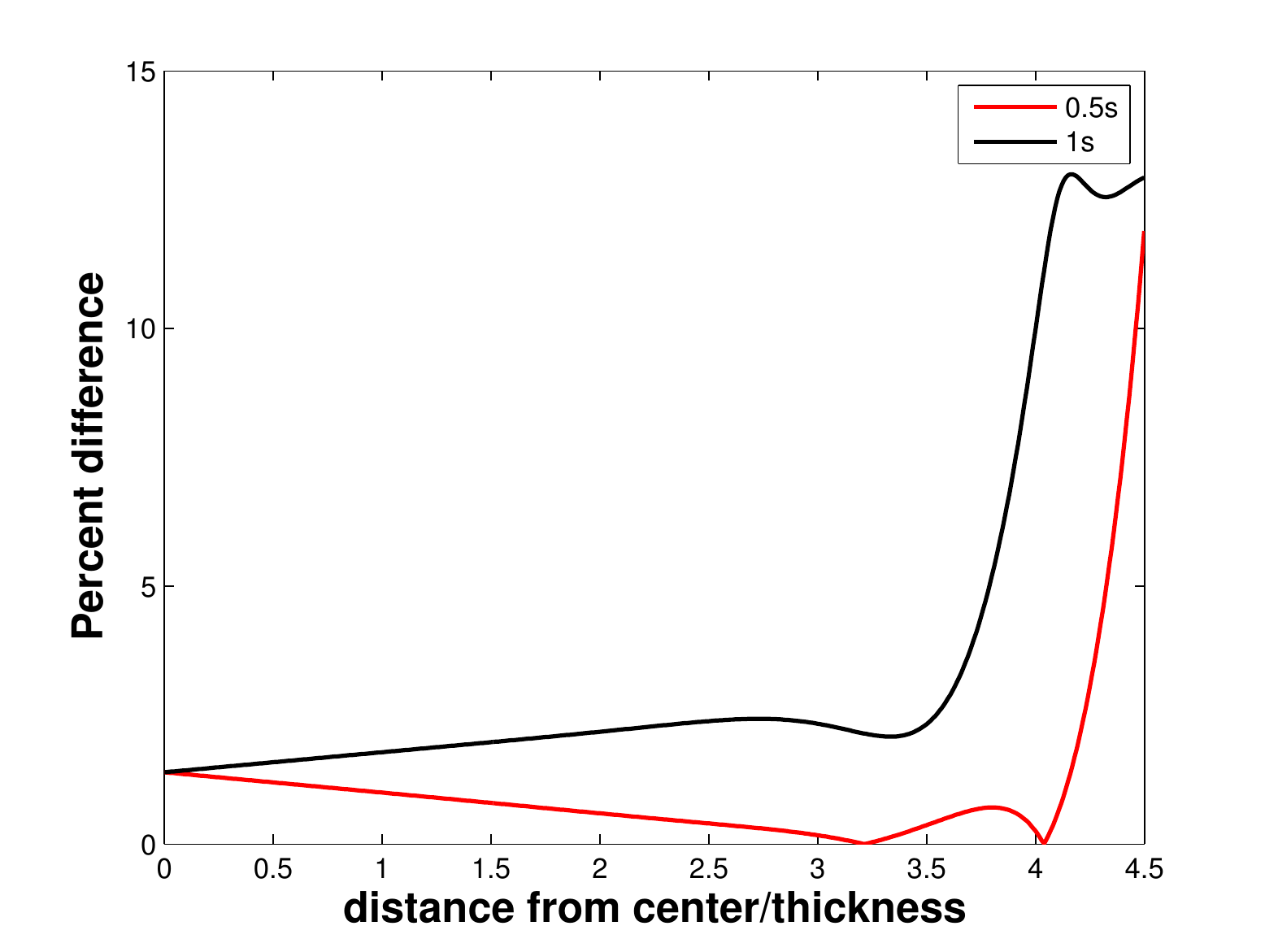}
\label{Ri_1_2_n_5_1_10_along_z}
\end{subfigure}
% =================================================
% Caption for entire figure
% =================================================
\caption{In figures \subref{Ri_10_n_5_1_10_time}.) and \subref{Ri_1_2_n_5_1_10_time}.) the percent difference in stress between the infinite and finite length solution at various distances from center on the inner surface is given as a function of time.  In figures \subref{Ri_10_n_5_1_10_along_z}.) and \subref{Ri_1_2_n_5_1_10_along_z}.) the percent difference in comparison to the infinite case at 0.5s and 1s along the length of the cylinder on the inner surface is given.  The percent difference is reported for n = 5.00, $\mu_{in}:\mu_{out}=1:10$}
\label{Ri_10_n_5_1_10}
%\vspace{-20pt}
% =================================================
\end{figure}

Motion was studied for three different values of $n$, (0.55, 1.0, 5.0) and $b=1$. The inner layer's material constant $\mu_{in}$ is maintained at 50 kPa in all the simulations whereas the outer layer's material constant $\mu_{out}$ was set at 5 kPa or 500 kPa depending on the simulation.  In figure \ref{Ri_10_n_5_1_10} we show the percent difference in shear stress between the finite solution and the infinite solutions as a function of time and axial position (Z) respectively for the strain hardening case ($n$ = 5).  The total length of the finite annulus is ($2a$) with a measured distance of ($a$) from the center of the annulus to the end.  

Similar to the neo-Hookean case, stiffening materials with n = 5 show negligible deviation from the infinite case after 0.4a away from the ends even for an annulus with (TR = 10). It can also be observed that the deviation of the finite length solution from the infinite length solution is less in an annulus with (TR = 10) than one with (TR = 1.2). Similar results (not shown here) are obtained for ($n$ = 1,5) and when $\mu_{in}/\mu_{out}$ is changed to 10:1.

% Discussion
\section{Discussion}

In this manuscript, we studied the motion of a generalized neo-Hookean annular cylinder of infinite and finite length, comprised of a single layer or two layers subjected to longitudinal shearing on the inner surface.  The power-law exponent $n$ represents shear softening or shear stiffening behavior.  The parameter $\mu$ represents the small strain modulus.  The parameter $b$ represents the degree of shear softening or shear stiffening.

For the infinite case, geometry played a substantial role in the development of stress boundary layers.  A stress boundary layer is formed in the thick-walled cylinders (TR > 2) for strain hardening as well as strain softening materials.  Displacements displayed boundary-layer type solutions for the strain softening case, (n < 1.0). Strain boundary layers were introduced in thick walled cylinders for strain softening conditions and in thin walled cylinders when the parameter $b$ became very large.  Our investigation has shown that the norm of the shear strain $\vert\vert\mathbf{E}\vert\vert$ becomes asymptotic as the exponent $n$ tends to a limiting value of 0.5 (see figure \ref{SD:D}).  Strain and displacement (see figure \ref{SD:B}) boundary layers form for values of $n$ approaching 0.5 for thick walled cylinders ($R_o/R_i$=50).  

In a region that is two times the thickness (2t) from the cylinder ends the solution to the infinite annular cylinder and the solutions to the finite annular cylinder do not match.  However, beyond this distance the solutions to the finite annular cylinder is within 2\% of the solution to the infinite annular cylinder.  This corresponds to a region that is 60\% of the length for a cylinder with a length to thickness (LTR) of 10:1.  For a cylinder with an LTR of 5:1 this corresponds to a region that is 20\% of the cylinder length.  So as the cylinder gets longer the region of applicability increases.

From this study, several conclusions can be drawn.  It was seen in the quasi-static case and in the dynamic case that stress boundary layers only formed for very thick walled cylinders.  Strain boundary layers occurred for hardening and softening materials alike.  The most striking observation is that the solution for the infinite case has little error associated with it for 60\% of the cylinder length measured from the center of the cylinder when the length of the cylinder is at least ten times the thickness cylinder.  The solutions only seem to differ for a region of a length twice the thickness from the cylinder ends.

\section*{Acknowledgments}

Chandler Benjamin would like to acknowledge helpful conversations with my colleagues Dr. Alan Freed.  This research was funded by the Texas A\&M Engineering Experiment Station.

\section*{References}

% ========================================================================
% References
% ========================================================================

%\bibliographystyle{unsrtnat}
%\bibliography{ArtBib}

\end{document}